\documentclass[prb,amssymb,aps,amsmath,amssymb,twocolumn,superscriptaddress,longbibliography]{revtex4-2}

\usepackage[colorlinks=true,citecolor=blue,linkcolor=blue,urlcolor=blue]{hyperref}
\usepackage{graphicx, nicefrac, textcomp}
\usepackage[utf8]{inputenc}
\usepackage{multirow}
\usepackage{amsmath}
\usepackage{xfrac}
\usepackage{graphicx}
\usepackage{tabularx}
\usepackage{comment}

\usepackage{xcolor}
\usepackage{xr}

\begin{document}

\title{From perovskite to infinite-layer nickelates: hole concentration from x-ray absorption}

\author{R.~Pons}
\affiliation{Max Planck Institute for Solid State Research, Heisenbergstraße 1, 70569 Stuttgart, Germany}

\author{M.~Flavenot}
\affiliation{Universite de Strasbourg, CNRS, IPCMS UMR 7504, F-67034 Strasbourg, France}

\author{K.~Fürsich}
\affiliation{Max Planck Institute for Solid State Research, Heisenbergstraße 1, 70569 Stuttgart, Germany}

\author{E.~Schierle}
\affiliation{Helmholtz-Zentrum Berlin für Materialien und Energie, Albert-Einstein-Straße 15, 12489 Berlin, Germany}

\author{E.~Weschke}
\affiliation{Helmholtz-Zentrum Berlin für Materialien und Energie, Albert-Einstein-Straße 15, 12489 Berlin, Germany}

\author{M.~R.~Cantarino}
\affiliation{European Synchrotron Radiation Facility, F-38043 Grenoble Cedex 9, France}

\author{E.~Goering}
\affiliation{Max Planck Institute for Solid State Research, Heisenbergstraße 1, 70569 Stuttgart, Germany}

\author{P.~Nagel}
\affiliation{Institute for Quantum Materials and Technologies (IQMT) and Karlsruhe Nano and Micro Facility (KNMFi), Karlsruhe Institute of Technology (KIT), 76021 Karlsruhe, Germany}

\author{S.~Schuppler}
\affiliation{Institute for Quantum Materials and Technologies (IQMT) and Karlsruhe Nano and Micro Facility (KNMFi), Karlsruhe Institute of Technology (KIT), 76021 Karlsruhe, Germany}

\author{G.~Kim}
\affiliation{Max Planck Institute for Solid State Research, Heisenbergstraße 1, 70569 Stuttgart, Germany}

\author{G.~Logvenov}
\affiliation{Max Planck Institute for Solid State Research, Heisenbergstraße 1, 70569 Stuttgart, Germany}

\author{B.~Keimer}
\affiliation{Max Planck Institute for Solid State Research, Heisenbergstraße 1, 70569 Stuttgart, Germany}
\affiliation{Center for Integrated Quantum Science and Technology (IQST), Stuttgart, Germany}

\author{R.~J.~Green}
\affiliation{Department of Physics and Engineering Physics, University of Saskatchewan, Saskatoon, SK, Canada S7N 5E2}
\affiliation{Stewart Blusson Quantum Matter Institute, University of British Columbia, Vancouver, BC, Canada V6T 1Z4}

\author{D.~Preziosi}
\affiliation{Universite de Strasbourg, CNRS, IPCMS UMR 7504, F-67034 Strasbourg, France}

\author{E.~Benckiser}
\email{E.Benckiser@fkf.mpg.de}
\affiliation{Max Planck Institute for Solid State Research, Heisenbergstraße 1, 70569 Stuttgart, Germany}
\affiliation{Center for Integrated Quantum Science and Technology (IQST), Stuttgart, Germany}

\date{\today}

\begin{abstract}
The difficulty of determining cation concentrations and oxygen stoichiometry in infinite-layer nickelate thin films has so far prevented clear experimental identification of the nickel electron configuration in the superconducting phase. We used soft x-ray absorption spectroscopy to study the successive changes in PrNiO$_x$ thin films at various intermediate stages of topotactic reduction with $x=2-3$. By comparing the Ni-$L$ edge spectra to single and double cluster ligand-field calculations, we find that none of our samples exhibit a pure $d^9$ configuration. Our quantitative analysis using the charge sum rule shows that even when films are maximally reduced, the averaged number of nickel $3d$ holes is 1.35. Superconducting samples have even higher values, calling into question the previously assumed limit of hole doping. Concomitant changes in the oxygen $K$-edge absorption spectra upon reduction indicate the presence of oxygen $2p$ holes, even in the most reduced films. Overall, our results suggest a complex interplay of hole doping mechanisms resulting from self-doping effects and oxygen non-stoichiometry.
\end{abstract}

\maketitle

\section{Introduction}

In recent years, nickelates have emerged as an important class of materials for the study of unconventional superconductivity. Analogous to cuprates, the prominent, common features considered for the occurrence of superconductivity in infinite-layer nickelates are the orbital polarization due to the two-dimensionality of the electronic structure and the dome-like hole doping dependence of the Ni/Cu-$d^9$ configuration. In contrast, the hybridization with the nearest neighbor O-$2p$ orbitals appears to be different between nickelates and cuprates \cite{Hepting2020}. 

\begin{figure*}[tb]
\center\includegraphics[width=0.99\linewidth]{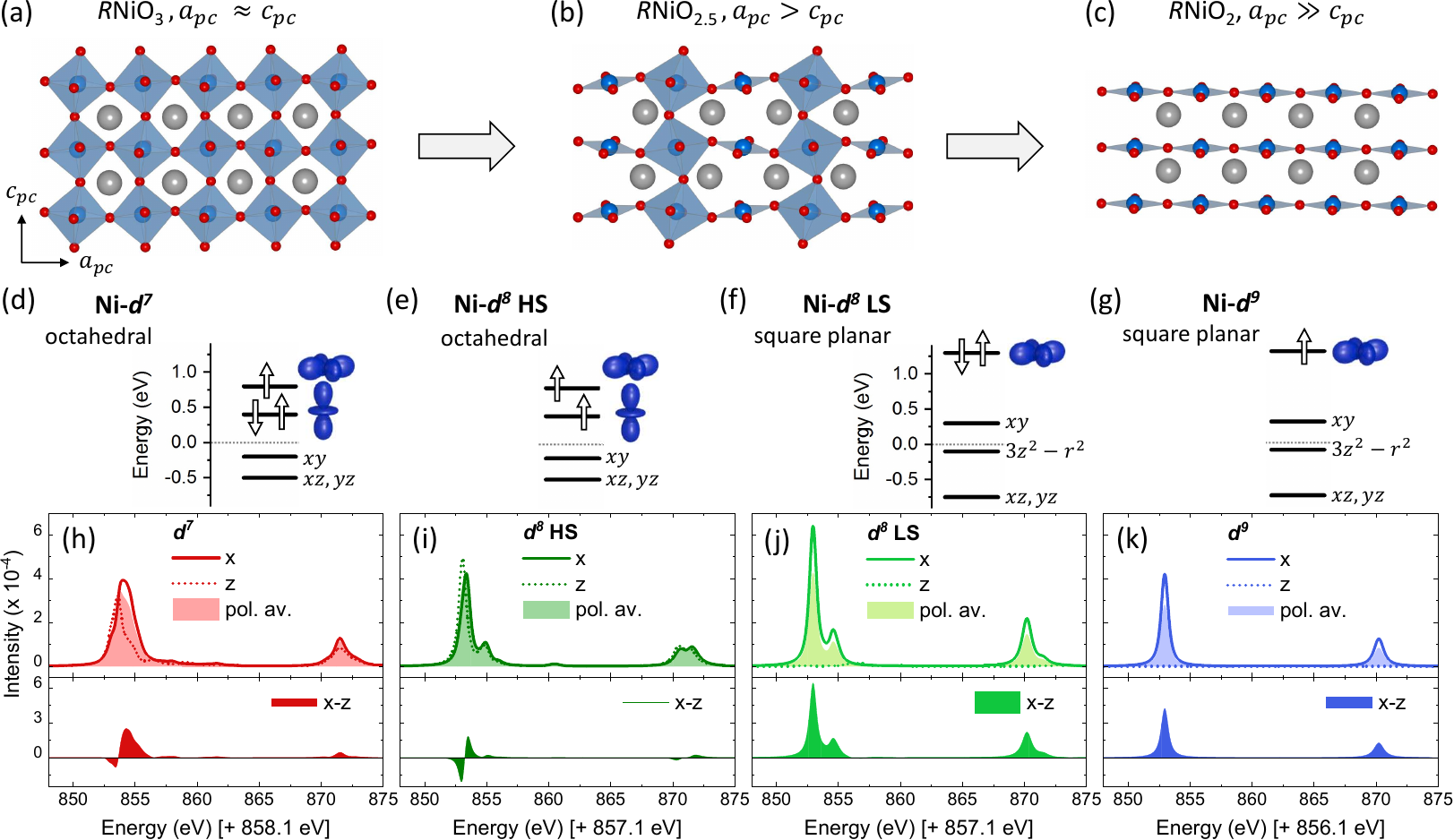}
\caption{Structure of (a) perovskite $R$NiO$_3$ \cite{Garcia1992}, (b) metastable, intermediate phase $R$NiO$_{2.5}$ with ordered oxygen vacancies \cite{Alonso1997}, and (c) infinite-layer nickelates $R$NiO$_2$ \cite{Crespin2005}. Panels (d-g) show the Ni-$3d$ orbital energies for the two sets of $D_{4h}$ ligand fields used in the model calculations together with their nominal hole spin occupation for the different Ni configurations considered in the single cluster calculations. (h-k) Resulting spectra $I_{\rm XAS}\propto f''/ {\rm Energy}$, where $f''$ is the imaginary part of the scattering factor that was obtained from single cluster ligand-field calculations for the four different configurations sketched in (d-g). Each top panel show the individual spectra for in-plane ($x$) and out-of-plane ($z$) polarization of the incoming x-rays, together with the polarization averaged spectrum calculated by $I_{\rm av} = (2I_x + I_z)/3$. The bottom panels show the linear dichroism defined as $I_x-I_z$. The energy values have been shifted for $d^8$ by 857.1~eV to match the experimental data and for $d^7$ ($d^9$) additionally by +1 (-1)~eV according to reference data \cite{Wang2000}. Note that the spectra satisfy the charge sum rule, i.e.\ the integral of the polarization-averaged spectrum is proportional to the number of $3d$ holes. For better visualization we plot the height of the legend symbols for $I_{\rm pol. av.}$ and $I_{x-z}$ proportional to the mathematical integral of the shaded areas.}\label{Fig_PNOxCluster}
\end{figure*}

Superconductivity was first observed in nominal stoichiometric Nd$_{1-y}$Sr$_y$NiO$_{2}$ infinite-layer nickelate thin films with optimal doping around $y=0.2$ \cite{Li2019}. The infinite-layer phase is synthesized through epitaxial growth of the perovskite, $R_{1-y}A_y$NiO$_3$ ($R$= rare-earth and $A$= alkaline-earth ions), as starting phase [Fig.~\ref{Fig_PNOxCluster}(a)] and a subsequent topotactic reduction, most commonly achieved by annealing with CaH$_2$. While superconducting samples could be produced by several research groups \cite{Zeng2020, Osada2020, Osada2021, Zeng2022, Wei2023, Pan2022}, the topotactic synthesis is not well understood. The reduction process does not proceed uniformly. In the first and fast step of the reduction procedure, intermediate phases are formed, e.g.\ the oxygen vacancy ordered compounds LaNiO$_{2.5}$ \cite{alonso1996} [Fig.~\ref{Fig_PNOxCluster}(b)] or PrNiO$_{2.34}$\cite{Moriga2002}. Further reduction towards the infinite-layer phase, $R$NiO$_{2.0}$ [Fig.~\ref{Fig_PNOxCluster}(c)], is significantly slower.

Interestingly, two recent publications reveal that superconductivity can be stabilized in infinite-layer nickelates in the parent compound, i.e.\ without alkaline-earth doping ($y=0$) \cite{Sahib2025, Parzyck2024}. This raises new questions about the electronic configuration relevant for superconductivity and/or what kind of doping mechanism is at play. Three scenarios have been discussed so far. First, cation non-stoichiometry could inadvertently introduce $R$O-$R$O fluoride blocks, known as Ruddlesden-Popper (RP) faults, which could potentially act as a source of doping. The arrangement and density of these defects depends strongly on the method and conditions of thin film growth of the initial perovskite phase, in particular on the control of the cation stoichiometry \cite{Lee2020,Wrobel2017}.

A second source of doping in infinite-layer nickelates arises from the reduction step, where it is difficult to control and measure oxygen off-stoichiometry. This necessary synthesis step leads to disorder in the anion sublattice being added to the disorder in the cation lattice. 

The third doping mechanism is intrinsic to the infinite-layer compound and arises from $R-5d$ hybridization with Ni-$3d$ states. Many-body electronic structure calculations revealed that the $R-5d$ states give rise to partially occupied electron pockets at the $\Gamma$ point, which hybridize with $Ni-d$ states and provide self-doped holes in the Ni-$3d_{x^2-y^2}$ bands \cite{Karp2020}. A recent theoretical study found that holes are primarily doped to the Ni-$d_{x^2-y^2}$ and "interstitial-$s$" orbitals, while the Ni-$d_{3z^2-r^2}$ and O-$2p$ orbitals are less affected \cite{Qu2025}, in agreement with previously proposed scenarios \cite{Foyevtsova2023}. Experimentally, this has been verified recently by an angle-resolved photoemission spectroscopy study, which shows that the Ni-$d_{x^2-y^2}$ and "interstitial-$s$" states primarily contribute to the Fermi surface in infinite-layer nickelates \cite{Li2025}.

In the complex interplay of doping mechanisms in currently available thin-film samples, the respective domain sizes in which different types of doping and disorder occur play an important role. This is because a continuous superconducting path only forms in electrical resistance measurements below the percolation threshold between these domains. The size of the domains of the different phases is also important when interpreting experiments that provide averaged information about different volumes.

X-ray absorption spectroscopy (XAS) is a powerful experimental technique for investigating the local electronic structure in an element sensitive fashion. The fine structure of the Ni-$L$ and O-$K$ absorption spectra, measured with linearly polarized x-rays, provides information about the valence state, the orbital polarization and the strength of the hybridization of Ni-$d$ and O-$p$ orbitals. However, different results may occur due to the fact that different XAS measurements probe very different sample volumes. One example is the observed correlation between the complete Ni-$3d$ $e_g$ orbital polarization and the onset of superconductivity \cite{Parzyck2024,Zeng2024}. In XAS measurements with very small beam spot sizes ($\sim 30\mu m$), nearly complete orbital polarization was observed \cite{Zeng2024}, but the spectra of superconducting and non-superconducting samples were indistinguishable. This suggests that the length scale of the disorder relevant for electrical transport is larger than the volume probed by the x-ray beam. Furthermore, the domain sizes for the cation and anion sublattice disorder are likely to be different and may interact differently with the clean, potentially self-doped regions. A recent study on (LaSr)$_2$CuO$_4$ thin films showed that doping-induced disorder is the leading cause of the degradation of superconductivity for large carrier concentrations \cite{Kim2021}.

The final goal would be to quantify the length scales and types of different domains, but to this end information on the signature of different Ni valence states in different ligand fields is needed. In the present study, we aim to provide answers to these questions by providing a detailed analysis of the XAS at the Ni-$L$ and O-$K$ edges measured on step-wise reduced PrNiO$_3$ thin films and heterostructures, grown by either molecular beam epitaxy (MBE) or pulsed laser deposition (PLD). Furthermore, different methods for topotactic reduction were applied and compared. This set of samples, which includes superconducting ones, allows for a comparison of spectra from samples with varying degrees of disorder in their cation and anion lattices. 

\section{Model calculations}
In order to provide an intuitive understanding of the x-ray absorption spectra and their linear dichroism, we will first consider model calculations. At the Ni-$L$ edge we measure transitions from $2p$ core levels into unoccupied $3d$ states. Starting from the perovskite $R$NiO$_3$ with octahedrally coordinated Ni$^{3+}$, there is nominally a $d^7$ valence electron configurations [Fig.~\ref{Fig_PNOxCluster}(d)]. In the cubic crystal field of the surrounding six oxygen ions the atomic $3d$ level split into a three-fold degenerate $t_{2g}$ level at lower energies and a two-fold degenerate $e_g$ level at higher energies. For all $R$NiO$_x$ compounds considered here, the $t_{2g}$ orbitals are fully occupied, while the occupation of the $e_g$ orbitals varies with $x$. In most perovskites, the local symmetry is lower, leading to a lifting of the $e_g$ orbital degeneracy. Furthermore, charge transfer from oxygen $2p$ states cause a significant deviation from the nominal $d^7$ filling \cite{Green2016}. In our model calculations, we consider a $D_{4h}$ ligand field with two sets of orbital energies shown in Fig.~\ref{Fig_PNOxCluster}(d-g) (see also the Supplemental Material, Tab.~S2). The smaller crystal field for the $d^7$ configuration leads to linearly polarized spectra that are very similar for $x$ and $z$ polarization with a small difference in the integrated intensities due to the imbalance of the nominal two to one hole in orthogonal $x^2-y^2$ and $3z^2-r^2$ orbitals [Fig.~\ref{Fig_PNOxCluster}(h)].

In the intermediate phase with $x=2.5$ a structure with ordered oxygen vacancies with NiO$_6$ octahedra sharing apical oxygen along $c$ and alternate with NiO$_4$ square-planar polyhedra in the $ab$ plane was determined from powder neutron diffraction for $R$= La \cite{Alonso1997}[Fig.~\ref{Fig_PNOxCluster}(b)]. The averaged valence is Ni$^{2+}$. Recent density functional theory (DFT) calculations for LaNiO$_{2.5}$ indicated that the columnar arranged NiO$_4$ square planar units adopt a low-spin (LS) Ni$^{2+}$-$d^8$ configurations ($S = 0$, empty $x^2-y^2$ orbital), while the interconnected NiO$_6$ octahedra with high-spin (HS) Ni$^{2+}$-$d^8$ ($S = 1$) couple in antiferromagnetic chains \cite{Shin2022}. 

The $d^8$~HS configuration was modeled with the same crystal field used for $d^7$ [Fig.~\ref{Fig_PNOxCluster}(e)]. The finite energy splitting of $x^2-y^2$ and $3z^2-r^2$ orbitals gives rise to spectra for $x$ and $z$ polarization that show different intensity at different energies, resulting in a down-up wiggle in the $x-z$ spectrum [Fig.~\ref{Fig_PNOxCluster}(i)]. However, the energy-integrated intensities of $x$ and $z$ spectra are almost the same, evident of no orbital polarization, i.e.\ a balanced hole occupation of the $e_g$ orbitals which are split in energy. An analogous interpretation was obtained from the analysis of the x-ray absorption spectra of NiO \cite{Haverkort2004}.

To model a $d^8$~LS ground state, the crystal field splitting had to be increased to overcome the Hund's coupling and stabilize two holes in the $x^2-y^2$ orbital [Fig.~\ref{Fig_PNOxCluster}(f)]. This results in spectra with maximum dichroism, i.e.\ a full spectrum for $x$ polarization with an energy integral proportional to two holes in the $d$ states, and no intensity for $z$ polarization [Fig.~\ref{Fig_PNOxCluster}(j)]. 

In infinite-layer nickelates with $x=2.0$, all Ni$^{1+}$-$d^9$ ions are in a square-planar ligand field of the four nearest-neighbor oxygen ions [Fig.~\ref{Fig_PNOxCluster}(c)]. The single $3d$ hole occupies the planar $x^2-y^2$ orbital analogous to Cu$^{2+}$ in high-$T_c$ cuprates. Accordingly, there is a pronounced linear dichroism with full orbital polarization. The line shape of the spectrum for the $x$ polarization shows a single Lorentzian line at the Ni-$L_3$ and $L_2$ edge, while there is no intensity in $z$ polarization [Fig.~\ref{Fig_PNOxCluster}(k)]. We emphasize that all calculated spectra satisfy the charge sum rule, so that for the $d^9$ state, the energy-integrated intensity in the $x$ polarization is only half as large as in the $d^8$~LS configuration.

The effective occupation of the $e_g$ orbitals is influenced by the strength of hybridization with the oxygen ligands and the associated charge transfer, which change upon reduction from the perovskite to the infinite-layer compound. For the perovskite phase with $x=3$, several previous studies argued that due to the negative charge transfer nature, the ground state is dominated by a $3d^8\underline{L}$ configuration, where $\underline{L}$ denotes a ligand hole \cite{Green2016}. In contrast, the infinite-layer compounds were found to have a Mott-Hubbard-type electronic structure, where the charge transfer from oxygen is strongly reduced. 

With these single cluster model calculations as basis we can now roughly classify our experimental results at different reduction stages. In particular, we note that the charge sum rule implies that the integral area of the polarization-averaged spectrum, defined by $I_{av}=(2I_x+I_z)/3$ and shown as shaded areas is Fig.~\ref{Fig_PNOxCluster}(h-k), is proportional to the number of $3d$ holes given by $10-n_{3d}$ where $n_{3d}$ is listed in the Supplemental Material, Table~S2. We also note here that for the perovskite in its bond-disproportionated ground state, the experimental spectra are better described by a double cluster model in which charge transfer between the two Ni sites is allowed by the coupling of two octahedra \cite{Green2016}. A finite coupling between Ni sites may also be relevant in oxygen-reduced structures \cite{Li2021}. In particular, at medium oxygen content, the crystal fields of two neighboring sites differ significantly due to their different ligand fields. Coupling between the sites has a significant influence on the ground state, as discussed below.

\section{Sample Details}

\begin{figure*}[tb]
\center\includegraphics[width=0.99\textwidth]{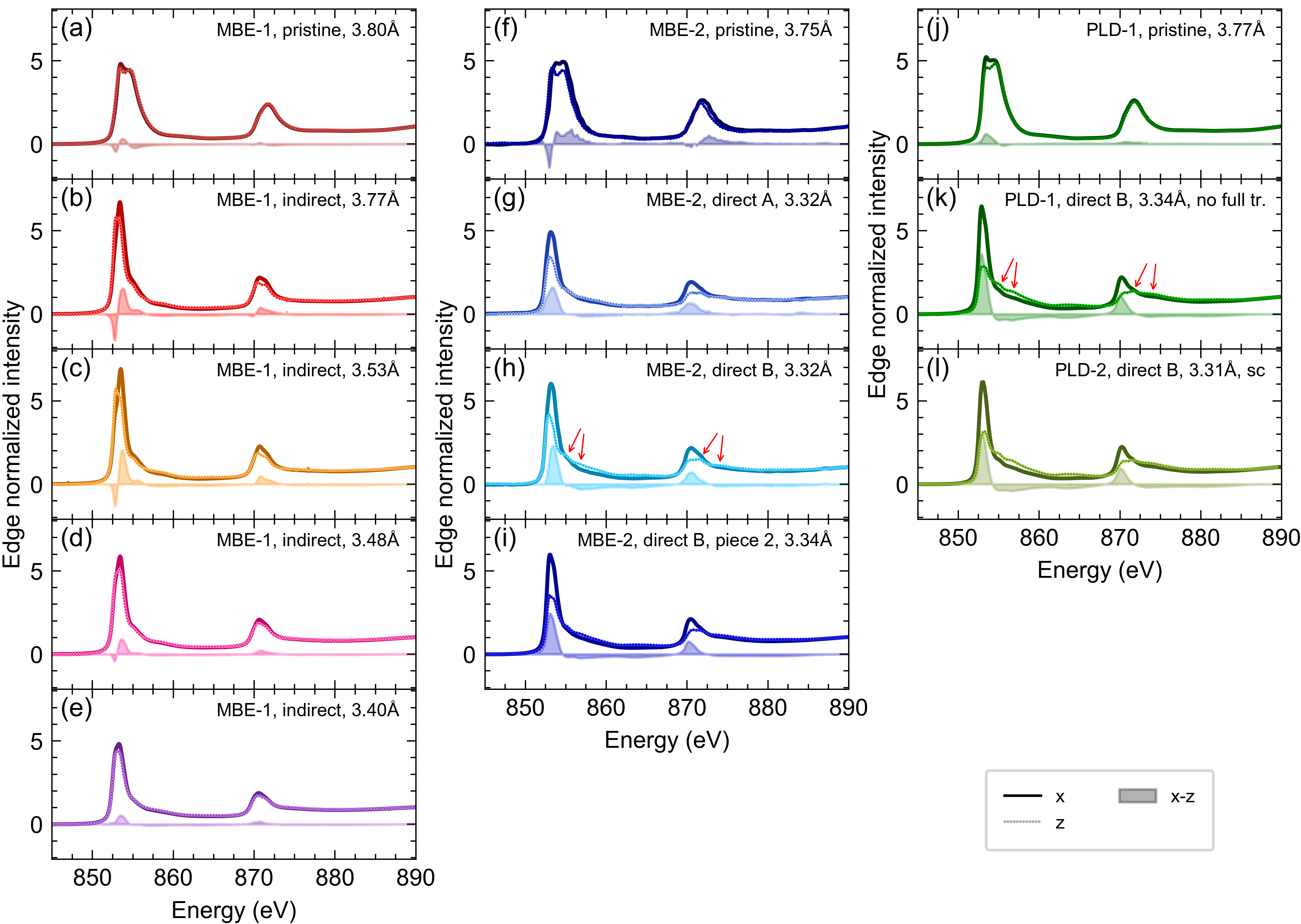}
\caption{Ni-$L$ edge XAS for in-plane ($x$) and out-of-plane ($z$) polarization for MBE-grown PNO on NGO at several intermediate reduction steps (left column), characterized by the given $c$, and, in the middle and right column, PNO on STO with STO capping comparing  superconducting ("sc") PLD samples (right column) and non-superconducting MBE samples (middle column) obtained by two different reduction procedures (see text for details). All spectra have been post-edge normalized at 889~eV.}\label{Fig_LDIC_NiL}
\end{figure*}

To shed light on the changes of the Ni electronic state during the reduction and to what characterizes the "fully reduced" state, we studied three sets of samples, all heterostructures with a 6-10~nm thick PrNiO$_{x}$ layer. The details of all samples are summarized in the Supplemental Material, Tab.~S1. The x-ray diffraction (XRD) patterns are shown in the Methods section, Fig.~\ref{Fig_XRD}.

The first set are all pieces of the same atomic layer-by-layer MBE-grown PrNiO$_3$ epitaxial film on NdGaO$_3$ (110) (NGO), named MBE-1 from here on, but reduced to different extents. Although the reduction temperatures and times varied for this sample group, all reductions were carried out in such a way that the sample was not in direct contact with the CaH$_2$ powder. The out-of-plane Ni-Ni distance $c$, estimated from the shift of the pseudo-cubic (001) and (002) Bragg peaks [Fig.~\ref{Fig_XRD}(a)], is used to label the reaction progress. Interrupting the reduction process results in inhomogeneous intermediate phases, evident from the broadening and intensity reduction of the Bragg peaks that indicates a spread of out-of-plane distances. Furthermore, this contactless method only allows to produce maximally reduced samples with lower crystallinity. Despite the overall lower crystalline order of these samples, they provide valuable insights into intermediate reduction states.

The second set of samples consists of pieces of a thinner PrNiO$_3$ film that was grown by MBE under conditions that were nominally identical to those of MBE-1, but on STO substrate and with an STO cap layer. Previous studies have also reported improved sample quality for thinner films under large tensile strain when protected by a capping layer \cite{Parzyck2024b}. The three studied pieces are considered maximally reduced with the similar $c$ values of 3.32, 3.32 and 3.34~\AA. All three were reduced with the CaH$_2$ in direct contact with the film surface, but the procedure was performed in two different laboratories under similar, indicated by labels A and B (for details see \ref{App:SampleSynthesis}). While both sample pieces that underwent procedure B exhibit very good crystallinity, as visible by the Laue fringes in Fig.~\ref{Fig_XRD}(b), this is not the case for the other piece that was reduced in setup A.

The third set of samples consists of heterostructures that have the same composition as MBE-2 but were grown by PLD and are named PLD-1 and PLD-2. Upon direct contact reduction with setup and procedure B, PLD-2 becomes superconducting and PLD-1 shows a downturn in resistance at low temperatures (see Supplemental Material Fig.~S1). In order to make the presentation of the results and discussions easier to follow, colors and symbols are kept as consistent as possible throughout the work.

\section{Results}

\subsection{Nickel \textit{L} edge x-ray absorption}
The linear polarized Ni-$L$ edge x-ray absorption spectra at several intermediate reduction steps, labeled by their $c$ parameter are shown in Fig.~\ref{Fig_LDIC_NiL}(a-e). Starting from the as grown perovskite phase [Fig.~\ref{Fig_LDIC_NiL}(a)] we observe the characteristic double-peak structure of nickelates at the $L_3$ edge. The small dichroism shows as a down-up structure in the dichroic difference spectrum $I_{x-z}(E)$. We relate this to either small oxygen non-stoichiometry, a distortion of the NiO$_6$ octahedra due to the epitaxy with the underlying NGO substrate, or a combination of both. After the first reduction step, resulting in $c=3.77$~\AA\ the spectral shape is altered towards a more dominant low energy shoulder and the x-ray linear dichroism is clearly enhanced [Fig.~\ref{Fig_LDIC_NiL}(b)]. These changes become more pronounced after the second reduction step yielding $c=3.53$~\AA\ [Fig.~\ref{Fig_LDIC_NiL}(c)], but reduces again in the next step with $c=3.48$~\AA\ [Fig.~\ref{Fig_LDIC_NiL}(d)]. In the spectra of the maximally reduced samples with $c=3.40$~\AA\ the intensity at the right shoulder of the $L_3$ line for $x$ polarization continues to approach that measured with $z$ polarization, i.e.\ the dip in the dichroism $I_{x-z}$ disappears, while the peak remains at slightly higher energies [Fig.~\ref{Fig_LDIC_NiL}(e)]. Qualitatively, the single-peak-like spectral shape and a positive dichroism is as expected for Ni-$d^9$ in the infinite-layer phase [Fig.~\ref{Fig_PNOxCluster}(k)], but the quantitative deviation is enormous, because we observe small instead of complete orbital polarization. We will come back to this point in the discussion. 

Significant changes upon reduction are already observed in the polarization averaged spectra $I_{\rm av}$ (see Supplemental Material Fig.~S2). Comparing to the pristine state the intensity maximum of the white lines shifts to lower energies. From the perovskite to the maximally reduced phase, we observe a shift towards lower energies of the intensity center of mass of the Ni-$L_2$ peak by 1.2~eV, whose fine structure is broader than that of $L_3$ due to the shorter core hole lifetime. Most of this shift already occurs with the first reduction step.

The spectra shown in Fig.~\ref{Fig_LDIC_NiL}(g-i,k,l) are from samples that have nearly the same Ni-Ni out-of-plane distance $c\approx 3.33$\AA, within our experimental error. All of those samples were grown on STO substrates, have been synthesized through direct contact reduction, and have an STO capping layer. Although only some of those samples become superconducting, their spectral features are very similar, but with quantitative differences in the overall intensity and the dichroism.

A discernible difference are features on the right shoulders of the L$_3$ and L$_2$, at 855, 856.8, 871.6 and 874~eV, whose intensity in the $z$ polarization exceeds that in the $x$ polarization (marked by arrows in panels  Fig.~\ref{Fig_LDIC_NiL}(h,k)). These features are reminiscent on the XAS spectra of multi-layer square-planar Pr$_4$Ni$_3$O$_8$ single crystals \cite{Zhang2017}. For those single crystals only the L$_2$ edge has been measured, but the similarity of the PLD samples in both polarizations to the Pr$_4$Ni$_3$O$_8$ results is remarkable. Comparing our single cluster calculations for $d^8$~LS with $d^9$ [Fig.~\ref{Fig_PNOxCluster}(j,k)], the additional structure above the main lines is a characteristic of the $d^8$~LS configuration. Similar additional features in the $z$ polarization also appear for reduced NdNiO$_2$-SrTiO$_3$ superlattices \cite{Ortiz2025} and possibly originate from a reconstructed nickelate-STO interface, where one apical oxygen remains.

\begin{figure}[tb]
\center\includegraphics[width=0.9\linewidth]{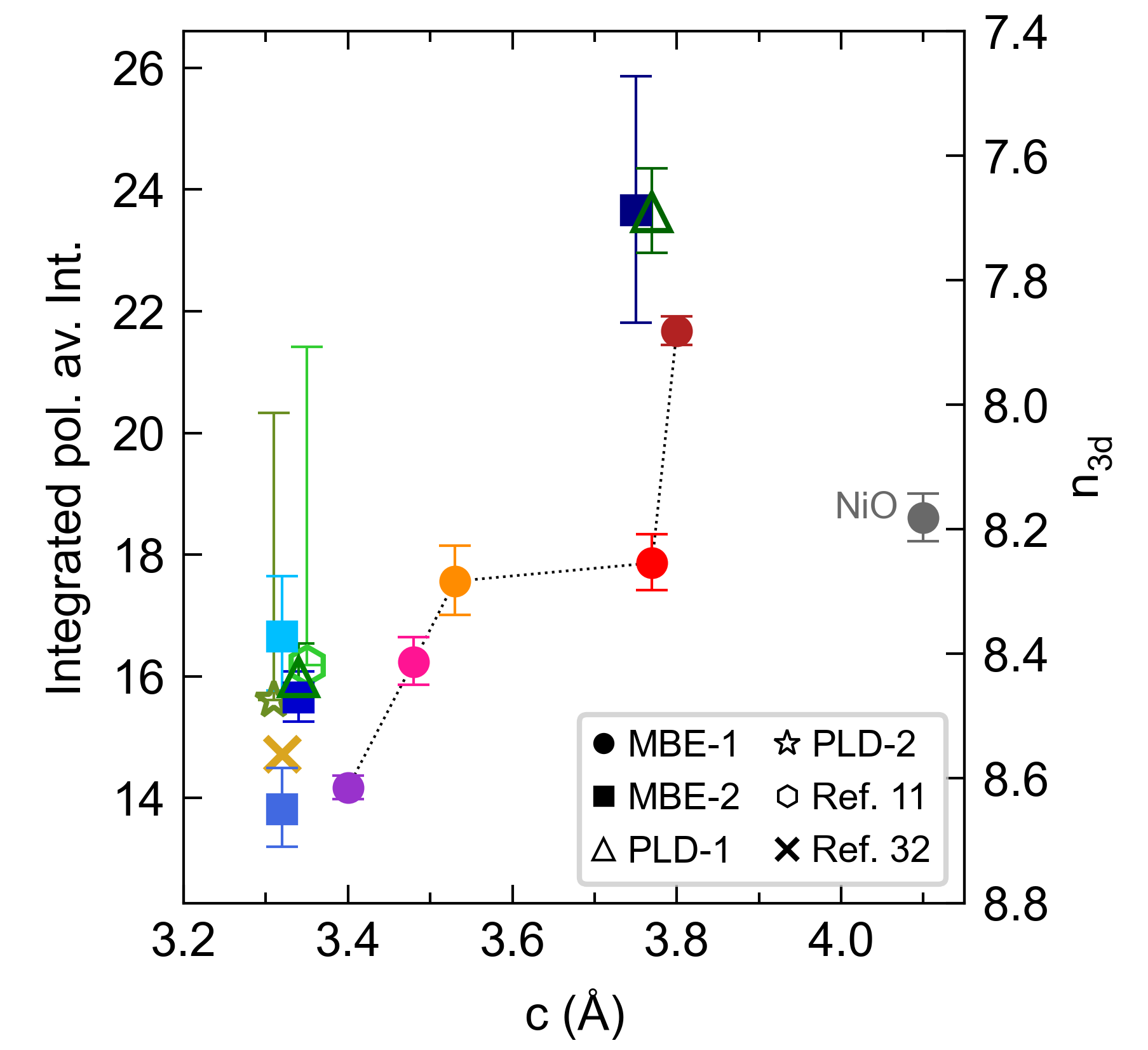}
\caption{Results from the charge sum rule analysis comparing the values of the integrated area over the $L_{3,2}$ region (845-880~eV) for the different samples as a function of out-of-plane Ni-Ni distance $c$. The colored point symbols show the results of the different pieces of MBE-1. The gray point was obtained from data measured on a NiO reference crystal. The filled, blue squares belong to samples of MBE-2. The dark yellow cross is obtained from analyzing published data of a MBE-grown NdNiO$_2$ film \cite{Parzyck2024a}. The four unfilled markers correspond to the PLD-grown samples, where the hexagon indicates the result for a similar sample measured in Ref.~\cite{Sahib2025}. As reference values to calculate the $3d$ occupation, we used cluster calculations for NiO and the MBE-1 sample with $c$=3.77\AA, as well as the zero for the $d10$ state, which leads to a proportionality factor of 0.0977 holes per unit of integral. For details on how the errors were estimated, see Methods section.}\label{Fig_SumRule_vsC}
\end{figure}

To determine the Ni-$d$ hole concentration in the various samples, we apply the charge sum rule to the polarization-averaged spectrum $I_{\rm av}$. To this end, a background originating from the edge jumps was estimated using a step function at $L_3$ and $L_2$ with the steps centered underneath L$_3$ and the L$_2$ maximum (see Supplemental Material, Fig.~S2). The integrated area of $I_{\rm av}-I_{\rm background}$ over the entire Ni-$L$ energy range is then proportional to the number of Ni-$3d$ holes $h_{3d}$ \cite{Stoehr2006}. For the MBE-1 samples, we observe a gradual decrease in the integral value with decreasing $c$ parameter [Fig.~\ref{Fig_SumRule_vsC}]. Starting from perovskite with $c=3.80$~\AA, we observe a drop in the integrated intensity to $c=3.77$~\AA\ and a nearly constant value up to $c=3.53$~\AA. After that, a gradual further decrease to $c=3.40$~\AA\ is observed.

Since the PLD and MBE-2 samples were grown on STO substrates with a larger tensile lattice mismatch for the perovskite phase, their pristine $c$ values are slightly smaller than that for the MBE-1 samples on NGO. The reduced PLD and MBE-2 samples, as well as the NdNiO$_2$ film measured in Ref.~\cite{Parzyck2024a} also have a slightly smaller $c$ value of 3.32~\AA. Although all "fully reduced" samples on STO exhibit almost the same $c$ value, their integral values are distributed over a range extending from the MBE-1 sample $c=3.53$~\AA\ to $c=3.40$~\AA. At this point, we note two observations: first, the parameter $c$ is not sufficient to characterize the Ni-$d$ electronic configuration. Second, we find that the superconducting samples are not those with the lowest integral values. This means that, with a constant parameter $c$, Ni can be further reduced, as shown by one of the MBE-2 samples, but this does not lead to the superconducting phase.

\begin{figure}[tb]
\center\includegraphics[width=0.9\linewidth]{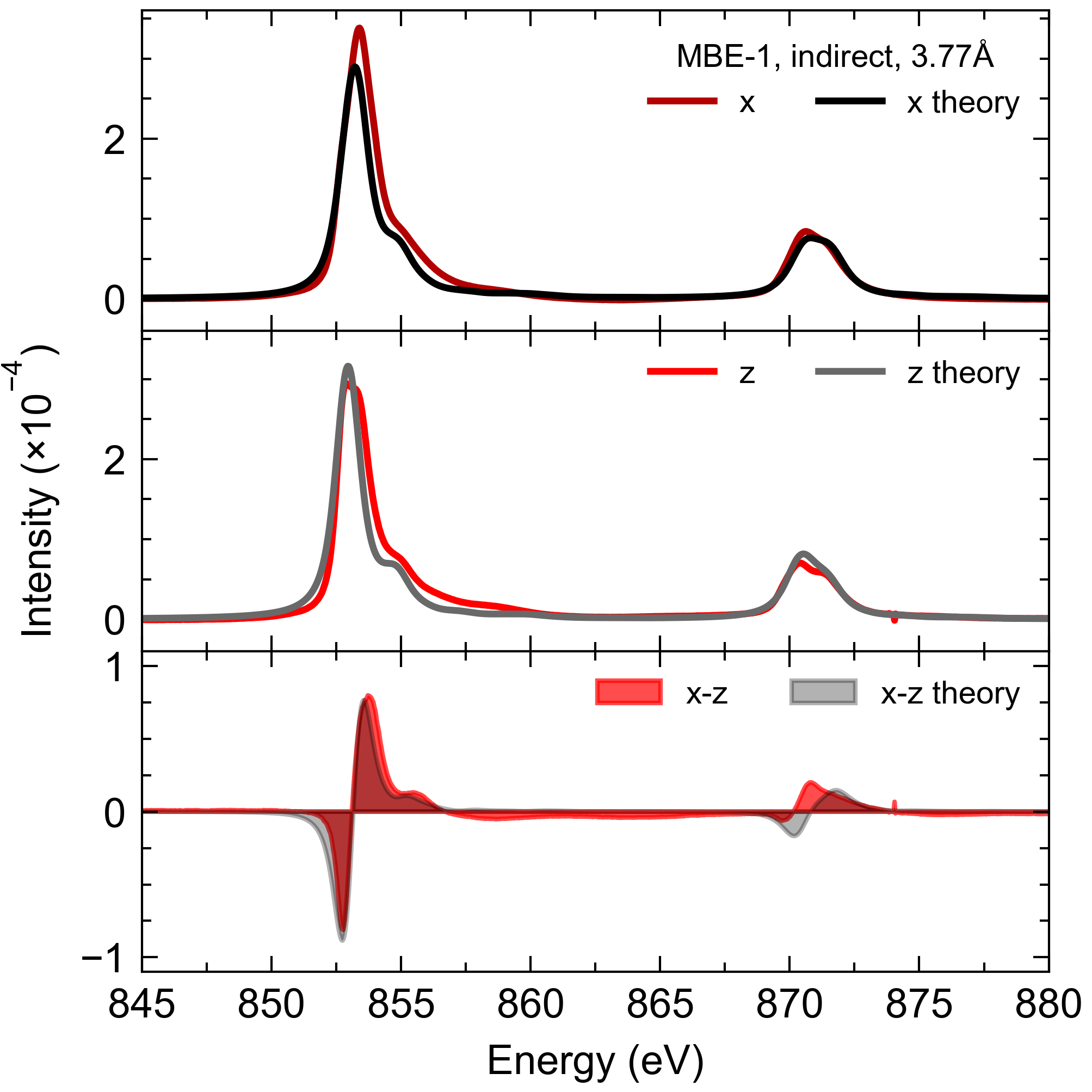}
\caption{The polarization dependent Ni-$L$ spectra and dichroism of the MBE-1 sample piece with $c=3.77$~\AA\ can be well described by a double cluster calculation with both sites in a Ni 3$d^8$ high-spin configuration and some coupling between the sites.}\label{Fig_theory_3.77}
\end{figure}

To convert the integral values in Fig.~\ref{Fig_SumRule_vsC} into the $3d$ electron occupation number $n_{3d}$, a reference value to determine the proportionality factor is required. First, we exploited the fact that the integral is zero for a filled shell, corresponding to the $d^{10}$ state. To obtain a second reference point, we measured multiple spectra of NiO and averaged their energy integrated intensity (). Due to the highly insulating and ionic nature of NiO, the $L$-edge XAS can already be reproduced well at the crystal field level. However, earlier work has shown that the agreement between theory and experiment can be significantly improved by ligand field calculations with a charge transfer of about 0.2 electrons from the O-$2p$ to the Ni-$3d$ states \cite{Haverkort2012}. With the parameters from Ref.~\cite{Haverkort2012} we reproduced the \textsc{Quanty} calculations and achieve good agreement with our data, see Supplemental Material Fig.~S4, which leads to the value of $n_{3d}=8.178$. As a third reference point we used the result from the double cluster calculation of the MBE-1 sample with $c=3.77$~\AA. The choice of the MBE-1 reference value is justified by the excellent agreement of the polarized experimental spectra with results from a double cluster calculation where two $d^8$-HS sites were coupled [Fig.~\ref{Fig_theory_3.77}], using the parameters listed in the Supplemental Material, Table~S2 and S3. The simulation yields an occupation of $n_{3d}=8.257$.

Finally for each reference point a proportionality factor to the integral was computed and we took the average, 0.0977 holes per unit of integral ($\pm0.0016$ from our rough estimate of errors on the integrals), to convert the integral values into Ni-$d$ occupation numbers for all other samples (right axis of Fig.~\ref{Fig_SumRule_vsC}). With this calibration we find $n_{3d}\sim 7.8$ for the pristine perovskite samples (dark red dot, dark blue square, and dark green triangle in Fig.~\ref{Fig_SumRule_vsC}), which is consistent with $n_{3d}\sim 8$ of the ground state configuration of $R$NiO$_3$ that was found for a similar set of parameters in Ref.~\cite{Green2016}.

\begin{figure}[tb]
\center\includegraphics[width=\linewidth]{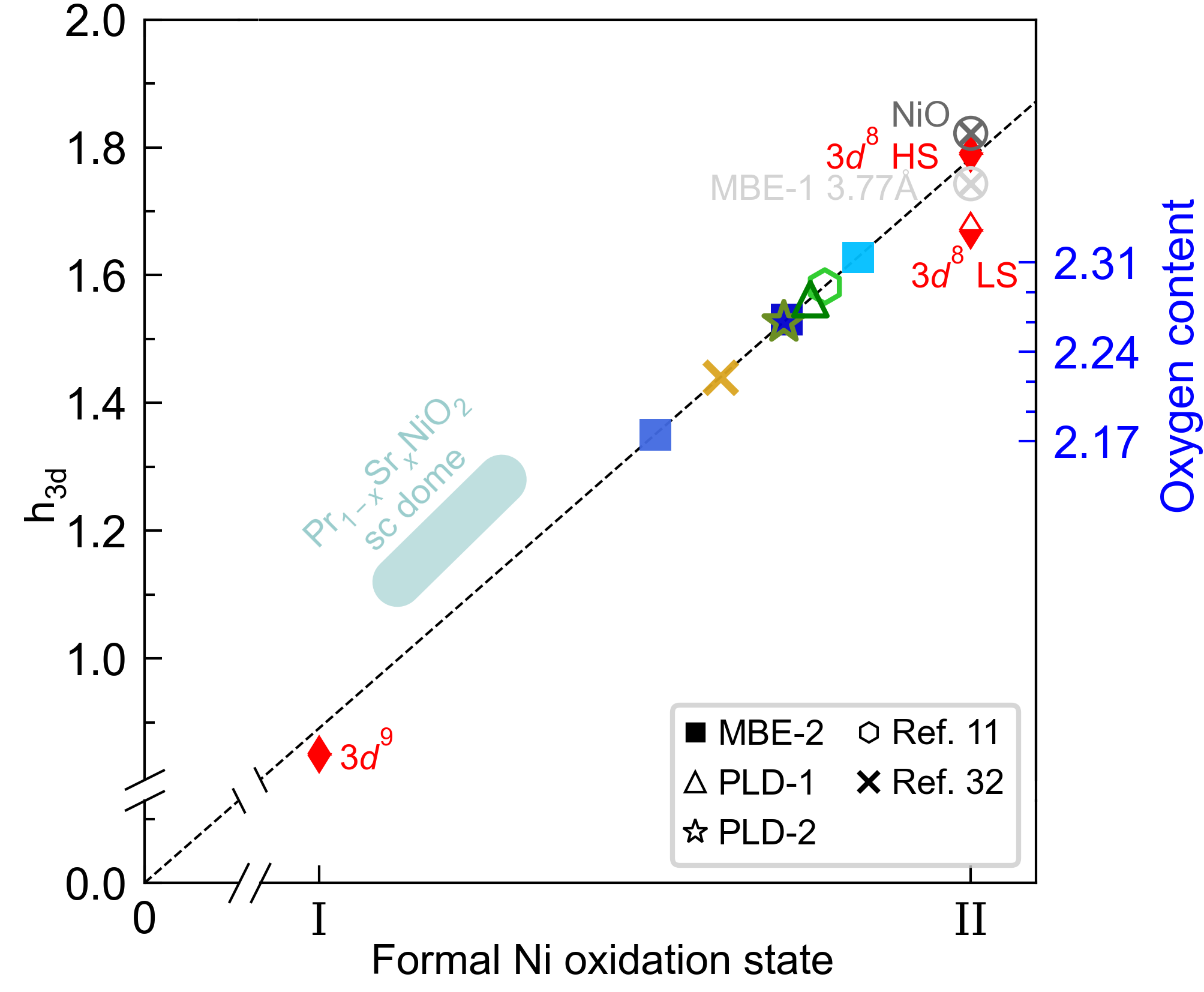}
\caption{Results of the sum rule analysis, in which the number of $d$ holes, as determined from the integrated spectra, is plotted against the formal Ni oxidation state from zero to II. The number of $d$ holes is calculated as 10 minus the number of $d$ electrons from Fig.~\ref{Fig_SumRule_vsC}. The two reference values from  the 3.77\AA\ sample and NiO (circles with crosses inside) were set to Ni(II) and used to determine the proportionality relation indicated by a straight line passing through the origin. The values of all other samples were placed on the line according to their $h_{3d}$ values (left vertical axis). For the oxygen reduced nickelates a low covalence and almost no charge transfer from O to Ni is assumed \cite{Li2021}, therefore the number of $d$ holes can be translated into an oxygen content (right vertical axis). The light teal area shows where the dome of Pr$_{1-x}$Sr$_x$NiO$_2$ lies\cite{Osada2020} with a proportionality factor of one since previous papers calculated the number of $d$ electrons/holes from the formal stoichiometry \cite{Pan2022}. The red diamonds show the results of our single cluster calculations for $d^8$~HS and $d^8$~LS (II) and the $d^9$ (I) configuration.}\label{Fig_SumRule_Ni_oxi}
\end{figure}

Interestingly, the overall lowest integral value determined in one of the MBE-2 samples corresponds only to a number of $n_{3d}=8.65$. 
This value is significantly lower than the value of $n_{3d}=8.8$, which is calculated for optimal doping at $y=0.2$ and $O_{2.0}$ oxygen stoichiometry in alkaline-earth-doped infinite-layer nickelates or the superconducting Nd$_6$Ni$_5$O$_{12}$ compound. Furthermore, the average electron occupation determined for the superconducting PLD-2 sample is notably not the highest in the set of considered infinite-layer samples. To make a comparison with another superconducting sample that was grown by PLD and prepared in the same way, the XAS data from Ref.~\cite{Sahib2025} were analyzed in the same way, yielding a similar value of $n_{3d}~\sim 8.5$. Analysis of literature data from Ref.~\cite{Parzyck2024a} for a non-superconducting NdNiO$_2$ infinite-layer sample yielded a slightly smaller $n_{3d}$, but did not reach the value of the most reduced sample in our study. Due to possible inhomogeneities in the electronic structure and different domain distributions in various samples, the XAS detection method used can be relevant. Among the data we analyzed, the data provided in Supplementary Information Fig.~S2 from Ref.~\cite{Parzyck2024a} were measured using partial fluorescence yield. All other data were measured using the total electron yield. Although the probed volumes differ between the two detection methods, all data are consistent and no relevant differences are observed for maximally reduced STO-capped films resulting from the type of detection \cite{Krieger2024}.

To put our results in context with other data on nickelates, we plot the number of $d$ holes, $h_{3d}$, over the formal Ni oxidation state in Fig.~\ref{Fig_SumRule_Ni_oxi}. To this end we set the formal valence for NiO, the 3.77\AA\ sample, and the $d^8$~HS and LS single cluster scenarios to II. The $d^9$ single cluster value is assigned to Ni-I. Based on the linear relationship between formal valence and $h_{3d}$, the valence state of the other samples with unknown stoichiometry is defined. This form of presentation shows that samples we analyzed here were only reduced to $h_{3d}=1.4$ holes and superconductivity in the non alkaline-earth doped films can occur at least up to $h_{3d}=1.6$ holes per Ni, which places all samples well above the superconducting dome of Pr$_{1-x}$Sr$_x$NiO$_2$ \cite{Osada2020} as indicated by the teal bar in Fig.~\ref{Fig_SumRule_Ni_oxi}. The dome's position is based on the nominal Sr and O stoichiometry. The expected 1~eV shift of the $L_3$ per oxidation state \cite{Wang2000} has not been observed in the XAS spectra for the doped samples, a contradiction that can be resolved by assuming a deviation in oxygen stoichiometry from exactly two oxygen per formula unit. 

The authors of reference~\cite{Li2021} studied XAS signatures of oxygen reduced nickelates creating oxygen vacancies through vacuum annealing and observed that with more reduction/$e^-$-doping there was less charge transfer from the ligands and covalency became significantly smaller. With CaH$_2$, which was used in this study, a further reduction can be achieved, resulting in a change from a charge-transfer to a Mott-like electronic structure. Electronic structure calculations predict the O-$2p$ orbitals in the infinite-layer nickelates to lie significantly lower than in the cuprates and suggests them to be irrelevant for describing the low-energy physics \cite{Held2022}. With no charge transfer from the ligands, every removed oxygen results in two more electrons on Ni, i.e.\ PrNiO$_{2.0}$ has a Ni-$d$ occupation of nine electrons and PrNiO$_{2.25}$ of 8.5, respectively. Figure~\ref{Fig_SumRule_Ni_oxi} displays the thus obtained oxygen content on the right vertical axis. If all holes doped onto Ni are due to residual oxygen, the most reduced sample is off-stoichiometric by 0.17 oxygen per formula unit, compared to the ideal infinite-layer structure. Taking into account the general difficulty to quantify oxygen in thin films and previous work indicating ordered residual oxygen \cite{Parzyck2024a}, such a large number might not be surprising.

\subsection{Oxygen \textit{K}-edge x-ray absorption}

\begin{figure*}[tb]
\center\includegraphics[width=0.9\textwidth]{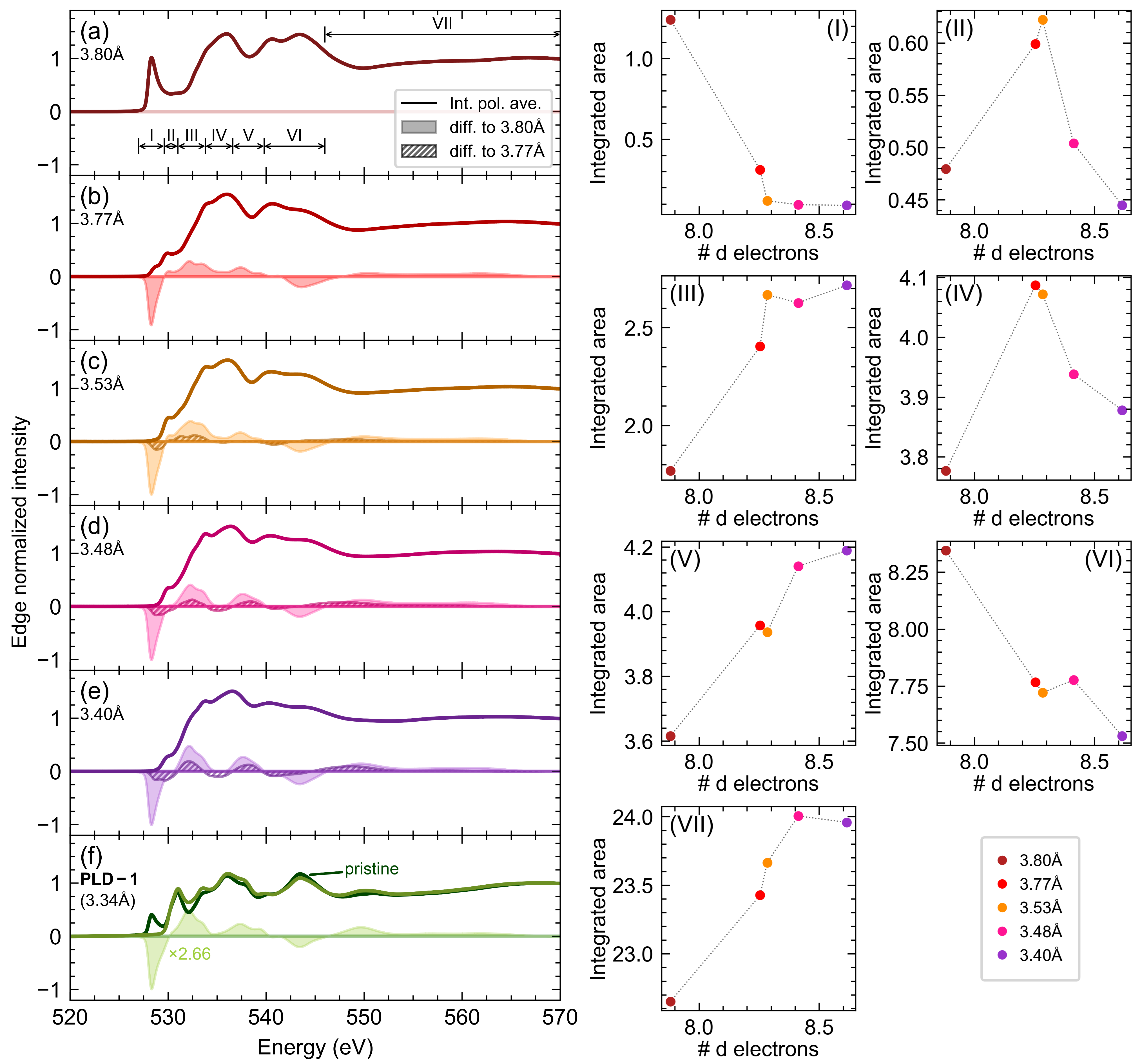}
\caption{Changes in O-$K$ edge features at different reduction stages: on the left side (panels a-e) XAS at the O-$K$ edge of sample set MBE-1. The panels are labeled by the corresponding out-of-plane lattice parameter and the solid lines show the polarization averaged intensity for each (all edge normalized at 569~eV). Additionally the reduced samples are compared to the pristine (3.8~\AA) and the least reduced (3.77~\AA) and their intensity minus these spectra is also plotted. In (e) the pristine and reduced PLD-1 sample are included and the difference between them scaled by 2.66 for ease of comparison (more details see Supplemental Material Fig.~S7). On the right side the panels I-VII display changes of individual features at different reduction stages as characterized by their difference in number of d electrons. The selected energy intervals of the features are marked in panel (a) and labeled the same as the corresponding panel in the right half of this figure, in which the plotted points mark the integral of the polarization averaged O-$K$ edge spectrum over this interval for each of the MBE-1 samples.}\label{Fig_OK_all}
\end{figure*}

In order to experimentally study the hybridization strength of the nickel ions with the neighboring oxygen-$2p$ states at different reduction stages, we measured O-$K$ XAS data. In the analysis that is described in the following we focused on the polarization averaged data, of the reduction series [Fig.~\ref{Fig_OK_all} left (a-e)]. The polarization dependent spectra are provided in the Supplemental Material Fig.~S5 and S6. The spectrum of the pristine sample, shown in Fig.~\ref{Fig_OK_all}(a), exhibits a characteristic pre-peak at 528~eV, which has been assigned with strongly hybridized Ni-O states in the fully oxygenated perovskite \cite{Abbate2002}. This peak is strongly suppressed already after the first reduction step [Fig.~\ref{Fig_OK_all}(b)] and has completely disappeared in the second step [Fig.~\ref{Fig_OK_all}(c)]. This is in agreement with the observations in oxygen deficient bulk, where the intensity of the pre-peak observed in LaNiO$_3$ decreases in the spectrum of LaNiO$_{2.75}$ and disappears almost completely for LaNiO$_{2.50}$ \cite{Abbate2002}. In the higher energy range of the spectrum, broad features centered at 535 and 542~eV were assigned to oxygen hybridization with La-$5d$ and Ni-$4sp$ states, respectively \cite{Abbate2002}. To better perceive changes in this energy range, we calculated the difference spectra to the phases with well-defined Ni valence states, namely the 3.80 and 3.77~\AA\ samples with nominally Ni$^{3+}$ and Ni$^{2+}$. These difference spectra clearly show that the largest change at the O-$K$ edge occurs in the first reduction step and is accompanied by a similar energy-dependent loss and gain of spectral weight. The loss in the region from 527-530~eV and 540-546~eV match with the assignment of $3d^8\underline{L} \rightarrow \underline{c}\,3d^8$ transitions, where $\underline{c}$ denotes a O-$1s$ core hole, and transitions to Ni-$4sp$ states. The gain of spectral weight in the region 530-540~eV and above 546~eV gradually increases upon further reduction. In particular there is additional loss of spectral weight around 530, 535, and 540~eV for Ni oxidation states below $2+$. One should note that the same changes between pristine and most reduced sample occur also for the PLD-1 samples with STO capping on STO substrate, as apparent when subtracting the pristine from the reduced sample's spectrum [Fig.~\ref{Fig_OK_all}(f)]. The difference looks almost identical but scaled down by a factor of 2.66. This is due to the post-edge normalization which cannot distinguish signals from PNO film and STO capping layers, so for both the nickelate's signal is smaller than in the MBE-1 spectra. Since the STO capping layer does not change in the reduction process, its features disappear in the difference.

To track the observed changes, we integrated the edge-normalized spectra within the indicated energy ranges (labeled I-VII in the inset of the top left panel (a) in Fig.~\ref{Fig_OK_all}) and plotted the results in the right half of the same Fig.~\ref{Fig_OK_all}, labeled by the corresponding roman number (I-VII), as a function of the Ni-$3d$ electron filling determined by the sum rule analysis [Fig.~\ref{Fig_SumRule_vsC}]. 

Thus we find that previously unrecognized features II and IV are related to Ni$^{2+}$-O hybridized states in PrNiO$_{x}$, as their intensity increases after the first reduction step and then decreases again with further reduction towards Ni$^{1+}$. Regions III, V, and VII are characteristic of the most reduced samples, and as the spectral weight increases and saturates, this suggests finite hybridization of O-$2p$ with either Ni-$3d$ or Pr-$5d$ states. Further insights may be gained from comparison with corresponding site and symmetry-projected DFT calculations in future work.

\section{Discussion}

\begin{figure*}[tb]
\center\includegraphics[width=0.9\linewidth]{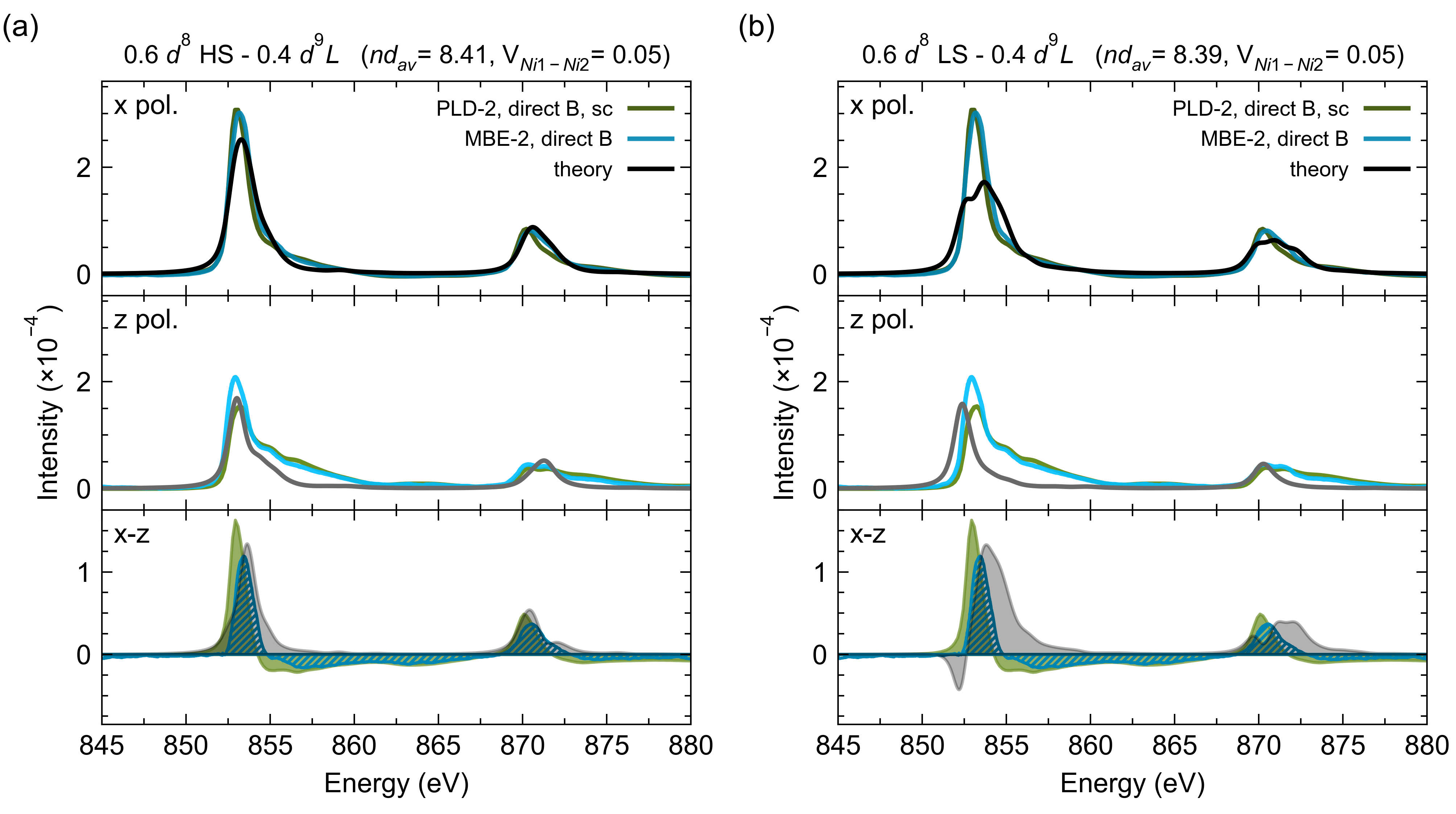}
\caption{(a) Comparison of the experimental linear dichroism of the non-superconducting MBE-2 direct B sample (light blue lines) and the superconducting PLD-2 sample (light green lines) compared with the model calculations for a mixture of coupled Ni1: $d^8$~HS -- Ni2: $d^9\underline{L}$ sites with coupling strength stated in the title and a ratio of 60:40 leading to an average nickel $d$ filling of $nd_{\rm av}=8.41$ electrons. (b)  Comparison of the experimental linear dichroism of the same two samples compared with the model calculations for coupled Ni1: $d^8$~LS -- Ni2: $d^9\underline{L}$ sites using the same coupling strength and the same ratio 60:40 with $nd_{\rm av}=8.39$ electrons.}
\label{Fig_DoubleClusterB}
\end{figure*}

We find that superconductivity in "parent-compound", non-alkaline earth doped infinite-layer samples appears at a hole doping level of $h_{3d}\sim 1.55$, clearly larger compared to what was reported in Sr or Ca-doped compounds (nominally $h_{3d}\sim 1.2$). This discrepancy could be due to the fact that the cation and oxygen stoichiometry in infinite-layer thin films has not yet been determined experimentally with sufficient accuracy. In alkaline-earth doped films, the cation stoichiometry was adopted from that of the respective PLD target stoichiometry. The oxygen stoichiometry was assumed to be O$_{2.0}$ based on the maximally reduced $c$-lattice constant, which is insufficient, since we observe the same low $c$ values for widely differing $3d$ fillings. If the oxygen stoichiometry indeed deviates from the nominal value, our results suggest that the oxygen stoichiometry is closer to O$_{2.2}$. In samples with nominal alkaline-earth ion doping of $y=0.2$ this would imply a significant shift of the superconducting dome to higher hole doping. However, if the oxygen stoichiometry was correctly assumed to be O$_{2.0}$, then the difference may lie in the doping mechanism resulting from cation substitution. 

Doping infinite-layer nickelates via oxygen stoichiometry is associated with the introduction of $d^8$~HS sites \cite{Ortiz2025}, since the partial occupation of apical oxygen positions reduces the $e_g$ crystal-field splitting such that the $d^8$~LS state becomes unfavorable. In the case of Sr or Ca cation substitution the crystal-field splitting of the square-planar coordinated Ni site is less strongly altered, and a stable $d^8$~LS state is possible. However, our double cluster calculations indicate that, when coupled to surrounding $d^9\underline{L}$ states, the $d^8$~LS state is unstable. Using the ligand field parameters from the single cluster calculations [Fig.~\ref{Fig_PNOxCluster}(f,j)] and coupling the $d^8$~LS to a neighboring $d^9\underline{L}$ site the $e_g$ orbital occupation numbers at the $d^8$~LS site become more uniform (see Methods Fig.~\ref{Fig_DoubleClusterA}). This is accompanied by a balancing of the ligand occupancy at both Ni sites and a change in the Ni $3d$ occupancy towards the ionic values of 8 and 9 at each Ni site. At finite coupling, the $d^8$~LS site becomes effectively a $d^8$~HS that leads to the contribution in XAS measured with $z$ polarization (Supplemental Material Fig.~S9), while the orbital polarization in XAS originates primarily from the $d^9\underline{L}$ site. Qualitatively, $d^8$~LS and $d^8$~HS coupled to $d^9\underline{L}$ therefore behave similar, but small quantitative difference can be noticed in the linear polarized spectra. In Figure~\ref{Fig_DoubleClusterB} we compare the calculated spectra for (a) $d^8$~HS--$d^9\underline{L}$ and (b) $d^8$~LS--$d^9\underline{L}$ sites with the polarization dependent Ni-$L$ experimental spectra of two samples, superconducting and not, with almost the same $d$-filling of $n_{3d}\sim 8.4$ (PLD-2 sc and MBE-2 non-sc, both reduced by method direct B). In order to obtain spectra from the double cluster calculation for the same values of $n_{3d}$ that we previously determined experimentally from the sum rule analysis, we calculated a weighted mixture of the spectra of 60$\%$ Ni1 and 40$\%$ Ni2 sites. We note that the PLD-2 and MBE-2 experimental spectra in the individual polarization channels are nearly indistinguishable. However, there is a small, but noticeable difference in the dichroic spectrum. The dichroism of the PLD-2 sample is slightly greater, primarily due to a weaker signal in the $z$ polarization peaks at the low-energy side of $L_3$ and $L_{2}$ lines [Fig.~\ref{Fig_DoubleClusterB} bottom panels]. We find that the spectral shape for both polarizations and the dichroism of both, the PLD-2 and MBE-2 sample is better described by the $d^8$~HS--$d^9\underline{L}$ coupled sites [Fig.~\ref{Fig_DoubleClusterB}(a)]. This observation would be consistent with a scenario of incomplete reduction, in which apical oxygen positions lead to Ni-$d^8$~HS sites in the infinite-layer structure.

One possible explanation for the difference in the linear dichroism between the superconducting PLD-2 and non-superconducting MBE-2 samples could be the size of the self-doped domains within the anion-disordered regions. In order to stabilize superconductivity, orbital polarization and hole doping of the Ni-$3d$ state must be coordinated on the scale that is probed by electric transport measurements. To measure a superconducting transition in resistance, there must be a connected path for the current between the superconducting regions.

As homogeneity issues arise on length scales larger than those typically probed in scanning transmission electron microscopy (STEM) and smaller than probed in XAS, it is not possible to unambiguously identify the electronic structure of the superconducting phase based on data available in the literature. The difficulty in distinguishing superconducting and non-superconducting samples based on their XAS spectra, as well as their sensitivity to spot size and reduction method, further points towards a difference in microstructure. Recent STEM measurements found that disordered or rotated boundary regions form \cite{Osada2021}, the length scales of which likely depend on the fault type and density in pristine films, and the reduction method. This would explain the fluctuations in the strength of orbital polarization that we [Fig.~\ref{Fig_DoubleClusterB}] and others observed at doping levels that should fall into the superconducting regime.

\section{Conclusion}

By comparing XAS measurements with ligand field calculations we provided insights into how the electronic configuration of Ni changes when oxygen is successively removed from the perovskite rare-earth nickelates towards the infinite-layer phase. Our quantitative analysis of Ni-$L$ edge spectra using the charge sum rule revealed that the most reduced films exhibit a valence state of Ni$^{1.35+}$ and that superconductivity occurs in samples with even higher valence states. Changes in the O-$K$ edge spectrum suggest a finite degree of hybridization with nickel states in the most reduced samples. The measured and simulated XAS spectra, along with the evaluation of the data presented, should serve as useful references for future investigations of oxygen-reduced nickelates.

Our results challenge previous findings regarding the doping range in which superconductivity occurs in infinite-layer nickelates. In earlier studies, doping-dependent superconducting domes were always determined based on the assumption of an exact $A$ cation and oxygen stoichiometry in $R_{1-y}A_y$NiO$_{2.0}$. However, our analysis suggest that this assumption might be incorrect and that deviations of up to O$_{2.3}$ are possible, at least in undoped ($y=0$) samples. The possibility to observe superconductivity at higher hole doping in nickelates does not necessarily contradict the analogy to cuprates. A recent study on La$_{1-y}$Ca$_y$CuO$_4$ has shown that the previously established stability range of superconductivity in the bulk material extends to significantly higher doping levels of $y=0.5$ in thin films \cite{Kim2021}.

In agreement with previous studies, we conclude that, while strong linear dichroism is evident in the Ni-$L$ spectra of all the superconducting samples, the magnitude of the dichroism is insufficient to distinguish between the superconducting and non-superconducting samples. We attribute this to a complex interplay of ordered, potentially self-doped regions, and disorder occurring on different length scales in both the cation and anion sublattices. Reducing this disorder through a targeted heterostructure design will enable a more reproducible synthesis of superconducting samples in future work.

\section*{Acknowledgments}
The authors gratefully acknowledge financial support by the German Research Foundation (DFG) via TRR~360-492547816. We thank Helmholtz-Zentrum Berlin for the allocation of synchrotron radiation beamtime at BESSY II. The Institute for Beam Physics and Technology (IBPT) at the Karlsruhe Institute of Technology (KIT) is acknowledged for operating the Karlsruhe Research Accelerator (KARA) and for provision of beamtime at the KIT Light Source. We acknowledge the European Synchrotron Radiation Facility (ESRF) for provision of synchrotron radiation beamtime under proposal number IHHC3995 at the ID32 beamline. We thank F.~Rosa for support during the beamtime, and E.~Brücher for supporting us with access to a PPMS measurement system. We thank P.~Puphal, B.~Goodge, G.~Sawatzky, K.~Zou, and G.~Khaliullin for helpful discussions.

\appendix
\renewcommand\thefigure{A\arabic{figure}}    
\setcounter{figure}{0}  
\renewcommand\thetable{A\arabic{table}}    
\setcounter{table}{0}

\section*{Appendix}
\subsection{Sample synthesis}
\label{App:SampleSynthesis}

The thin films of perovskite PrNiO$_3$ are grown on a (110)-oriented NdGaO$_3$ (NGO) and (100)-oriented SrTiO$_3$ (STO) substrates (supplied by Crystal GmbH) using ozone-assisted atomic layer-by-layer molecular beam epitaxy (MBE). With a background pressure of 2.4$\times$10$^{-5}\,$mbar ozone-rich atmosphere and the substrate heater temperature set to $\sim$550$\,^{\circ}$C the film is grown by shutter growth, alternatingly depositing atomic layers of Pr and Ni. The film on STO has 2\,nm of STO grown on top as a capping layer. To examine the influence of the topotactic reduction independently of possible (smallest) variations in the cation stoichiometry, we cut an optimally grown sample into several pieces of 2.5$\times$2.5\,mm$^2$ and reduced them at varying time intervals.

\begin{figure}[tb]
\center\includegraphics[width=0.90\linewidth]{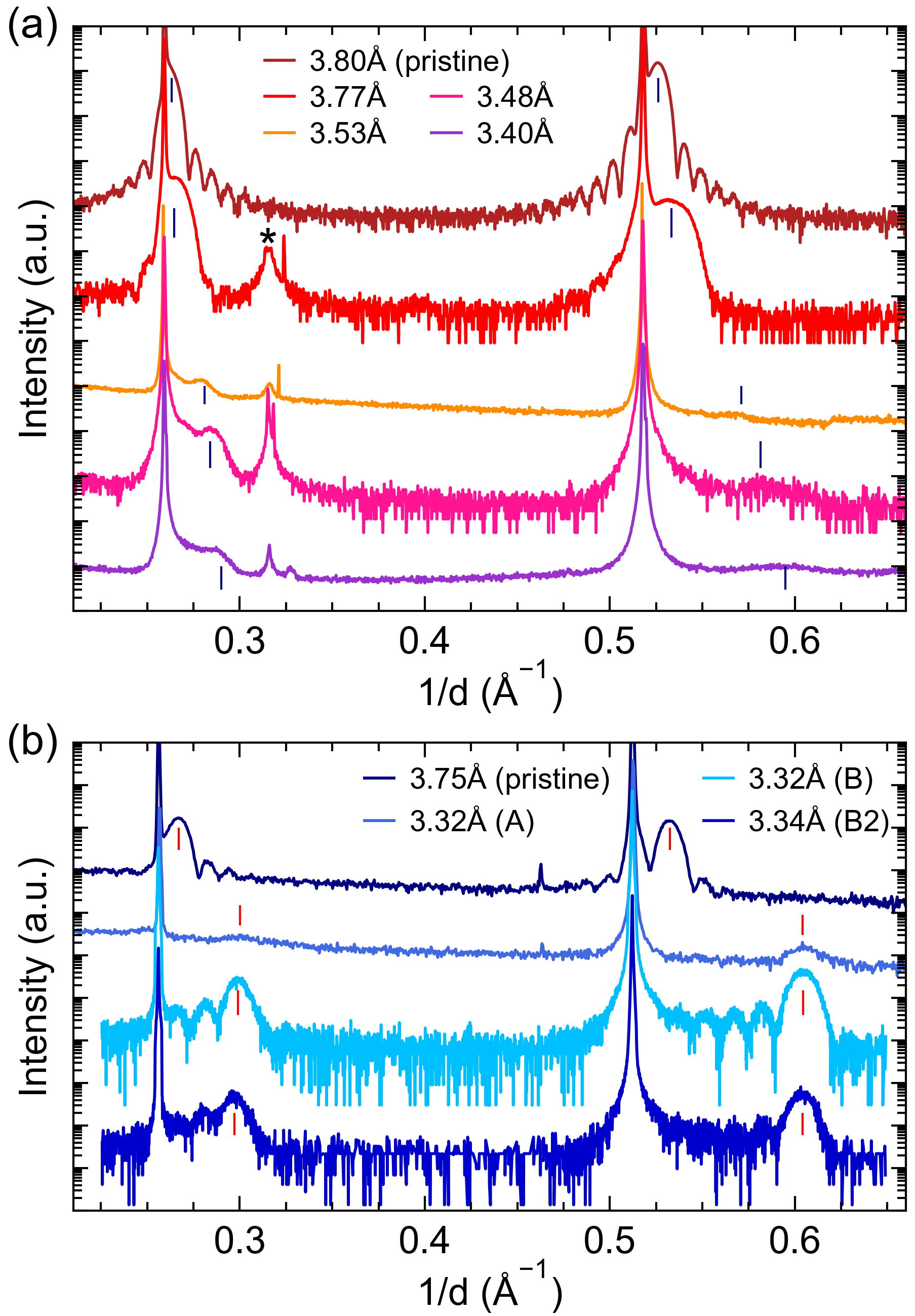}
\caption{XRD of (a) the sample set MBE-1 on NGO without STO capping and reduced without direct contact and (b) the samples MBE-2 on STO with STO capping reduced with direct contact. The star indicates the signal from the sample holder. Scans offset for clarity. The peak positions used to calculate $c$ for each scan are marked by small blue (a) and red (b) lines.}\label{Fig_XRD}
\end{figure}

Oxygen reduction was achieved by topotactic soft-chemistry reaction with CaH$_2$. We used two different approaches for the topotactic reduction. For all pieces of the film on NGO 100\,mg of CaH$_2$ were filled into a DURAN glass tube with the sample piece on top in a crucible folded from aluminum foil. There was no direct contact between the sample and the reducing reagent. The thus prepared glass tubes were evacuated to $\sim$10$^{-6}$\,mbar and sealed. Through heating to 240-260\,$^{\circ}$C (using a ramp of 30\,min) the reduction process was started. Slightly different temperatures were selected and the time they were maintained was varied from 12 to 84\,h to obtain samples reduced to different degrees.  

Figure~\ref{Fig_XRD}a shows the out-of-plane x-ray diffraction (XRD) scans around the substrates pseudocubic (001) and (002) peak which show that subsequent reduction steps cause a collapse of the structure along the pseudo-cubic $c$ direction as evident from the shift of the (001) and (002) film peaks to higher values of $1/d$. Along with this, we observe a reduction in Bragg peak intensities, indicating a loss of crystallinity. Therefore we used a second reduction method where a PrNiO$_3$ film with a STO capping grown on STO was reduced in direct contact with the CaH$_2$ powder. To this end 500\,mg CaH$_2$ and the sample piece were placed in a DURAN glass tube ("direct A") or a glass tube with a valve ("direct B", see Ref.~\cite{Krieger2023}) such that the powder completely covered the sample. The tube was evacuated and closed off with the valve, then it was placed in a box (A) or tube (B) furnace, heated to 320$\,^{\circ}$C (A) and 260$\,^{\circ}$C (B) respectively, and kept at the temperature for 2-4\,h (A and B), for most samples procedure B was repeated to improve XRD and transport. The corresponding XRD scans are plotted in Figure~\ref{Fig_XRD}b. In this case, especially for method "direct B", the peaks remain almost as sharp as for the perovskite film and also display Laue fringes, thus showing that by this method we synthesized an infinite-layer nickelate of much higher crystallinity. Comparing both topotactic reduction approaches, we conclude that avoiding direct contact with CaH$_2$ possibly leads to the formation of water at the film surface, which in this case was not protected by a capping layer. The water remaining on the surface can lead to partial reoxidation or hydroxide formation with substantial disorder in the anion sublattice. In the case of direct contact of the film with the CaH$_2$ and the presence of a surface-protecting capping layer, Ca(OH)$_2$ can form which is stable and does not react further with the nickelate film. The much faster reaction time when in direct contact with CaH$_2$ further supports this explanation. However, the significantly longer reaction time when avoiding direct contact allows to stabilize easily intermediate stages. Also, the absence of a STO capping layer makes the interpretation of the soft x-ray absorption spectra measured by means of total electron yield with a probing depth of the order of 5~nm, less controversial.

To characterize the samples at different stages of reduction, we can calculate the pseudo-cubic $c$ parameter from the peak centers of the (001) and (002) film peaks (marked by small lines in Fig.~\ref{Fig_XRD}) and then take the average of the two results. The results are shown in the legend of Fig.~\ref{Fig_XRD} and range from $c=3.80$~\AA\ in the pristine, perovskite film down to $c=3.32 - 3.40$~\AA\ in the maximally reduced samples.

\subsection{X-ray absorption measurements}
The XAS measurements have all been performed in total electron yield (TEY) mode. For the samples on NGO, the PLD-1 pair and the "MBE-2,direct B, piece 2"-sample measurements were done at beamline UE46 at BESSY, for the other MBE-2 samples at beamline WERA at KARA and PLD-2 at beamline ID32 at ESRF with the RIXS endstation. To measure polarization dependence the samples were tilted to $\theta$=10$^{\circ}$ for PLD-2 and $\theta$=30$^{\circ}$ for all other samples and a linear polarized beam was used. At BESSY and ESRF then the polarization was switched between horizontal and vertical while at KARA the x-ray polarization relative to the sample orientation was switched by rotating the sample by 90$^{\circ}$ around the beam k vector. The beam spotsizes were $\mathrm{100~\mu m \times 50~\mu m}$ (UE46, BESSY), $\mathrm{2~\mu m \times 10~\mu m}$ (RIXS branch, ID-32, ESRF), and 1.2~mm $\times$ 0.2~mm (WERA, KARA) respectively. First the obtained spectra for each polarization were scaled to match in the pre- and post-edge regions, corrected by fitting a line to the pre-edge region of both polarizations simultaneously and subtracting it, then normalizing so the post-edge intensity is 1. At both $\theta$=10$^{\circ}$ and $\theta$=30$^{\circ}$ one polarization has the E-field vector in-plane for the sample and therefore directly gives us $I_x$. The other polarization has the E-field vector mostly out-of-plane and the fully out-of-plane spectrum $I_z$ can be calculated from the measured $I(\theta)$ with the in-plane spectrum $I_x$ and the following formula $I_z = \frac{I(\theta) - I_x\sin^2\theta}{\cos^2\theta}$\cite{Stoehr2006}, respectively $I_z=1.03\, I(10^{\circ})-0.03\, I_x$ and $I_z=\frac{4}{3} I(30^{\circ})-\frac{1}{3} I_x$.
To estimate the errors of the integral of the polarization averaged spectra (Fig.~\ref{Fig_SumRule_vsC}) we assumed the selection of the linear background to produce the biggest difference since in all the following steps the samples were treated equal. The linear background is determined through a fit with the \textsc{python} Scipy \cite{2020SciPy-NMeth} \textit{curve\_fit} routine which gives a numeric error on the determined slope adding or subtracting 10-times this error was the point where it was visible that the linear background was selected wrongly but the resulting spectra did not look off. Thus we repeated our whole procedure with this linear background selection and determined the integrals which give the shown error estimates. For NiO the inter-dataset difference is larger than the changes due to background selection and thus their integrals are averaged and the corresponding standard deviation displayed as error. For the PLD samples a larger difference could be achieved by different selection of the scaling factor, so this error is used instead. These errorbars should mainly give an idea on how much the given numbers are influence by the correction procedure.

\subsection{Cluster calculations}\label{Sec:ClusterCalulations}
\begin{figure}[tb]
\center\includegraphics[width=0.99\linewidth]{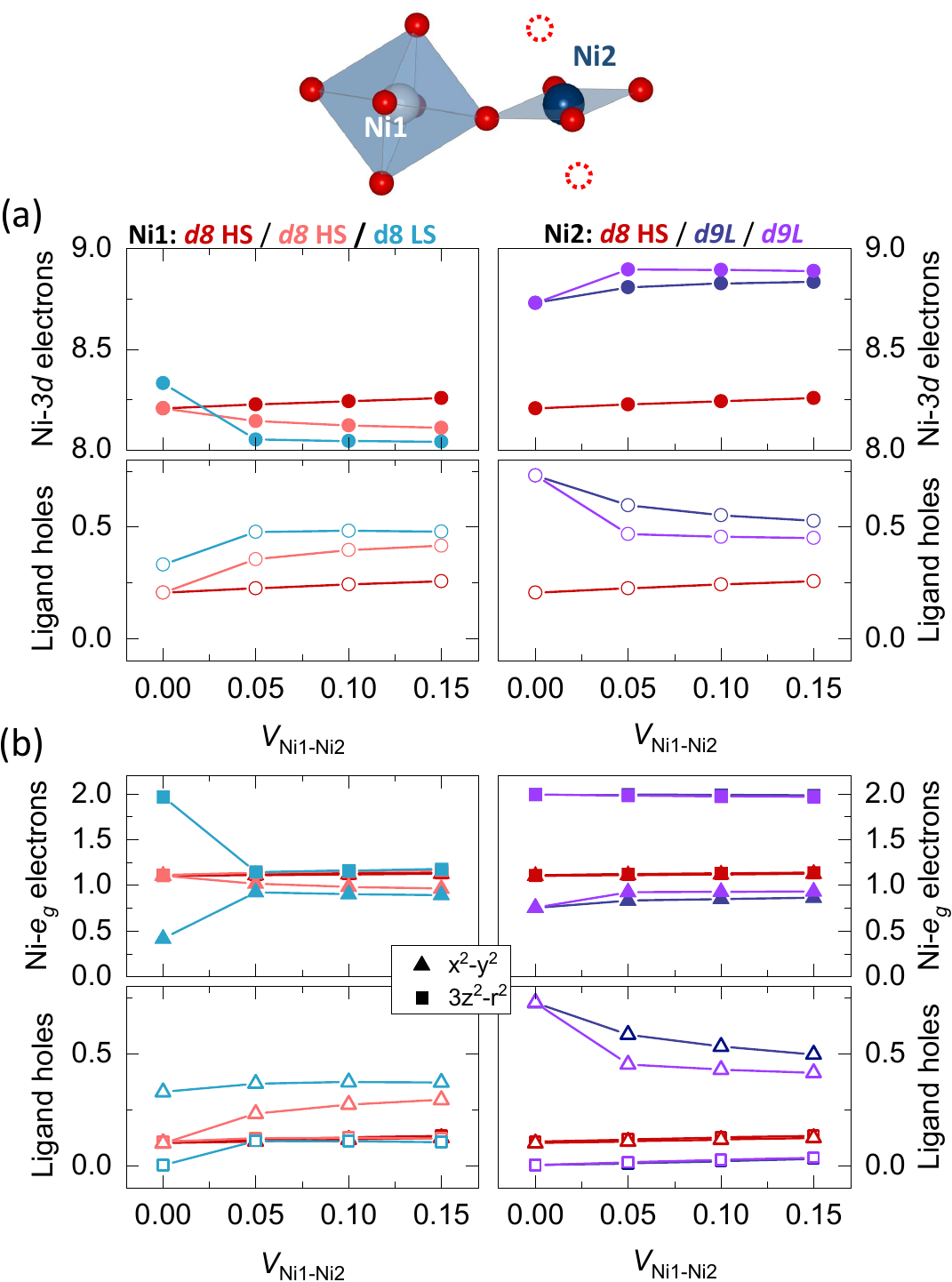}
\caption{Top: Sketch of the considered double cluster with two potentially inequivalent nickel sites Ni1 and Ni2. Bottom: (a) Ni-$3d$ orbital electron occupations for coupled Ni1--Ni2 clusters with $d^8$~HS -- $d^8$~HS (red symbols), $d^8$~HS -- $d^9\underline{L}$ (light red -- dark blue symbols), and  $d^8$~LS -- $d^9\underline{L}$ (light blue -- violet symbols) configuration as a function of the inter-cluster coupling strength $V_{\rm Ni1-Ni2}$. (b) Ni-$d$ orbital resolved occupations determined for the three different coupled Ni1--Ni2 clusters. The lower panels in (a) and (b) show the corresponding O-$2p$ ligand holes.}\label{Fig_DoubleClusterA}
\end{figure}

The cluster calculations were performed using the quantum many body script language \textsc{Quanty} \cite{Lu2014, Haverkort2014, Haverkort2012}. We considered a $D_{4h}$ ligand field where the crystal field energies of the $d$ orbitals were parametrized by $10Dq$, $D_s$, $D_t$. Since the aim was to use these model calculations to demonstrate qualitative differences, we set $10Dq$ = 1.0~eV and $D_t$ = 0~eV for all configurations and only varied $D_s$ (see Supplemental Material, Tab.~S2) in order to stabilize the $d^8$~LS in the single cluster calculation. 

The orbital energy values are provided in Fig.~\ref{Fig_PNOxCluster}(d-g). The ground state orbital occupations of the different configurations used in the single cluster calculations are summarized in the right part of Supplemental Material, Tab.~S2. To economize the calculations, a linear combination of the O-$2p$ orbitals is used to form the ligand orbitals. Of the 6 $\times$ 6 O-$2p_{x,y,z}$ spin orbitals for octahedrally coordinated Ni, only 10 interact with the Ni-$d$ orbitals. The size of the electron hopping between the oxygen and nickel sites can be expressed by the parameters $pd\sigma$ and $pd\pi$, which can be calculated from the Ni--O bond distances (Supplemental Material, caption of Tab.~S2).

To calculate XAS spectra, \textsc{Quanty} uses Green's functions to calculate the excitation probability for all final states for a given initial state Hamiltonian. The resulting set of delta functions can then be compared with the experiment after broadening. To obtain the model spectra from single cluster calculations shown in Fig.~\ref{Fig_PNOxCluster}(h-k) we applied moderate energy broadening of 0.3 for the Gaussian full-width-at-half-maximum (FWHM) and energy dependent Lorentzian FWHM of 0.4 over the $L_3$ energies and 0.6 over $L_2$ energies. For the double cluster spectra shown in Fig.~\ref{Fig_theory_3.77} and Fig.~\ref{Fig_DoubleClusterB} energy broadening was increased to 0.6 for the Gaussian FWHM, and energy dependent Lorenzian FWHM of 0.5-0.8 over the $L_3$ energies and 1.0-1.2 over $L_2$ energies. In addition, for calculated XAS spectra, the energy axis has to be shifted in order to match the one of the experiment. Since the absolute energy measured in the experiment depends on the calibration, these values are known only relatively. Previous work has found an energy shift of about 1~eV per Ni valence state change \cite{Wang2000}. We therefore determined the energy shift applied to the single cluster data shown in Fig.~\ref{Fig_PNOxCluster}(h-k) using the 3.77~\AA\ sample as a $d^8$ reference and assumed +1(-1)~eV for the calculated $d^7$ ($d^9$) spectra. Further parameters used to describe the Coulomb interaction and spin-orbit coupling are summarized in the Supplemental Material, Tab.~S3.

In the double cluster calculations, we included coupling in the form of intercluster Ni-ligand hopping $V_{{\rm Ni1-Ni2}}$ in addition to the standard ligand field Ni-ligand hopping in the individual clusters \cite{Green2016} and considered different combinations of the various configurations for the coupled Ni1--Ni2 sites [Fig.~\ref{Fig_DoubleClusterA}]. 

To circumvent the problem of introducing valence state-dependent energy shifts in XAS in the double cluster calculation, we modeled the $d^9$ state by a $d^8$ configuration with negative charge transfer energy, resulting in a dominant contribution of $d^9\underline{L}$ to the ground state (see Ni2 occupations in Fig.~\ref{Fig_DoubleClusterA}(a)). Although the $d^8$~LS can be stabilized in a single cluster with the chosen crystal field parameters (Supplemental Material, Tab.~S2), a coupling to a neighboring $d^9\underline{L}$ site causes the charge to equalize between the sites [Fig.~\ref{Fig_DoubleClusterA}(a)]. This charge transfer destabilizes the $d^8$~LS and yields an almost equal occupation of Ni1 $d_{x^2-y^2}$ and $d_{3z^2-r^2}$ orbitals for coupling strengths $V_{{\rm Ni1-Ni2}} \geq 0.05$ (Fig.~\ref{Fig_DoubleClusterA}(b), light blue symbols). Similar observations are made for other configurations: For coupled $d^8$~LS--$d^8$~LS sites with slightly larger $V_{{\rm Ni1-Ni2}} \geq 0.1$ and $d^8$~LS--$d^8$~HS with $V_{{\rm Ni1-Ni2}} \geq 0.05$ our results show low-spin to high-spin transitions at the Ni1 site. In other words, our calculations indicate that the $d^8$~LS state is unstable as the coupling between the sites becomes finite. To compare our experimental spectra of samples reduced to $nd_{\rm av}\sim 8.4$ [Fig.~\ref{Fig_DoubleClusterB}], we used a proportional mixture of Ni1-$d^8$~HS($d^8$~LS) sites with Ni2-$d^9\underline{L}$ sites, i.e.\ we considered an incoherent superposition of $60\%$~Ni1 and $40\%$~Ni2 sites.

\clearpage
\newpage
\onecolumngrid
\pagestyle{empty}
\renewcommand\thefigure{S\arabic{figure}}    
\setcounter{figure}{0}  
\renewcommand\thetable{S\arabic{table}}    
\setcounter{table}{0}
\section*{Supplemental Material:\\From perovskite to infinite-layer nickelates: hole concentration from x-ray absorption}

\section*{Samples details}

\begin{table*}[htb]
\setlength{\tabcolsep}{5pt}
\renewcommand{\arraystretch}{1.2}
\centering
\begin{tabular}{l | c c c c}
\hline\hline
sample name & substrate & STO cap & reduction method & $c$ $\left[\mathrm{\AA}\right]$\\
\hline
MBE-1, pristine, 3.80 \AA & NGO & no & \textemdash & 3.80\\
MBE-1, indirect, 3.77 \AA & NGO & no & separated from CaH$_2$ & 3.77\\
MBE-1, indirect, 3.53 \AA & NGO & no & separated from CaH$_2$ & 3.53\\
MBE-1, indirect, 3.48 \AA & NGO & no & separated from CaH$_2$ & 3.48\\
MBE-1, indirect, 3.40 \AA & NGO & no & separated from CaH$_2$ & 3.40\\
\hline
MBE-2, pristine & STO & yes & \textemdash & 3.75\\
MBE-2, direct A & STO & yes & direct contact to CaH$_2$, setup \textbf{A} & 3.32\\
MBE-2, direct B & STO & yes & direct contact to CaH$_2$, setup \textbf{B} & 3.32\\
MBE-2, direct B, piece 2 & STO & yes & direct contact to CaH$_2$, setup \textbf{B} & 3.34\\
\hline
PLD-1, pristine & STO & yes & \textemdash & 3.77\\
PLD-1, direct B, no full tr. & STO & yes & direct contact to CaH$_2$, setup \textbf{B} & 3.34\\
PLD-2, direct B, sc & STO & yes & direct contact to CaH$_2$, setup \textbf{B} & 3.31\\
\hline\hline
\end{tabular}
\caption{List of samples and details of their synthesis. NGO (STO) is the abbreviation for NdGaO$_3$ (SrTiO$_3$).}\label{Tab_Samplelist}
\end{table*}

\begin{figure*}[htb]
\center\includegraphics[width=0.9\textwidth]{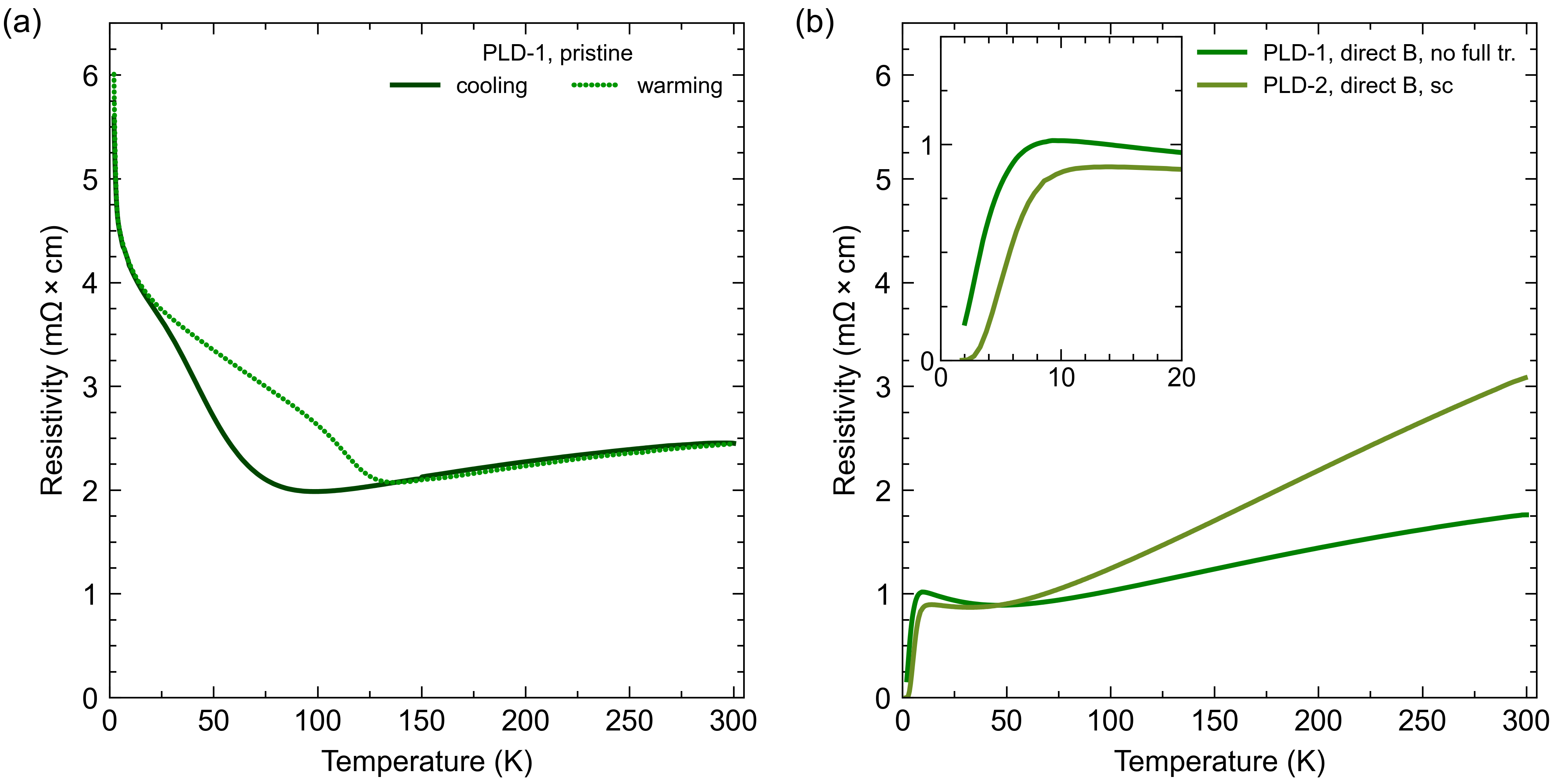}
\caption{Temperature-dependent resistivity of the PLD-grown PNO films on STO with STO capping. (a) The measured pristine PLD-1 sample shows the expected metal-to-insulator transition of perovskite PrNiO$_3$ and (b) all reduced infinite-layer samples exhibit metallic behavior with a steep downturn at low temperatures. The enlarged view of the low temperature region indicates that only PLD-2 reaches the zero-resistance state within the measured temperature range.}\label{Fig_transport_pld}
\end{figure*}

\clearpage
\section*{Additional x-ray absorption data}

\begin{figure*}[htb]
\center\includegraphics[width=0.99\textwidth]{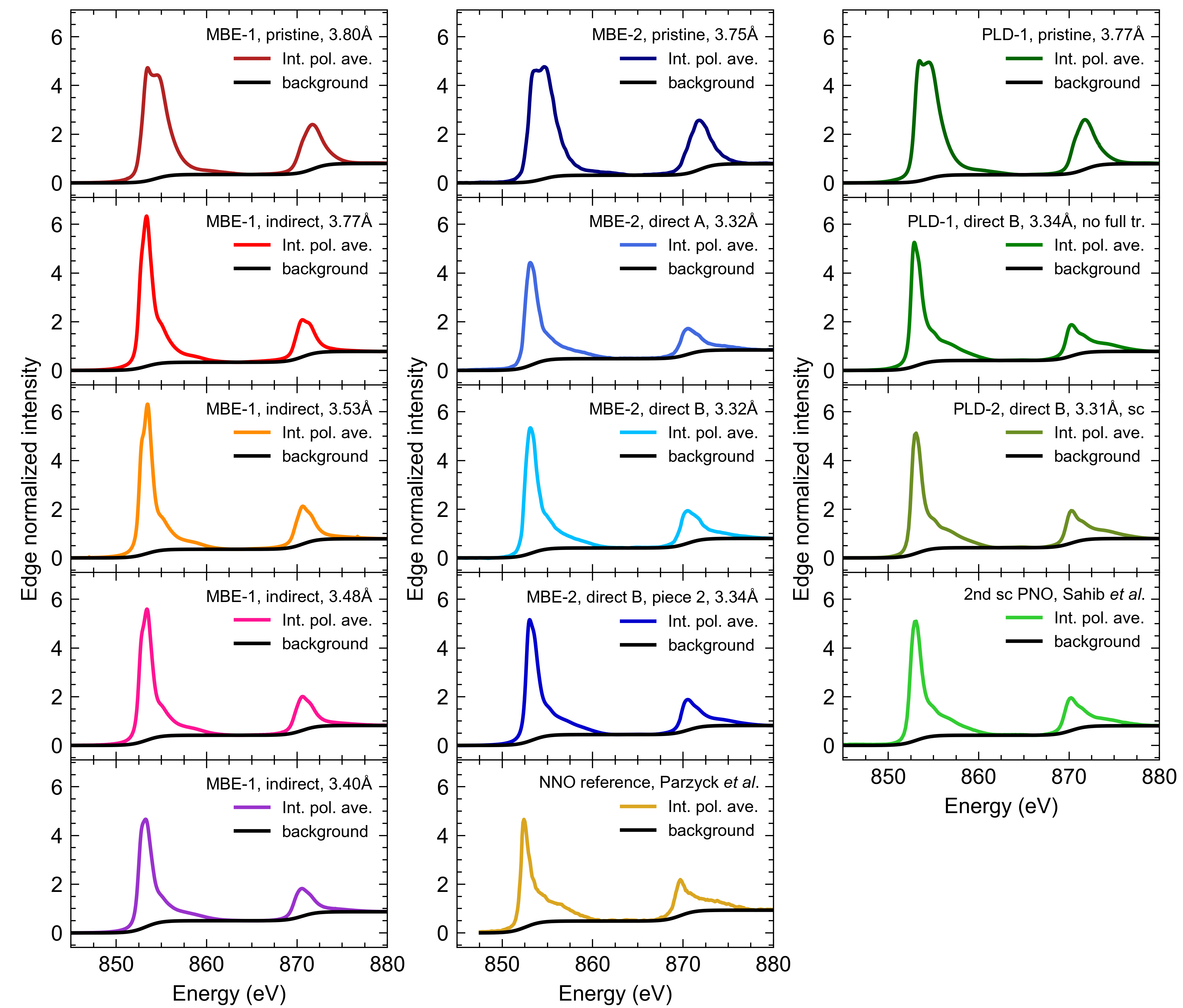}
\caption{Polarization averaged Ni-$L$ spectra of the data shown in Fig.~2 of the main text. Left column: MBE-1 samples at several intermediate reduction steps, labeled by their out-of-plane lattice parameter $c$. Middle column: Pristine and reduced MBE-2 samples obtained by two different reduction procedures. Right column: Pristine and reduced PLD-grown samples, comparing superconducting ("sc") and "no full transition" samples. To gain another data point for a superconducting PrNiO$_2$ film, the data from Sahib $et$ $al.$, Ref.~\cite{Sahib2025S}, is also analyzed here, since PLD-growth and reduction procedure are the same as for PLD-1 and PLD-2. To also include a very different infinite-layer sample the data for a MBE-grown reduced NdNiO$_2$ sample on STO with STO capping is taken from literature. Its polarization averaged spectrum, shown in the bottom panel of the middle column, is computed from the data published by Parzyck $et$ $al.$ in Ref.~\cite{Parzyck2024aS}'s Supplementary Information labeled "sample F" }\label{Fig_polave_NiL}
\end{figure*}

\begin{figure}[htb]
\center\includegraphics[width=0.5\linewidth]{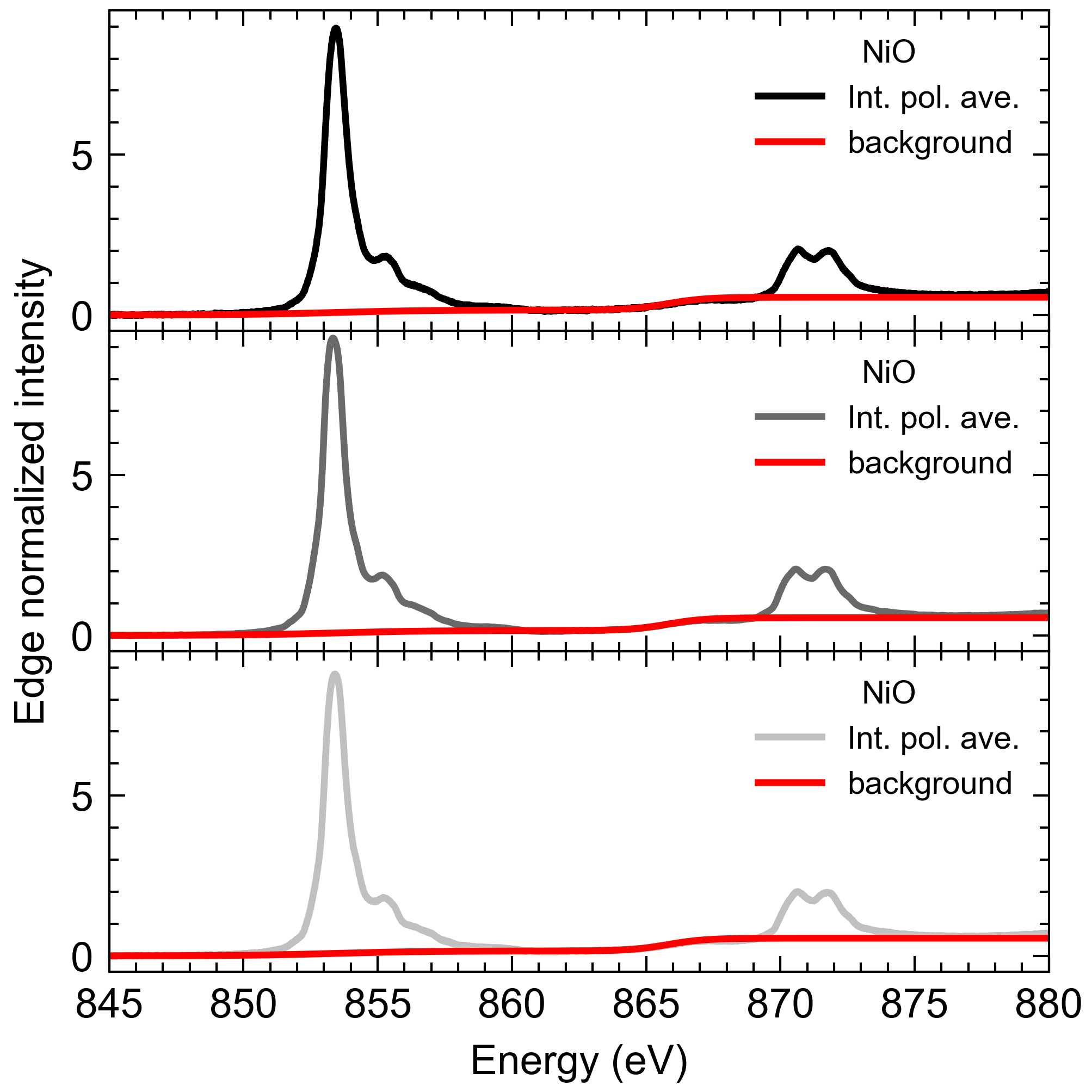}
\caption{Polarization averaged Ni-$L$ spectra for the three data sets measured on the NiO reference crystal.}\label{Fig_NiO_polave}
\end{figure}

\begin{figure}[htb]
\center\includegraphics[width=0.5\linewidth]{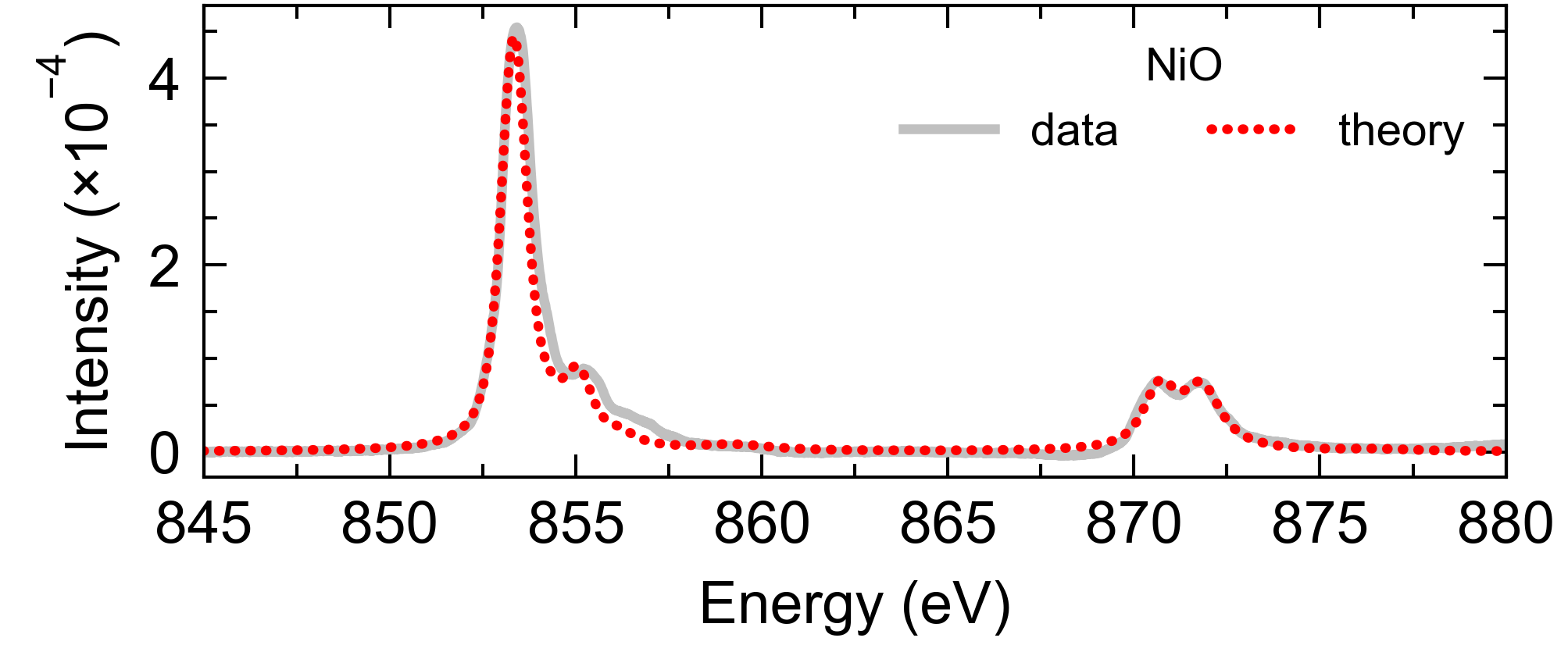}
\caption{Measured data on NiO compared to the single cluster ligand-field calculation using the parameters from Ref.~\cite{Haverkort2012S} and the software package \textsc{Quanty}.}\label{Fig_NiOtheory}
\end{figure}

\begin{figure}[htb]
\center\includegraphics[width=0.6\linewidth]{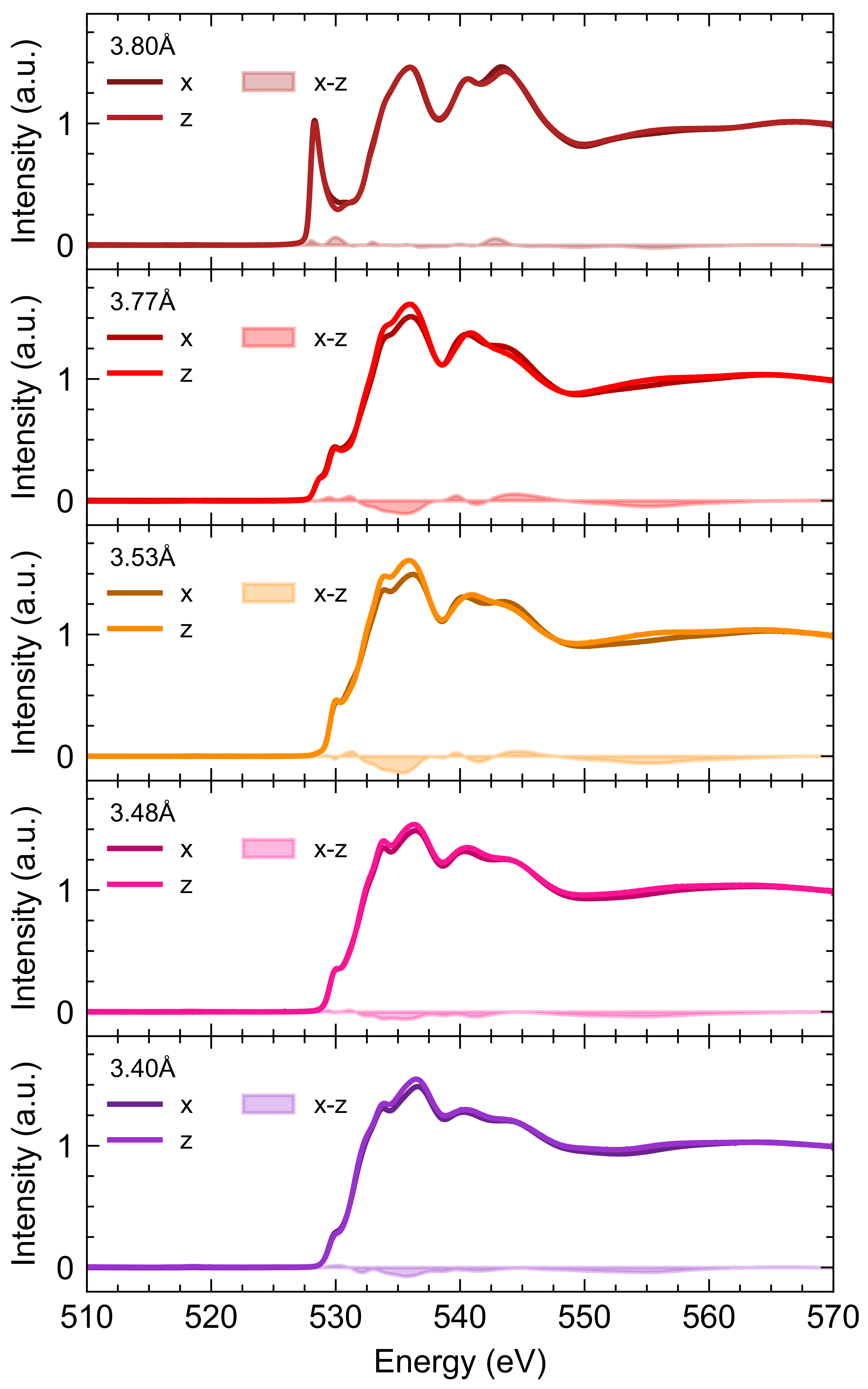}
\caption{O-$K$ edge polarization dependent spectra of the MBE-1 samples. The corresponding panels are labeled by their out-of-plane Ni-Ni distance $c$. }\label{Fig_oxygen_pol}
\end{figure}
\begin{figure}[htb]
\center\includegraphics[width=0.55\linewidth]{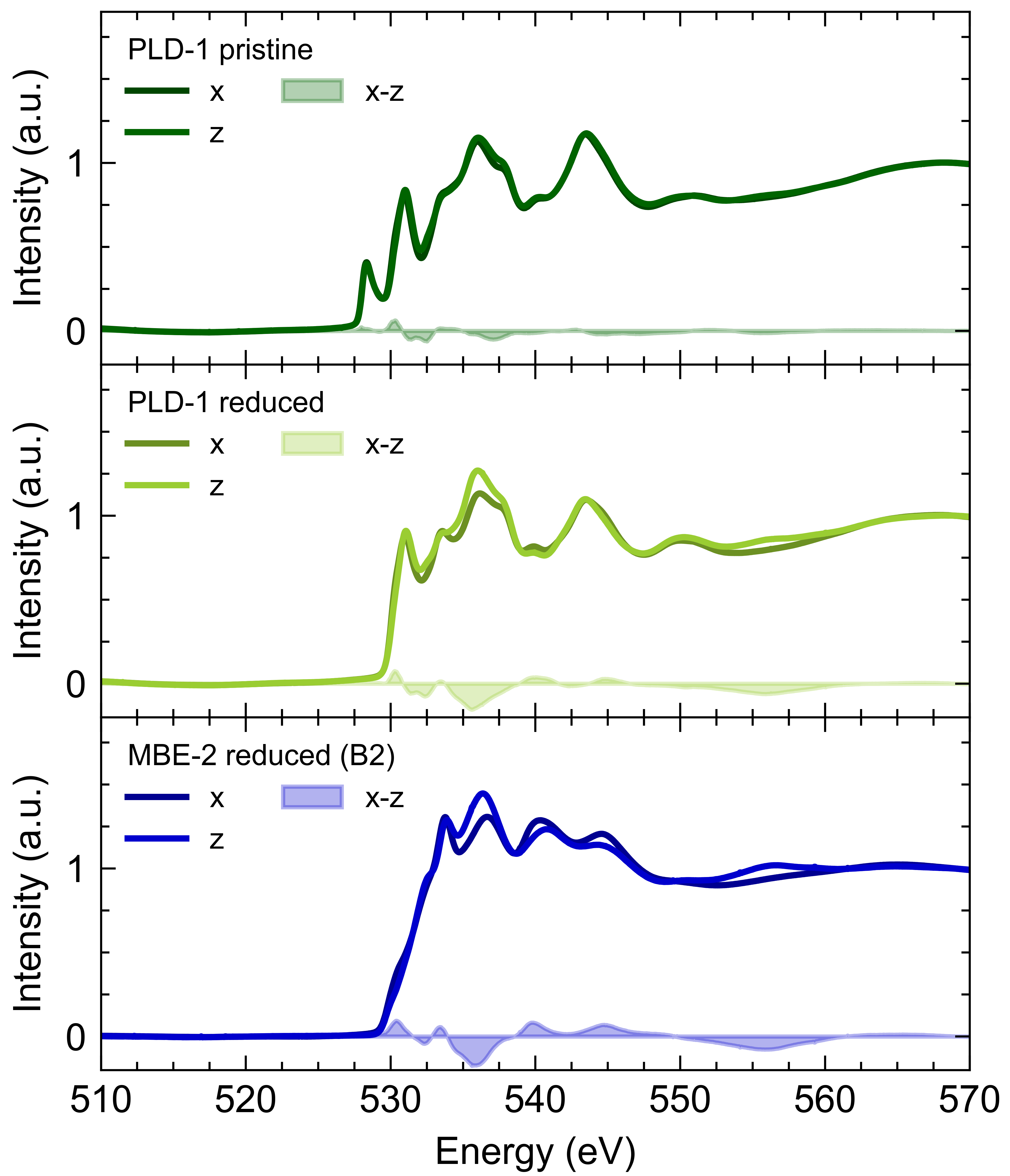}
\caption{O-$K$ edge polarization dependent spectra of the PLD-1 sample pair and the second MBE-2 "direct B" piece. }\label{Fig_oxygen_pol_part2}
\end{figure}

\begin{figure}[htb]
\center\includegraphics[width=0.55\linewidth]{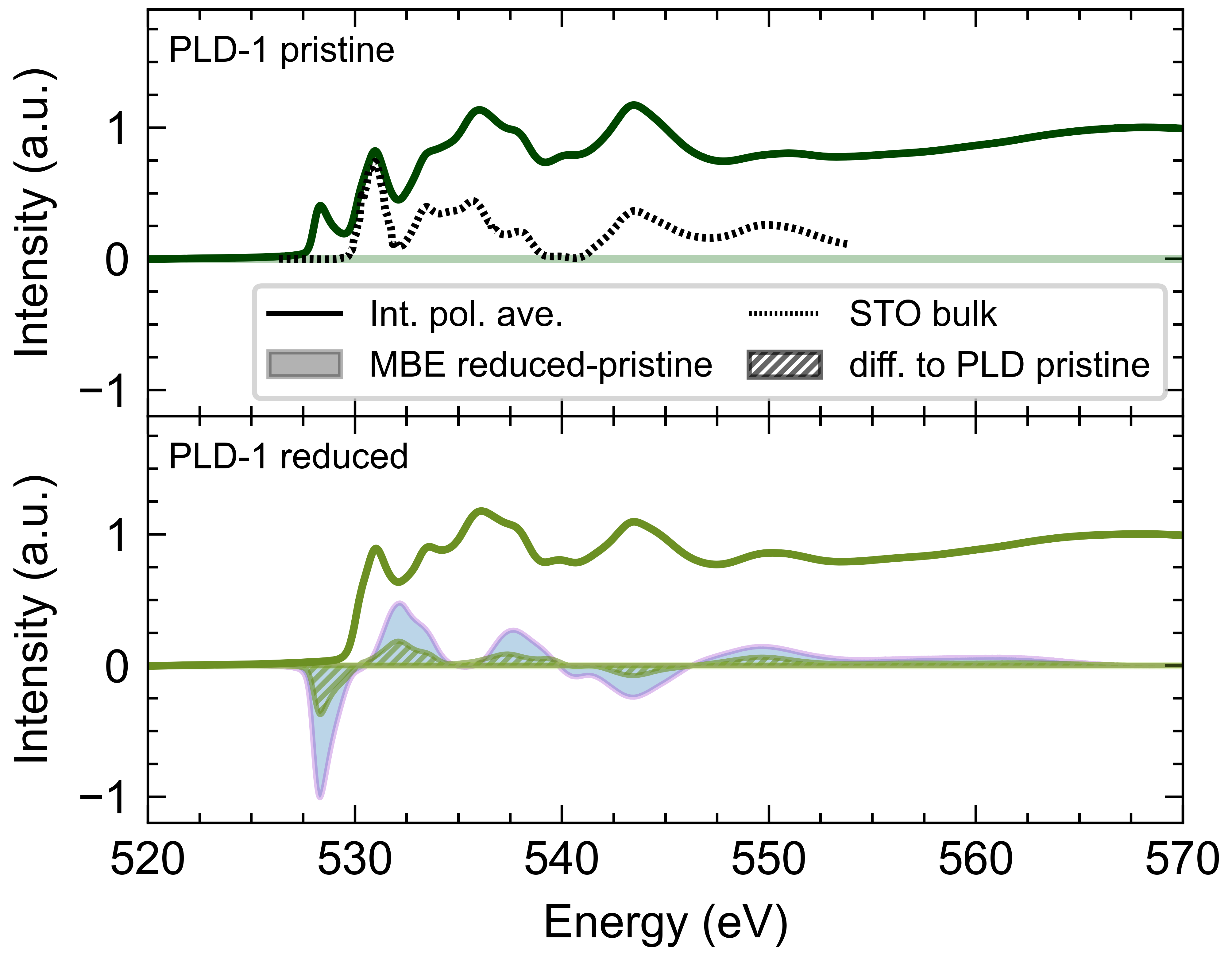}
\caption{O-$K$ edge polarization averaged spectra of the PLD-1 sample pair. As a reference the spectrum of bulk STO measured by Ref.~\cite{BegumHudde2023S} is shown. The sample has a STO capping layer on top of the PNO which, while irrelevant at the Ni-$L$ edge, does produce features at the O-$K$ edge. Since both pristine and reduced sample pieces have this cap layer, its effect disappears in the difference spectrum (green hashed area). This very similar to the difference spectrum of the pristine and most reduced MBE-1 sample (violet area), except for a scaling factor of 2.66.}\label{Fig_oxygen_diff_PLD}
\end{figure}

\clearpage
\section*{Parameters used in the cluster calculations}

The Ni-$3d$ orbital energies for the $D_{4h}$ ligand field where the crystal field energies parametrized by $10Dq$, $D_s$, $D_t$ follow from the following set of equations:
\begin{eqnarray*}
    E_{x^2-y^2}=&6 Dq + 2 D_s - D_t\\
    E_{3z^2-r^2}=&6 Dq - 2 D_s - 6D_t\\
    E_{xy}=&-4 Dq + 2 D_s - D_t\\
    E_{xz, yz}=&-4 Dq - D_s + 4 D_t
\end{eqnarray*}

\begin{table*}[htb]
\setlength{\tabcolsep}{5pt}
\renewcommand{\arraystretch}{1.2}
\centering
\begin{tabular}{l | c c c c c c | c c c c c c}
\hline\hline
& $D_s$ & $Vpd_{x^2-y^2}$ & $Vpd_{3z^2-r^2}$ & $Vpd_{xz, yz}$ & $Vpd_{xy}$ & $\Delta$ & $n_{3d}$ & $n_{L}$ & $n_{x^2-y^2}$ & $n_{3z^2-r^2}$ & $n_{xy}$ & $n_{xz/yz}$\\
\hline
$3d^7$ ($d^8L$) & 0.1 & 2.591 & 2.558 & 1.711 & 1.727 & -0.5 & 7.79  & 9.21 & 0.71 & 1.19 & 1.92 & 3.96\\
$3d^8$ HS & 0.1 & 2.591 & 2.558 & 1.711 & 1.727 & 4.5 & 8.21  & 9.79 & 1.10 & 1.11 & 2.00 & 4.00 \\
$3d^8$ LS & 0.35 & 3.043 & 1.757 & 1.435 & 2.029 & 4.5 & 8.33  & 9.67 & 0.41 & 1.97 & 1.98 & 3.98 \\
$3d^9L$ & 0.35 & 3.043 & 1.757 & 1.435 & 2.029 & -0.5 & 8.73  & 9.27 & 0.75 & 1.99 & 2.00 & 4.00 \\
$3d^9$ & 0.35 & 3.043 & 1.757 & 1.435 & 2.029 & 4.5 & 9.15  & 9.85 & 1.16 & 2.00 & 2.00 & 4.00 \\
\hline\hline
\end{tabular}
\caption{Left: $D_{4h}$ ligand field parameters (all values in eV). O-$2p$ Ni-$3d$ hopping integrals in $D_{4h}$ are given by $V_{a1g} = -(pd\sigma_x+2pd\sigma_z)/\sqrt{3}$, $V_{b1g} = -\sqrt{3} pd\sigma_x$, $V_{eg} = pd\pi_x + pd\pi_z$, $V_{b2g} = 2 pd\pi$ with $pd\pi_i = -pd\sigma_i / 2.17$ for $i=x, z$ \cite{Harrison1987S}. $pd\sigma_i$ and $pd\pi_i$ were calculated using the formulas given in (B7) in Ref.~\cite{Harrison1987S} with $r_p^{\rm O}=4.41$~\AA\ and $r_d^{\rm Ni}=0.767$~\AA\ and Ni-O distances of LaNiO$_{2.5}$ of $d_x=2.04$~\AA, $d_z=2.03$~\AA\ (HS) and $d_x=1.929$~\AA, $d_z=\infty$ (LS). For the $d^9$ configuration we used the values reported for LaNiO$_2$ of $d_x=1.980$~\AA, $d_z=\infty$. Right: Electron occupation numbers in the ground state of the different nickel configurations obtained from the single cluster ligand-field  calculations. Here $n_{3d}$ denotes the total $3d$ occupation and $n_{L}$ is the electron occupation of the five relevant O-$2p$ ligands which deviate form their purely ionic values of $n_{3d}$= 7, 8, or 9 and $n_{L}$=10 due to charge transfer.
}\label{Tab_Cluster1}
\end{table*}

\begin{table*}[hbt]
\setlength{\tabcolsep}{10pt}
\renewcommand{\arraystretch}{1.2}
\centering
\begin{tabular}{l | c c c c c c}
\hline\hline
& $U_{dd}$ (eV) & $F^{2}_{dd}$ / $F^{4}_{dd}$ (eV)* & $\zeta_{3d}$ (eV) & $U_{pd}$ (eV) & $F^2_{pd}$ / $G^1_{pd}$ / $G^3_{pd}$ (eV) * & $\zeta_{core}$ (eV)\\
\hline
$3d^7$ & 6.0  & 10.621 / 6.635 & 0.091 & 7.0 & 6.679 / 5.063 / 2.882 & 11.506\\
$3d^8$ & 6.0  & 9.786 / 6.076 & 0.083 & 7.0 & 6.176 / 4.626 / 2.632 & 11.507\\
$3d^9$ & 6.0  & 8.867 / 5.468 & 0.074 & 7.0 & 5.678 / 4.210 / 2.394 & 11.509\\
\hline\hline
\end{tabular}
\caption{Parameters used in the cluster calculations: $U_{dd}$ denotes the Hubbard Coulomb repulsion between electrons in the $d$ shell and $\Delta$ is the charge-transfer energy. For the Coulomb repulsion of the excited core electron $U_{pd}$ we used the values of NiO. The Slater integrals for the ground state $2p^6 3d^n$ ($F^{2}_{dd}$ / $F^{4}_{dd}$) and the state with excited core electron $2p^5 3d^{n+1}$ ($F^2_{pd}$ / $G^1_{pd}$ / $G^3_{pd}$), as well as the spin-orbit coupling in the $3d$ ($\zeta_{3d}$) and $2p$ core-level ($\zeta_{core}$) were calculated within the Hartree-Fock approximation \cite{Haverkort2005S}. Values marked with an asterisk were reduced by a factor of 0.8 from their atomic values (Refs.~\onlinecite{deGroot1990S,Mizokawa1996S}).}\label{Tab_Cluster2}
\end{table*}

\begin{figure*}[htb]
\center\includegraphics[width=0.9\textwidth]{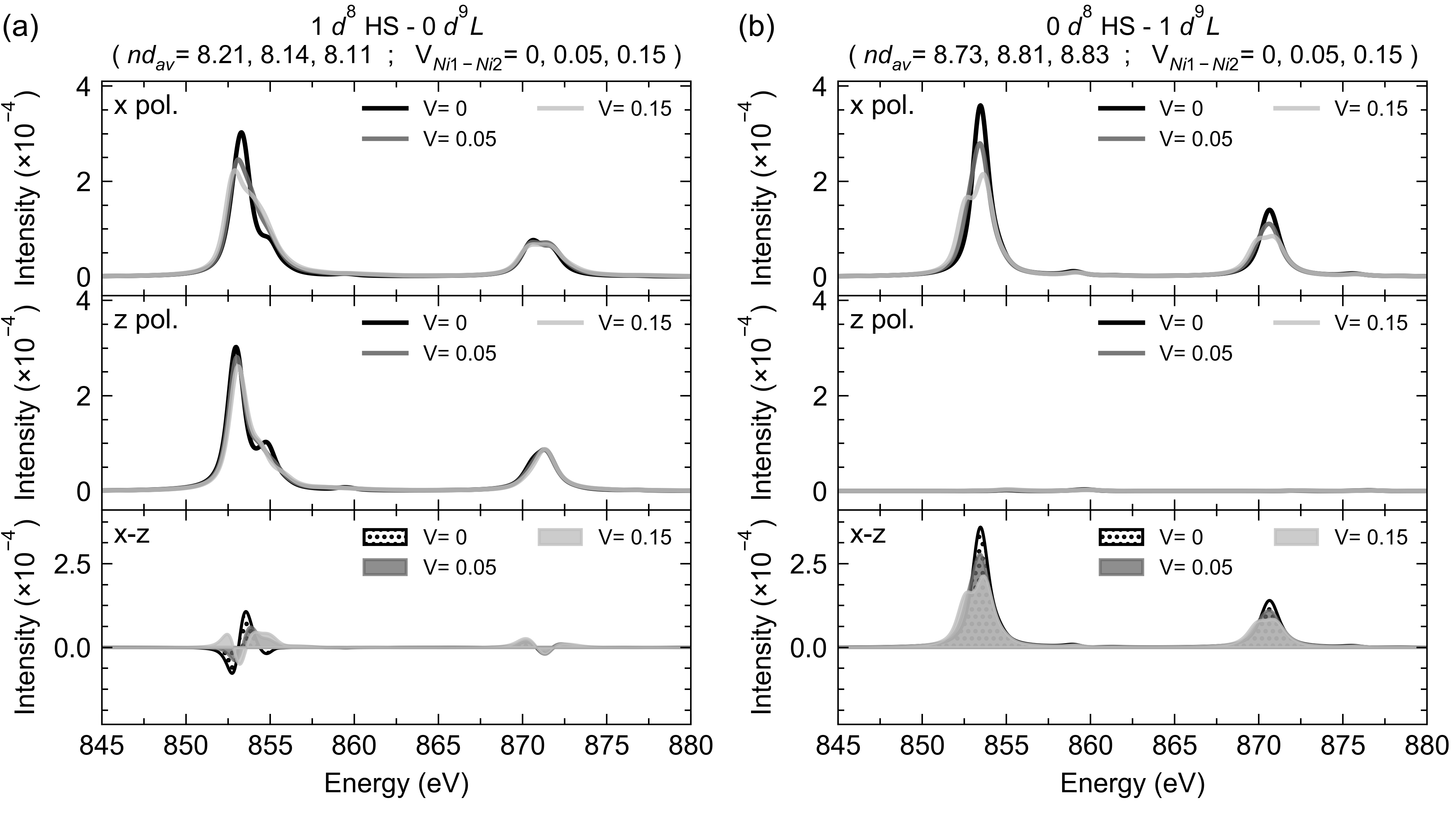}
\caption{Double cluster calculation for Ni1 site with $d^8$~HS parameters and Ni2 with $d^9\underline{L}$ at three different inter cluster coupling strengths $V_{Ni1-Ni2}$ from uncoupled ($V=0$) to $V=0.15$. The left and right panels show how each site contributes to the polarization-dependent Ni-$L$ edge XAS. Using the nomenclature of the main text, they correspond to 1.0~Ni1--0.0~Ni2 and 0.0~Ni1--1.0~Ni2, respectively. Subfigure (a) displays the $d^8$~HS sites contribution and (b) the $d^9\underline{L}$. (a) and (b) title also list the site's Ni occupation $nd_{av}$ for the corresponding parameter sets, ordered from weakest to strongest coupling.}\label{Fig_dc-theory_AC}
\end{figure*}

\begin{figure*}[hbt]
\center\includegraphics[width=0.9\textwidth]{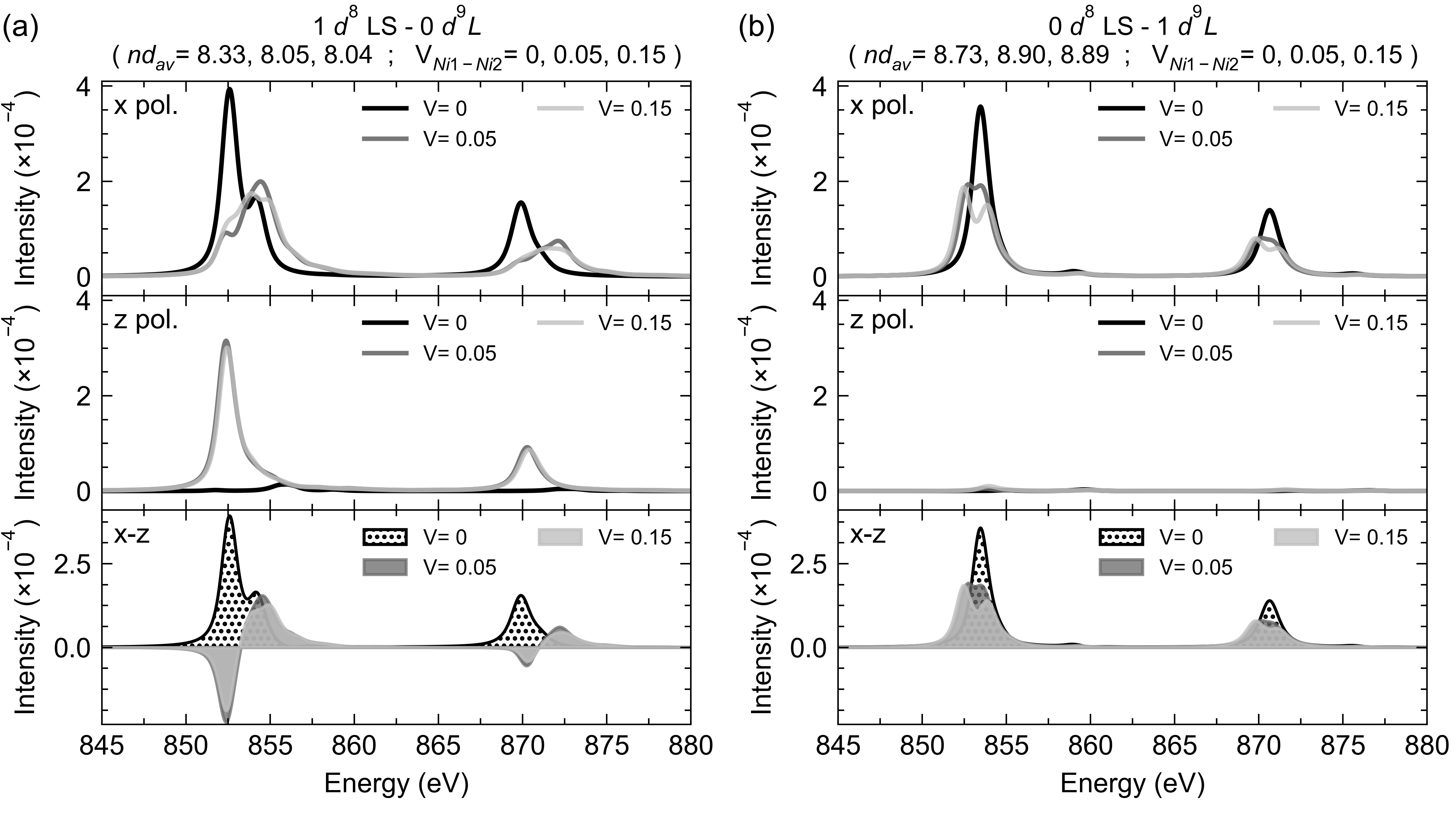}
\caption{Double cluster results similarly presented as Fig.~\ref{Fig_dc-theory_AC} but now for the low-spin parameters on site Ni1: Double cluster calculation for Ni1 site with $d^8$~LS parameters and Ni2 with $d^9\underline{L}$ at three different inter cluster coupling strengths $V_{Ni1-Ni2}$ from uncoupled ($V=0$) to $V=0.15$. Again, the left and right panels show how each site contributes to the polarization-dependent Ni-$L$ edge XAS. Using the nomenclature of the main text, they correspond to 1.0~Ni1--0.0~Ni2 and 0.0~Ni1--1.0~Ni2, respectively. Subfigure (a) displays the $d^8$~LS sites contribution and (b) the $d^9\underline{L}$. (a) and (b) title also list the site's Ni occupation $nd_{av}$ for the corresponding parameter sets, ordered from weakest to strongest coupling.}\label{Fig_dc-theory_BC}
\end{figure*}


\begin{thebibliography}{41}
\makeatletter
\providecommand \@ifxundefined [1]{%
 \@ifx{#1\undefined}
}%
\providecommand \@ifnum [1]{%
 \ifnum #1\expandafter \@firstoftwo
 \else \expandafter \@secondoftwo
 \fi
}%
\providecommand \@ifx [1]{%
 \ifx #1\expandafter \@firstoftwo
 \else \expandafter \@secondoftwo
 \fi
}%
\providecommand \natexlab [1]{#1}%
\providecommand \enquote  [1]{``#1''}%
\providecommand \bibnamefont  [1]{#1}%
\providecommand \bibfnamefont [1]{#1}%
\providecommand \citenamefont [1]{#1}%
\providecommand \href@noop [0]{\@secondoftwo}%
\providecommand \href [0]{\begingroup \@sanitize@url \@href}%
\providecommand \@href[1]{\@@startlink{#1}\@@href}%
\providecommand \@@href[1]{\endgroup#1\@@endlink}%
\providecommand \@sanitize@url [0]{\catcode `\\12\catcode `\$12\catcode
  `\&12\catcode `\#12\catcode `\^12\catcode `\_12\catcode `\%12\relax}%
\providecommand \@@startlink[1]{}%
\providecommand \@@endlink[0]{}%
\providecommand \url  [0]{\begingroup\@sanitize@url \@url }%
\providecommand \@url [1]{\endgroup\@href {#1}{\urlprefix }}%
\providecommand \urlprefix  [0]{URL }%
\providecommand \Eprint [0]{\href }%
\providecommand \doibase [0]{https://doi.org/}%
\providecommand \selectlanguage [0]{\@gobble}%
\providecommand \bibinfo  [0]{\@secondoftwo}%
\providecommand \bibfield  [0]{\@secondoftwo}%
\providecommand \translation [1]{[#1]}%
\providecommand \BibitemOpen [0]{}%
\providecommand \bibitemStop [0]{}%
\providecommand \bibitemNoStop [0]{.\EOS\space}%
\providecommand \EOS [0]{\spacefactor3000\relax}%
\providecommand \BibitemShut  [1]{\csname bibitem#1\endcsname}%
\let\auto@bib@innerbib\@empty
\bibitem [{\citenamefont {Hepting}\ \emph {et~al.}(2020)\citenamefont
  {Hepting}, \citenamefont {Li}, \citenamefont {Jia}, \citenamefont {Lu},
  \citenamefont {Paris}, \citenamefont {Tseng}, \citenamefont {Feng},
  \citenamefont {Osada}, \citenamefont {Been}, \citenamefont {Hikita},
  \citenamefont {Chuang}, \citenamefont {Hussain}, \citenamefont {Zhou},
  \citenamefont {Nag}, \citenamefont {Garcia-Fernandez}, \citenamefont {Rossi},
  \citenamefont {Huang}, \citenamefont {Huang}, \citenamefont {Shen},
  \citenamefont {Schmitt}, \citenamefont {Hwang}, \citenamefont {Moritz},
  \citenamefont {Zaanen}, \citenamefont {Devereaux},\ and\ \citenamefont
  {Lee}}]{Hepting2020}%
  \BibitemOpen
  \bibfield  {author} {\bibinfo {author} {\bibfnamefont {M.}~\bibnamefont
  {Hepting}}, \bibinfo {author} {\bibfnamefont {D.}~\bibnamefont {Li}},
  \bibinfo {author} {\bibfnamefont {C.~J.}\ \bibnamefont {Jia}}, \bibinfo
  {author} {\bibfnamefont {H.}~\bibnamefont {Lu}}, \bibinfo {author}
  {\bibfnamefont {E.}~\bibnamefont {Paris}}, \bibinfo {author} {\bibfnamefont
  {Y.}~\bibnamefont {Tseng}}, \bibinfo {author} {\bibfnamefont
  {X.}~\bibnamefont {Feng}}, \bibinfo {author} {\bibfnamefont {M.}~\bibnamefont
  {Osada}}, \bibinfo {author} {\bibfnamefont {E.}~\bibnamefont {Been}},
  \bibinfo {author} {\bibfnamefont {Y.}~\bibnamefont {Hikita}}, \bibinfo
  {author} {\bibfnamefont {Y.-D.}\ \bibnamefont {Chuang}}, \bibinfo {author}
  {\bibfnamefont {Z.}~\bibnamefont {Hussain}}, \bibinfo {author} {\bibfnamefont
  {K.~J.}\ \bibnamefont {Zhou}}, \bibinfo {author} {\bibfnamefont
  {A.}~\bibnamefont {Nag}}, \bibinfo {author} {\bibfnamefont {M.}~\bibnamefont
  {Garcia-Fernandez}}, \bibinfo {author} {\bibfnamefont {M.}~\bibnamefont
  {Rossi}}, \bibinfo {author} {\bibfnamefont {H.~Y.}\ \bibnamefont {Huang}},
  \bibinfo {author} {\bibfnamefont {D.~J.}\ \bibnamefont {Huang}}, \bibinfo
  {author} {\bibfnamefont {Z.~X.}\ \bibnamefont {Shen}}, \bibinfo {author}
  {\bibfnamefont {T.}~\bibnamefont {Schmitt}}, \bibinfo {author} {\bibfnamefont
  {H.~Y.}\ \bibnamefont {Hwang}}, \bibinfo {author} {\bibfnamefont
  {B.}~\bibnamefont {Moritz}}, \bibinfo {author} {\bibfnamefont
  {J.}~\bibnamefont {Zaanen}}, \bibinfo {author} {\bibfnamefont {T.~P.}\
  \bibnamefont {Devereaux}},\ and\ \bibinfo {author} {\bibfnamefont {W.~S.}\
  \bibnamefont {Lee}},\ }\bibfield  {title} {\bibinfo {title} {Electronic
  structure of the parent compound of superconducting infinite-layer
  nickelates},\ }\href {https://doi.org/10.1038/s41563-019-0585-z} {\bibfield
  {journal} {\bibinfo  {journal} {Nature Materials}\ }\textbf {\bibinfo
  {volume} {19}},\ \bibinfo {pages} {381} (\bibinfo {year} {2020})}\BibitemShut
  {NoStop}%
\bibitem [{\citenamefont {Li}\ \emph {et~al.}(2019)\citenamefont {Li},
  \citenamefont {Lee}, \citenamefont {Wang}, \citenamefont {Osada},
  \citenamefont {Crossley}, \citenamefont {Lee}, \citenamefont {Cui},
  \citenamefont {Hikita},\ and\ \citenamefont {Hwang}}]{Li2019}%
  \BibitemOpen
  \bibfield  {author} {\bibinfo {author} {\bibfnamefont {D.}~\bibnamefont
  {Li}}, \bibinfo {author} {\bibfnamefont {K.}~\bibnamefont {Lee}}, \bibinfo
  {author} {\bibfnamefont {B.~Y.}\ \bibnamefont {Wang}}, \bibinfo {author}
  {\bibfnamefont {M.}~\bibnamefont {Osada}}, \bibinfo {author} {\bibfnamefont
  {S.}~\bibnamefont {Crossley}}, \bibinfo {author} {\bibfnamefont {H.~R.}\
  \bibnamefont {Lee}}, \bibinfo {author} {\bibfnamefont {Y.}~\bibnamefont
  {Cui}}, \bibinfo {author} {\bibfnamefont {Y.}~\bibnamefont {Hikita}},\ and\
  \bibinfo {author} {\bibfnamefont {H.~Y.}\ \bibnamefont {Hwang}},\ }\bibfield
  {title} {\bibinfo {title} {Superconductivity in an infinite-layer
  nickelate},\ }\href {https://doi.org/10.1038/s41586-019-1496-5} {\bibfield
  {journal} {\bibinfo  {journal} {Nature}\ }\textbf {\bibinfo {volume} {572}},\
  \bibinfo {pages} {624} (\bibinfo {year} {2019})}\BibitemShut {NoStop}%
\bibitem [{\citenamefont {Zeng}\ \emph {et~al.}(2020)\citenamefont {Zeng},
  \citenamefont {Tang}, \citenamefont {Yin}, \citenamefont {Li}, \citenamefont
  {Li}, \citenamefont {Huang}, \citenamefont {Hu}, \citenamefont {Liu},
  \citenamefont {Omar}, \citenamefont {Jani}, \citenamefont {Lim},
  \citenamefont {Han}, \citenamefont {Wan}, \citenamefont {Yang}, \citenamefont
  {Pennycook}, \citenamefont {Wee},\ and\ \citenamefont {Ariando}}]{Zeng2020}%
  \BibitemOpen
  \bibfield  {author} {\bibinfo {author} {\bibfnamefont {S.}~\bibnamefont
  {Zeng}}, \bibinfo {author} {\bibfnamefont {C.~S.}\ \bibnamefont {Tang}},
  \bibinfo {author} {\bibfnamefont {X.}~\bibnamefont {Yin}}, \bibinfo {author}
  {\bibfnamefont {C.}~\bibnamefont {Li}}, \bibinfo {author} {\bibfnamefont
  {M.}~\bibnamefont {Li}}, \bibinfo {author} {\bibfnamefont {Z.}~\bibnamefont
  {Huang}}, \bibinfo {author} {\bibfnamefont {J.}~\bibnamefont {Hu}}, \bibinfo
  {author} {\bibfnamefont {W.}~\bibnamefont {Liu}}, \bibinfo {author}
  {\bibfnamefont {G.~J.}\ \bibnamefont {Omar}}, \bibinfo {author}
  {\bibfnamefont {H.}~\bibnamefont {Jani}}, \bibinfo {author} {\bibfnamefont
  {Z.~S.}\ \bibnamefont {Lim}}, \bibinfo {author} {\bibfnamefont
  {K.}~\bibnamefont {Han}}, \bibinfo {author} {\bibfnamefont {D.}~\bibnamefont
  {Wan}}, \bibinfo {author} {\bibfnamefont {P.}~\bibnamefont {Yang}}, \bibinfo
  {author} {\bibfnamefont {S.~J.}\ \bibnamefont {Pennycook}}, \bibinfo {author}
  {\bibfnamefont {A.~T.~S.}\ \bibnamefont {Wee}},\ and\ \bibinfo {author}
  {\bibfnamefont {A.}~\bibnamefont {Ariando}},\ }\bibfield  {title} {\bibinfo
  {title} {Phase diagram and superconducting dome of infinite-layer
  $\mathrm{Nd_{1-x}Sr_xNiO_2}$ thin films},\ }\href
  {https://doi.org/10.1103/PhysRevLett.125.147003} {\bibfield  {journal}
  {\bibinfo  {journal} {Phys. Rev. Lett.}\ }\textbf {\bibinfo {volume} {125}},\
  \bibinfo {pages} {147003} (\bibinfo {year} {2020})}\BibitemShut {NoStop}%
\bibitem [{\citenamefont {Osada}\ \emph {et~al.}(2020)\citenamefont {Osada},
  \citenamefont {Wang}, \citenamefont {Lee}, \citenamefont {Li},\ and\
  \citenamefont {Hwang}}]{Osada2020}%
  \BibitemOpen
  \bibfield  {author} {\bibinfo {author} {\bibfnamefont {M.}~\bibnamefont
  {Osada}}, \bibinfo {author} {\bibfnamefont {B.~Y.}\ \bibnamefont {Wang}},
  \bibinfo {author} {\bibfnamefont {K.}~\bibnamefont {Lee}}, \bibinfo {author}
  {\bibfnamefont {D.}~\bibnamefont {Li}},\ and\ \bibinfo {author}
  {\bibfnamefont {H.~Y.}\ \bibnamefont {Hwang}},\ }\bibfield  {title} {\bibinfo
  {title} {Phase diagram of infinite layer praseodymium nickelate
  $\mathrm{Pr_{1-x}Sr_xNiO_2}$ thin films},\ }\href
  {https://doi.org/10.1103/PhysRevMaterials.4.121801} {\bibfield  {journal}
  {\bibinfo  {journal} {Phys. Rev. Materials}\ }\textbf {\bibinfo {volume}
  {4}},\ \bibinfo {pages} {121801} (\bibinfo {year} {2020})}\BibitemShut
  {NoStop}%
\bibitem [{\citenamefont {Osada}\ \emph {et~al.}(2021)\citenamefont {Osada},
  \citenamefont {Wang}, \citenamefont {Goodge}, \citenamefont {Harvey},
  \citenamefont {Lee}, \citenamefont {Li}, \citenamefont {Kourkoutis},\ and\
  \citenamefont {Hwang}}]{Osada2021}%
  \BibitemOpen
  \bibfield  {author} {\bibinfo {author} {\bibfnamefont {M.}~\bibnamefont
  {Osada}}, \bibinfo {author} {\bibfnamefont {B.~Y.}\ \bibnamefont {Wang}},
  \bibinfo {author} {\bibfnamefont {B.~H.}\ \bibnamefont {Goodge}}, \bibinfo
  {author} {\bibfnamefont {S.~P.}\ \bibnamefont {Harvey}}, \bibinfo {author}
  {\bibfnamefont {K.}~\bibnamefont {Lee}}, \bibinfo {author} {\bibfnamefont
  {D.}~\bibnamefont {Li}}, \bibinfo {author} {\bibfnamefont {L.~F.}\
  \bibnamefont {Kourkoutis}},\ and\ \bibinfo {author} {\bibfnamefont {H.~Y.}\
  \bibnamefont {Hwang}},\ }\bibfield  {title} {\bibinfo {title} {Nickelate
  superconductivity without rare-earth magnetism: $\mathrm{(La,Sr)NiO_2}$},\
  }\href {https://doi.org/https://doi.org/10.1002/adma.202104083} {\bibfield
  {journal} {\bibinfo  {journal} {Advanced Materials}\ }\textbf {\bibinfo
  {volume} {33}},\ \bibinfo {pages} {2104083} (\bibinfo {year}
  {2021})}\BibitemShut {NoStop}%
\bibitem [{\citenamefont {Zeng}\ \emph {et~al.}(2022)\citenamefont {Zeng},
  \citenamefont {Li}, \citenamefont {Chow}, \citenamefont {Cao}, \citenamefont
  {Zhang}, \citenamefont {Tang}, \citenamefont {Yin}, \citenamefont {Lim},
  \citenamefont {Hu}, \citenamefont {Yang},\ and\ \citenamefont
  {Ariando}}]{Zeng2022}%
  \BibitemOpen
  \bibfield  {author} {\bibinfo {author} {\bibfnamefont {S.}~\bibnamefont
  {Zeng}}, \bibinfo {author} {\bibfnamefont {C.}~\bibnamefont {Li}}, \bibinfo
  {author} {\bibfnamefont {L.~E.}\ \bibnamefont {Chow}}, \bibinfo {author}
  {\bibfnamefont {Y.}~\bibnamefont {Cao}}, \bibinfo {author} {\bibfnamefont
  {Z.}~\bibnamefont {Zhang}}, \bibinfo {author} {\bibfnamefont {C.~S.}\
  \bibnamefont {Tang}}, \bibinfo {author} {\bibfnamefont {X.}~\bibnamefont
  {Yin}}, \bibinfo {author} {\bibfnamefont {Z.~S.}\ \bibnamefont {Lim}},
  \bibinfo {author} {\bibfnamefont {J.}~\bibnamefont {Hu}}, \bibinfo {author}
  {\bibfnamefont {P.}~\bibnamefont {Yang}},\ and\ \bibinfo {author}
  {\bibfnamefont {A.}~\bibnamefont {Ariando}},\ }\bibfield  {title} {\bibinfo
  {title} {Superconductivity in infinite-layer nickelate
  $\mathrm{La_{1-x}Ca_xNiO_2}$ thin films},\ }\href
  {https://doi.org/10.1126/sciadv.abl9927} {\bibfield  {journal} {\bibinfo
  {journal} {Science Advances}\ }\textbf {\bibinfo {volume} {8}},\ \bibinfo
  {pages} {eabl9927} (\bibinfo {year} {2022})}\BibitemShut {NoStop}%
\bibitem [{\citenamefont {Wei}\ \emph {et~al.}(2023)\citenamefont {Wei},
  \citenamefont {Vu}, \citenamefont {Zhang}, \citenamefont {Walker},\ and\
  \citenamefont {Ahn}}]{Wei2023}%
  \BibitemOpen
  \bibfield  {author} {\bibinfo {author} {\bibfnamefont {W.}~\bibnamefont
  {Wei}}, \bibinfo {author} {\bibfnamefont {D.}~\bibnamefont {Vu}}, \bibinfo
  {author} {\bibfnamefont {Z.}~\bibnamefont {Zhang}}, \bibinfo {author}
  {\bibfnamefont {F.~J.}\ \bibnamefont {Walker}},\ and\ \bibinfo {author}
  {\bibfnamefont {C.~H.}\ \bibnamefont {Ahn}},\ }\bibfield  {title} {\bibinfo
  {title} {Superconducting $\mathrm{Nd_{1-x}Eu_xNiO_2}$ thin films using in
  situ synthesis},\ }\href {https://doi.org/10.1126/sciadv.adh3327} {\bibfield
  {journal} {\bibinfo  {journal} {Science Advances}\ }\textbf {\bibinfo
  {volume} {9}},\ \bibinfo {pages} {eadh3327} (\bibinfo {year}
  {2023})}\BibitemShut {NoStop}%
\bibitem [{\citenamefont {Pan}\ \emph {et~al.}(2022)\citenamefont {Pan},
  \citenamefont {Ferenc~Segedin}, \citenamefont {LaBollita}, \citenamefont
  {Song}, \citenamefont {Nica}, \citenamefont {Goodge}, \citenamefont {Pierce},
  \citenamefont {Doyle}, \citenamefont {Novakov}, \citenamefont
  {C{\'o}rdova~Carrizales}, \citenamefont {N'Diaye}, \citenamefont {Shafer},
  \citenamefont {Paik}, \citenamefont {Heron}, \citenamefont {Mason},
  \citenamefont {Yacoby}, \citenamefont {Kourkoutis}, \citenamefont {Erten},
  \citenamefont {Brooks}, \citenamefont {Botana},\ and\ \citenamefont
  {Mundy}}]{Pan2022}%
  \BibitemOpen
  \bibfield  {author} {\bibinfo {author} {\bibfnamefont {G.~A.}\ \bibnamefont
  {Pan}}, \bibinfo {author} {\bibfnamefont {D.}~\bibnamefont {Ferenc~Segedin}},
  \bibinfo {author} {\bibfnamefont {H.}~\bibnamefont {LaBollita}}, \bibinfo
  {author} {\bibfnamefont {Q.}~\bibnamefont {Song}}, \bibinfo {author}
  {\bibfnamefont {E.~M.}\ \bibnamefont {Nica}}, \bibinfo {author}
  {\bibfnamefont {B.~H.}\ \bibnamefont {Goodge}}, \bibinfo {author}
  {\bibfnamefont {A.~T.}\ \bibnamefont {Pierce}}, \bibinfo {author}
  {\bibfnamefont {S.}~\bibnamefont {Doyle}}, \bibinfo {author} {\bibfnamefont
  {S.}~\bibnamefont {Novakov}}, \bibinfo {author} {\bibfnamefont
  {D.}~\bibnamefont {C{\'o}rdova~Carrizales}}, \bibinfo {author} {\bibfnamefont
  {A.~T.}\ \bibnamefont {N'Diaye}}, \bibinfo {author} {\bibfnamefont
  {P.}~\bibnamefont {Shafer}}, \bibinfo {author} {\bibfnamefont
  {H.}~\bibnamefont {Paik}}, \bibinfo {author} {\bibfnamefont {J.~T.}\
  \bibnamefont {Heron}}, \bibinfo {author} {\bibfnamefont {J.~A.}\ \bibnamefont
  {Mason}}, \bibinfo {author} {\bibfnamefont {A.}~\bibnamefont {Yacoby}},
  \bibinfo {author} {\bibfnamefont {L.~F.}\ \bibnamefont {Kourkoutis}},
  \bibinfo {author} {\bibfnamefont {O.}~\bibnamefont {Erten}}, \bibinfo
  {author} {\bibfnamefont {C.~M.}\ \bibnamefont {Brooks}}, \bibinfo {author}
  {\bibfnamefont {A.~S.}\ \bibnamefont {Botana}},\ and\ \bibinfo {author}
  {\bibfnamefont {J.~A.}\ \bibnamefont {Mundy}},\ }\bibfield  {title} {\bibinfo
  {title} {{Superconductivity in a quintuple-layer square-planar nickelate}},\
  }\href {https://doi.org/10.1038/s41563-021-01142-9} {\bibfield  {journal}
  {\bibinfo  {journal} {Nature Materials}\ }\textbf {\bibinfo {volume} {21}},\
  \bibinfo {pages} {160} (\bibinfo {year} {2022})}\BibitemShut {NoStop}%
\bibitem [{\citenamefont {Alonso}\ and\ \citenamefont
  {Mart{\'{\i}}nez-Lope}(1996)}]{alonso1996}%
  \BibitemOpen
  \bibfield  {author} {\bibinfo {author} {\bibfnamefont {J.~A.}\ \bibnamefont
  {Alonso}}\ and\ \bibinfo {author} {\bibfnamefont {M.}~\bibnamefont
  {Mart{\'{\i}}nez-Lope}},\ }\bibfield  {title} {\bibinfo {title} {Resolution
  of the crystal structure of the deficient perovskite LaNiO$_{2.5}$ from
  neutron powder diffraction data},\ }in\ \href
  {https://doi.org/10.4028/www.scientific.net/MSF.228-231.747} {\emph {\bibinfo
  {booktitle} {European Powder Diffraction 4}}},\ \bibinfo {series} {Materials
  Science Forum}, Vol.\ \bibinfo {volume} {228}\ (\bibinfo  {publisher} {Trans
  Tech Publications Ltd},\ \bibinfo {year} {1996})\ pp.\ \bibinfo {pages}
  {747--752}\BibitemShut {NoStop}%
\bibitem [{\citenamefont {Moriga}\ \emph {et~al.}(2002)\citenamefont {Moriga},
  \citenamefont {Hayashi}, \citenamefont {Sakamoto}, \citenamefont {Orihara},\
  and\ \citenamefont {Nakabayashi}}]{Moriga2002}%
  \BibitemOpen
  \bibfield  {author} {\bibinfo {author} {\bibfnamefont {T.}~\bibnamefont
  {Moriga}}, \bibinfo {author} {\bibfnamefont {M.}~\bibnamefont {Hayashi}},
  \bibinfo {author} {\bibfnamefont {T.}~\bibnamefont {Sakamoto}}, \bibinfo
  {author} {\bibfnamefont {M.}~\bibnamefont {Orihara}},\ and\ \bibinfo {author}
  {\bibfnamefont {I.}~\bibnamefont {Nakabayashi}},\ }\bibfield  {title}
  {\bibinfo {title} {Reduction processes of rare-earth nickelate perovskites
  LnNiO$_3$ (Ln=La, Pr, Nd)},\ }\href
  {https://doi.org/https://doi.org/10.1016/S0167-2738(02)00440-X} {\bibfield
  {journal} {\bibinfo  {journal} {Solid State Ionics}\ }\textbf {\bibinfo
  {volume} {154-155}},\ \bibinfo {pages} {251} (\bibinfo {year}
  {2002})}\BibitemShut {NoStop}%
\bibitem [{\citenamefont {Sahib}\ \emph {et~al.}(2025)\citenamefont {Sahib},
  \citenamefont {Raji}, \citenamefont {Rosa}, \citenamefont {Merzoni},
  \citenamefont {Ghiringhelli}, \citenamefont {Salluzzo}, \citenamefont
  {Gloter}, \citenamefont {Viart},\ and\ \citenamefont {Preziosi}}]{Sahib2025}%
  \BibitemOpen
  \bibfield  {author} {\bibinfo {author} {\bibfnamefont {H.}~\bibnamefont
  {Sahib}}, \bibinfo {author} {\bibfnamefont {A.}~\bibnamefont {Raji}},
  \bibinfo {author} {\bibfnamefont {F.}~\bibnamefont {Rosa}}, \bibinfo {author}
  {\bibfnamefont {G.}~\bibnamefont {Merzoni}}, \bibinfo {author} {\bibfnamefont
  {G.}~\bibnamefont {Ghiringhelli}}, \bibinfo {author} {\bibfnamefont
  {M.}~\bibnamefont {Salluzzo}}, \bibinfo {author} {\bibfnamefont
  {A.}~\bibnamefont {Gloter}}, \bibinfo {author} {\bibfnamefont
  {N.}~\bibnamefont {Viart}},\ and\ \bibinfo {author} {\bibfnamefont
  {D.}~\bibnamefont {Preziosi}},\ }\bibfield  {title} {\bibinfo {title}
  {Superconductivity in PrNiO$_2$ infinite-layer nickelates},\ }\href
  {https://doi.org/https://doi.org/10.1002/adma.202416187} {\bibfield
  {journal} {\bibinfo  {journal} {Advanced Materials}\ }\textbf {\bibinfo
  {volume} {37}},\ \bibinfo {pages} {2416187} (\bibinfo {year}
  {2025})}\BibitemShut {NoStop}%
\bibitem [{\citenamefont {Parzyck}\ \emph {et~al.}(2025)\citenamefont
  {Parzyck}, \citenamefont {Wu}, \citenamefont {Bhatt}, \citenamefont {Kang},
  \citenamefont {Arthur}, \citenamefont {Pedersen}, \citenamefont {Sutarto},
  \citenamefont {Fan}, \citenamefont {Pelliciari}, \citenamefont {Bisogni},
  \citenamefont {Herranz}, \citenamefont {Georgescu}, \citenamefont {Hawthorn},
  \citenamefont {Kourkoutis}, \citenamefont {Muller}, \citenamefont {Schlom},\
  and\ \citenamefont {Shen}}]{Parzyck2024}%
  \BibitemOpen
  \bibfield  {author} {\bibinfo {author} {\bibfnamefont {C.~T.}\ \bibnamefont
  {Parzyck}}, \bibinfo {author} {\bibfnamefont {Y.}~\bibnamefont {Wu}},
  \bibinfo {author} {\bibfnamefont {L.}~\bibnamefont {Bhatt}}, \bibinfo
  {author} {\bibfnamefont {M.}~\bibnamefont {Kang}}, \bibinfo {author}
  {\bibfnamefont {Z.}~\bibnamefont {Arthur}}, \bibinfo {author} {\bibfnamefont
  {T.~M.}\ \bibnamefont {Pedersen}}, \bibinfo {author} {\bibfnamefont
  {R.}~\bibnamefont {Sutarto}}, \bibinfo {author} {\bibfnamefont
  {S.}~\bibnamefont {Fan}}, \bibinfo {author} {\bibfnamefont {J.}~\bibnamefont
  {Pelliciari}}, \bibinfo {author} {\bibfnamefont {V.}~\bibnamefont {Bisogni}},
  \bibinfo {author} {\bibfnamefont {G.}~\bibnamefont {Herranz}}, \bibinfo
  {author} {\bibfnamefont {A.~B.}\ \bibnamefont {Georgescu}}, \bibinfo {author}
  {\bibfnamefont {D.~G.}\ \bibnamefont {Hawthorn}}, \bibinfo {author}
  {\bibfnamefont {L.~F.}\ \bibnamefont {Kourkoutis}}, \bibinfo {author}
  {\bibfnamefont {D.~A.}\ \bibnamefont {Muller}}, \bibinfo {author}
  {\bibfnamefont {D.~G.}\ \bibnamefont {Schlom}},\ and\ \bibinfo {author}
  {\bibfnamefont {K.~M.}\ \bibnamefont {Shen}},\ }\bibfield  {title} {\bibinfo
  {title} {Superconductivity in the parent infinite-layer nickelate
  ${\mathrm{NdNiO}}_{2}$},\ }\href {https://doi.org/10.1103/PhysRevX.15.021048}
  {\bibfield  {journal} {\bibinfo  {journal} {Phys. Rev. X}\ }\textbf {\bibinfo
  {volume} {15}},\ \bibinfo {pages} {021048} (\bibinfo {year}
  {2025})}\BibitemShut {NoStop}%
\bibitem [{\citenamefont {Lee}\ \emph {et~al.}(2020)\citenamefont {Lee},
  \citenamefont {Goodge}, \citenamefont {Li}, \citenamefont {Osada},
  \citenamefont {Wang}, \citenamefont {Cui}, \citenamefont {Kourkoutis},\ and\
  \citenamefont {Hwang}}]{Lee2020}%
  \BibitemOpen
  \bibfield  {author} {\bibinfo {author} {\bibfnamefont {K.}~\bibnamefont
  {Lee}}, \bibinfo {author} {\bibfnamefont {B.~H.}\ \bibnamefont {Goodge}},
  \bibinfo {author} {\bibfnamefont {D.}~\bibnamefont {Li}}, \bibinfo {author}
  {\bibfnamefont {M.}~\bibnamefont {Osada}}, \bibinfo {author} {\bibfnamefont
  {B.~Y.}\ \bibnamefont {Wang}}, \bibinfo {author} {\bibfnamefont
  {Y.}~\bibnamefont {Cui}}, \bibinfo {author} {\bibfnamefont {L.~F.}\
  \bibnamefont {Kourkoutis}},\ and\ \bibinfo {author} {\bibfnamefont {H.~Y.}\
  \bibnamefont {Hwang}},\ }\bibfield  {title} {\bibinfo {title} {{Aspects of
  the synthesis of thin film superconducting infinite-layer nickelates}},\
  }\href {https://doi.org/10.1063/5.0005103} {\bibfield  {journal} {\bibinfo
  {journal} {APL Materials}\ }\textbf {\bibinfo {volume} {8}},\ \bibinfo
  {pages} {041107} (\bibinfo {year} {2020})}\BibitemShut {NoStop}%
\bibitem [{\citenamefont {Wrobel}\ \emph {et~al.}(2017)\citenamefont {Wrobel},
  \citenamefont {Mark}, \citenamefont {Christiani}, \citenamefont {Sigle},
  \citenamefont {Habermeier}, \citenamefont {van Aken}, \citenamefont
  {Logvenov}, \citenamefont {Keimer},\ and\ \citenamefont
  {Benckiser}}]{Wrobel2017}%
  \BibitemOpen
  \bibfield  {author} {\bibinfo {author} {\bibfnamefont {F.}~\bibnamefont
  {Wrobel}}, \bibinfo {author} {\bibfnamefont {A.~F.}\ \bibnamefont {Mark}},
  \bibinfo {author} {\bibfnamefont {G.}~\bibnamefont {Christiani}}, \bibinfo
  {author} {\bibfnamefont {W.}~\bibnamefont {Sigle}}, \bibinfo {author}
  {\bibfnamefont {H.-U.}\ \bibnamefont {Habermeier}}, \bibinfo {author}
  {\bibfnamefont {P.~A.}\ \bibnamefont {van Aken}}, \bibinfo {author}
  {\bibfnamefont {G.}~\bibnamefont {Logvenov}}, \bibinfo {author}
  {\bibfnamefont {B.}~\bibnamefont {Keimer}},\ and\ \bibinfo {author}
  {\bibfnamefont {E.}~\bibnamefont {Benckiser}},\ }\bibfield  {title} {\bibinfo
  {title} {Comparative study of LaNiO$_3$/LaAlO$_3$ heterostructures grown by
  pulsed laser deposition and oxide molecular beam epitaxy},\ }\href
  {https://doi.org/10.1063/1.4975005} {\bibfield  {journal} {\bibinfo
  {journal} {Applied Physics Letters}\ }\textbf {\bibinfo {volume} {110}},\
  \bibinfo {pages} {041606} (\bibinfo {year} {2017})}\BibitemShut {NoStop}%
\bibitem [{\citenamefont {Karp}\ \emph {et~al.}(2020)\citenamefont {Karp},
  \citenamefont {Botana}, \citenamefont {Norman}, \citenamefont {Park},
  \citenamefont {Zingl},\ and\ \citenamefont {Millis}}]{Karp2020}%
  \BibitemOpen
  \bibfield  {author} {\bibinfo {author} {\bibfnamefont {J.}~\bibnamefont
  {Karp}}, \bibinfo {author} {\bibfnamefont {A.~S.}\ \bibnamefont {Botana}},
  \bibinfo {author} {\bibfnamefont {M.~R.}\ \bibnamefont {Norman}}, \bibinfo
  {author} {\bibfnamefont {H.}~\bibnamefont {Park}}, \bibinfo {author}
  {\bibfnamefont {M.}~\bibnamefont {Zingl}},\ and\ \bibinfo {author}
  {\bibfnamefont {A.}~\bibnamefont {Millis}},\ }\bibfield  {title} {\bibinfo
  {title} {Many-body electronic structure of ${\mathrm{NdNiO}}_{2}$ and
  ${\mathrm{CaCuO}}_{2}$},\ }\href {https://doi.org/10.1103/PhysRevX.10.021061}
  {\bibfield  {journal} {\bibinfo  {journal} {Phys. Rev. X}\ }\textbf {\bibinfo
  {volume} {10}},\ \bibinfo {pages} {021061} (\bibinfo {year}
  {2020})}\BibitemShut {NoStop}%
\bibitem [{\citenamefont {Qu}\ \emph {et~al.}(2025)\citenamefont {Qu},
  \citenamefont {Zhang},\ and\ \citenamefont {Li}}]{Qu2025}%
  \BibitemOpen
  \bibfield  {author} {\bibinfo {author} {\bibfnamefont {H.}~\bibnamefont
  {Qu}}, \bibinfo {author} {\bibfnamefont {G.-M.}\ \bibnamefont {Zhang}},\ and\
  \bibinfo {author} {\bibfnamefont {G.}~\bibnamefont {Li}},\ }\bibfield
  {title} {\bibinfo {title} {Hole distribution and self-doping enhanced
  electronic correlation in hole-doped infinite-layer nickelates},\ }\Eprint
  {https://arxiv.org/abs/2507.10364} {arXiv:2507.10364 [cond-mat.str-el]}
  (\bibinfo {year} {2025})\BibitemShut {NoStop}%
\bibitem [{\citenamefont {Foyevtsova}\ \emph {et~al.}(2023)\citenamefont
  {Foyevtsova}, \citenamefont {Elfimov},\ and\ \citenamefont
  {Sawatzky}}]{Foyevtsova2023}%
  \BibitemOpen
  \bibfield  {author} {\bibinfo {author} {\bibfnamefont {K.}~\bibnamefont
  {Foyevtsova}}, \bibinfo {author} {\bibfnamefont {I.}~\bibnamefont
  {Elfimov}},\ and\ \bibinfo {author} {\bibfnamefont {G.~A.}\ \bibnamefont
  {Sawatzky}},\ }\bibfield  {title} {\bibinfo {title} {Distinct electridelike
  nature of infinite-layer nickelates and the resulting theoretical challenges
  to calculate their electronic structure},\ }\href
  {https://doi.org/10.1103/PhysRevB.108.205124} {\bibfield  {journal} {\bibinfo
   {journal} {Phys. Rev. B}\ }\textbf {\bibinfo {volume} {108}},\ \bibinfo
  {pages} {205124} (\bibinfo {year} {2023})}\BibitemShut {NoStop}%
\bibitem [{\citenamefont {Li}\ \emph {et~al.}(2025)\citenamefont {Li},
  \citenamefont {Chen}, \citenamefont {Ding}, \citenamefont {Zhuang},
  \citenamefont {Guo}, \citenamefont {Chen}, \citenamefont {Fan}, \citenamefont
  {Ye}, \citenamefont {An}, \citenamefont {Sangphet}, \citenamefont {Tang},
  \citenamefont {Wang}, \citenamefont {Huang}, \citenamefont {Xu},
  \citenamefont {Feng},\ and\ \citenamefont {Peng}}]{Li2025}%
  \BibitemOpen
  \bibfield  {author} {\bibinfo {author} {\bibfnamefont {C.}~\bibnamefont
  {Li}}, \bibinfo {author} {\bibfnamefont {Y.}~\bibnamefont {Chen}}, \bibinfo
  {author} {\bibfnamefont {X.}~\bibnamefont {Ding}}, \bibinfo {author}
  {\bibfnamefont {Y.}~\bibnamefont {Zhuang}}, \bibinfo {author} {\bibfnamefont
  {N.}~\bibnamefont {Guo}}, \bibinfo {author} {\bibfnamefont {Z.}~\bibnamefont
  {Chen}}, \bibinfo {author} {\bibfnamefont {Y.}~\bibnamefont {Fan}}, \bibinfo
  {author} {\bibfnamefont {J.}~\bibnamefont {Ye}}, \bibinfo {author}
  {\bibfnamefont {Z.}~\bibnamefont {An}}, \bibinfo {author} {\bibfnamefont
  {S.}~\bibnamefont {Sangphet}}, \bibinfo {author} {\bibfnamefont
  {S.}~\bibnamefont {Tang}}, \bibinfo {author} {\bibfnamefont {X.}~\bibnamefont
  {Wang}}, \bibinfo {author} {\bibfnamefont {H.}~\bibnamefont {Huang}},
  \bibinfo {author} {\bibfnamefont {H.}~\bibnamefont {Xu}}, \bibinfo {author}
  {\bibfnamefont {D.}~\bibnamefont {Feng}},\ and\ \bibinfo {author}
  {\bibfnamefont {R.}~\bibnamefont {Peng}},\ }\bibfield  {title} {\bibinfo
  {title} {Observation of electridelike $s$ states coexisting with correlated
  $d$ electrons in ${\mathrm{NdNiO}}_{2}$},\ }\href
  {https://doi.org/10.1103/tptb-8hb4} {\bibfield  {journal} {\bibinfo
  {journal} {Phys. Rev. Lett.}\ }\textbf {\bibinfo {volume} {135}},\ \bibinfo
  {pages} {116501} (\bibinfo {year} {2025})}\BibitemShut {NoStop}%
\bibitem [{\citenamefont {Garc\'{\i}a-Mu\~noz}\ \emph
  {et~al.}(1992)\citenamefont {Garc\'{\i}a-Mu\~noz}, \citenamefont
  {Rodr\'{\i}guez-Carvajal}, \citenamefont {Lacorre},\ and\ \citenamefont
  {Torrance}}]{Garcia1992}%
  \BibitemOpen
  \bibfield  {author} {\bibinfo {author} {\bibfnamefont {J.~L.}\ \bibnamefont
  {Garc\'{\i}a-Mu\~noz}}, \bibinfo {author} {\bibfnamefont {J.}~\bibnamefont
  {Rodr\'{\i}guez-Carvajal}}, \bibinfo {author} {\bibfnamefont
  {P.}~\bibnamefont {Lacorre}},\ and\ \bibinfo {author} {\bibfnamefont {J.~B.}\
  \bibnamefont {Torrance}},\ }\bibfield  {title} {\bibinfo {title}
  {Neutron-diffraction study of ${\mathrm{RNiO}}_{3}$ (R=La, Pr, Nd, Sm):
  Electronically induced structural changes across the metal-insulator
  transition},\ }\href {https://doi.org/10.1103/PhysRevB.46.4414} {\bibfield
  {journal} {\bibinfo  {journal} {Phys. Rev. B}\ }\textbf {\bibinfo {volume}
  {46}},\ \bibinfo {pages} {4414} (\bibinfo {year} {1992})}\BibitemShut
  {NoStop}%
\bibitem [{\citenamefont {Alonso}\ \emph {et~al.}(1997)\citenamefont {Alonso},
  \citenamefont {Mart\'{\i}nez-Lope}, \citenamefont {{n}oz},\ and\
  \citenamefont {Fern\'{a}ndez}}]{Alonso1997}%
  \BibitemOpen
  \bibfield  {author} {\bibinfo {author} {\bibfnamefont {J.}~\bibnamefont
  {Alonso}}, \bibinfo {author} {\bibfnamefont {M.}~\bibnamefont
  {Mart\'{\i}nez-Lope}}, \bibinfo {author} {\bibfnamefont {J.~G.-M.}\
  \bibnamefont {{n}oz}},\ and\ \bibinfo {author} {\bibfnamefont
  {M.}~\bibnamefont {Fern\'{a}ndez}},\ }\bibfield  {title} {\bibinfo {title}
  {Crystal structure and magnetism in the defect perovskite LaNiO$_{2.5}$},\
  }\href {https://doi.org/https://doi.org/10.1016/S0921-4526(96)00863-0}
  {\bibfield  {journal} {\bibinfo  {journal} {Physica B: Condensed Matter}\
  }\textbf {\bibinfo {volume} {234-236}},\ \bibinfo {pages} {18} (\bibinfo
  {year} {1997})},\ \bibinfo {note} {proceedings of the First European
  Conference on Neutron Scattering}\BibitemShut {NoStop}%
\bibitem [{\citenamefont {Crespin}\ \emph {et~al.}(2005)\citenamefont
  {Crespin}, \citenamefont {Isnard}, \citenamefont {Dubois}, \citenamefont
  {Choisnet},\ and\ \citenamefont {Odier}}]{Crespin2005}%
  \BibitemOpen
  \bibfield  {author} {\bibinfo {author} {\bibfnamefont {M.}~\bibnamefont
  {Crespin}}, \bibinfo {author} {\bibfnamefont {O.}~\bibnamefont {Isnard}},
  \bibinfo {author} {\bibfnamefont {F.}~\bibnamefont {Dubois}}, \bibinfo
  {author} {\bibfnamefont {J.}~\bibnamefont {Choisnet}},\ and\ \bibinfo
  {author} {\bibfnamefont {P.}~\bibnamefont {Odier}},\ }\bibfield  {title}
  {\bibinfo {title} {LaNiO$_2$: Synthesis and structural characterization},\
  }\href {https://doi.org/https://doi.org/10.1016/j.jssc.2005.01.023}
  {\bibfield  {journal} {\bibinfo  {journal} {Journal of Solid State
  Chemistry}\ }\textbf {\bibinfo {volume} {178}},\ \bibinfo {pages} {1326}
  (\bibinfo {year} {2005})}\BibitemShut {NoStop}%
\bibitem [{\citenamefont {Wang}\ \emph {et~al.}(2000)\citenamefont {Wang},
  \citenamefont {Ralston}, \citenamefont {Patil}, \citenamefont {Jones},
  \citenamefont {Gu}, \citenamefont {Verhagen}, \citenamefont {Adams},
  \citenamefont {Ge}, \citenamefont {Riordan}, \citenamefont {Marganian},
  \citenamefont {Mascharak}, \citenamefont {Kovacs}, \citenamefont {Miller},
  \citenamefont {Collins}, \citenamefont {Brooker}, \citenamefont {Croucher},
  \citenamefont {Wang}, \citenamefont {Stiefel},\ and\ \citenamefont
  {Cramer}}]{Wang2000}%
  \BibitemOpen
  \bibfield  {author} {\bibinfo {author} {\bibfnamefont {H.}~\bibnamefont
  {Wang}}, \bibinfo {author} {\bibfnamefont {C.~Y.}\ \bibnamefont {Ralston}},
  \bibinfo {author} {\bibfnamefont {D.~S.}\ \bibnamefont {Patil}}, \bibinfo
  {author} {\bibfnamefont {R.~M.}\ \bibnamefont {Jones}}, \bibinfo {author}
  {\bibfnamefont {W.}~\bibnamefont {Gu}}, \bibinfo {author} {\bibfnamefont
  {M.}~\bibnamefont {Verhagen}}, \bibinfo {author} {\bibfnamefont
  {M.}~\bibnamefont {Adams}}, \bibinfo {author} {\bibfnamefont
  {P.}~\bibnamefont {Ge}}, \bibinfo {author} {\bibfnamefont {C.}~\bibnamefont
  {Riordan}}, \bibinfo {author} {\bibfnamefont {C.~A.}\ \bibnamefont
  {Marganian}}, \bibinfo {author} {\bibfnamefont {P.}~\bibnamefont
  {Mascharak}}, \bibinfo {author} {\bibfnamefont {J.}~\bibnamefont {Kovacs}},
  \bibinfo {author} {\bibfnamefont {C.~G.}\ \bibnamefont {Miller}}, \bibinfo
  {author} {\bibfnamefont {T.~J.}\ \bibnamefont {Collins}}, \bibinfo {author}
  {\bibfnamefont {S.}~\bibnamefont {Brooker}}, \bibinfo {author} {\bibfnamefont
  {P.~D.}\ \bibnamefont {Croucher}}, \bibinfo {author} {\bibfnamefont
  {K.}~\bibnamefont {Wang}}, \bibinfo {author} {\bibfnamefont {E.~I.}\
  \bibnamefont {Stiefel}},\ and\ \bibinfo {author} {\bibfnamefont {S.~P.}\
  \bibnamefont {Cramer}},\ }\bibfield  {title} {\bibinfo {title} {Nickel L-edge
  soft x-ray spectroscopy of nickel-iron hydrogenases and model compounds -
  evidence for high-spin nickel(II) in the active enzyme},\ }\href
  {https://doi.org/10.1021/ja000945g} {\bibfield  {journal} {\bibinfo
  {journal} {Journal of the American Chemical Society}\ }\textbf {\bibinfo
  {volume} {122}},\ \bibinfo {pages} {10544} (\bibinfo {year}
  {2000})}\BibitemShut {NoStop}%
\bibitem [{\citenamefont {Zeng}\ \emph {et~al.}(2024)\citenamefont {Zeng},
  \citenamefont {Tang}, \citenamefont {Luo}, \citenamefont {Chow},
  \citenamefont {Lim}, \citenamefont {Prakash}, \citenamefont {Yang},
  \citenamefont {Diao}, \citenamefont {Yu}, \citenamefont {Xing}, \citenamefont
  {Ji}, \citenamefont {Yin}, \citenamefont {Li}, \citenamefont {Wang},
  \citenamefont {He}, \citenamefont {Breese}, \citenamefont {Ariando},\ and\
  \citenamefont {Liu}}]{Zeng2024}%
  \BibitemOpen
  \bibfield  {author} {\bibinfo {author} {\bibfnamefont {S.}~\bibnamefont
  {Zeng}}, \bibinfo {author} {\bibfnamefont {C.~S.}\ \bibnamefont {Tang}},
  \bibinfo {author} {\bibfnamefont {Z.}~\bibnamefont {Luo}}, \bibinfo {author}
  {\bibfnamefont {L.~E.}\ \bibnamefont {Chow}}, \bibinfo {author}
  {\bibfnamefont {Z.~S.}\ \bibnamefont {Lim}}, \bibinfo {author} {\bibfnamefont
  {S.}~\bibnamefont {Prakash}}, \bibinfo {author} {\bibfnamefont
  {P.}~\bibnamefont {Yang}}, \bibinfo {author} {\bibfnamefont {C.}~\bibnamefont
  {Diao}}, \bibinfo {author} {\bibfnamefont {X.}~\bibnamefont {Yu}}, \bibinfo
  {author} {\bibfnamefont {Z.}~\bibnamefont {Xing}}, \bibinfo {author}
  {\bibfnamefont {R.}~\bibnamefont {Ji}}, \bibinfo {author} {\bibfnamefont
  {X.}~\bibnamefont {Yin}}, \bibinfo {author} {\bibfnamefont {C.}~\bibnamefont
  {Li}}, \bibinfo {author} {\bibfnamefont {X.~R.}\ \bibnamefont {Wang}},
  \bibinfo {author} {\bibfnamefont {Q.}~\bibnamefont {He}}, \bibinfo {author}
  {\bibfnamefont {M.~B.~H.}\ \bibnamefont {Breese}}, \bibinfo {author}
  {\bibfnamefont {A.}~\bibnamefont {Ariando}},\ and\ \bibinfo {author}
  {\bibfnamefont {H.}~\bibnamefont {Liu}},\ }\bibfield  {title} {\bibinfo
  {title} {Origin of a topotactic reduction effect for superconductivity in
  infinite-layer nickelates},\ }\href
  {https://doi.org/10.1103/PhysRevLett.133.066503} {\bibfield  {journal}
  {\bibinfo  {journal} {Phys. Rev. Lett.}\ }\textbf {\bibinfo {volume} {133}},\
  \bibinfo {pages} {066503} (\bibinfo {year} {2024})}\BibitemShut {NoStop}%
\bibitem [{\citenamefont {Kim}\ \emph {et~al.}(2021)\citenamefont {Kim},
  \citenamefont {Rabinovich}, \citenamefont {Boris}, \citenamefont {Yaresko},
  \citenamefont {Suyolcu}, \citenamefont {Wu}, \citenamefont {van Aken},
  \citenamefont {Christiani}, \citenamefont {Logvenov},\ and\ \citenamefont
  {Keimer}}]{Kim2021}%
  \BibitemOpen
  \bibfield  {author} {\bibinfo {author} {\bibfnamefont {G.}~\bibnamefont
  {Kim}}, \bibinfo {author} {\bibfnamefont {K.~S.}\ \bibnamefont {Rabinovich}},
  \bibinfo {author} {\bibfnamefont {A.~V.}\ \bibnamefont {Boris}}, \bibinfo
  {author} {\bibfnamefont {A.~N.}\ \bibnamefont {Yaresko}}, \bibinfo {author}
  {\bibfnamefont {Y.~E.}\ \bibnamefont {Suyolcu}}, \bibinfo {author}
  {\bibfnamefont {Y.-M.}\ \bibnamefont {Wu}}, \bibinfo {author} {\bibfnamefont
  {P.~A.}\ \bibnamefont {van Aken}}, \bibinfo {author} {\bibfnamefont
  {G.}~\bibnamefont {Christiani}}, \bibinfo {author} {\bibfnamefont
  {G.}~\bibnamefont {Logvenov}},\ and\ \bibinfo {author} {\bibfnamefont
  {B.}~\bibnamefont {Keimer}},\ }\bibfield  {title} {\bibinfo {title} {Optical
  conductivity and superconductivity in highly overdoped
  $\mathrm{La_{2-x}Ca_xCuO_4}$ thin films},\ }\href
  {https://doi.org/10.1073/pnas.2106170118} {\bibfield  {journal} {\bibinfo
  {journal} {Proceedings of the National Academy of Sciences}\ }\textbf
  {\bibinfo {volume} {118}},\ \bibinfo {pages} {e2106170118} (\bibinfo {year}
  {2021})}\BibitemShut {NoStop}%
\bibitem [{\citenamefont {Green}\ \emph {et~al.}(2016)\citenamefont {Green},
  \citenamefont {Haverkort},\ and\ \citenamefont {Sawatzky}}]{Green2016}%
  \BibitemOpen
  \bibfield  {author} {\bibinfo {author} {\bibfnamefont {R.~J.}\ \bibnamefont
  {Green}}, \bibinfo {author} {\bibfnamefont {M.~W.}\ \bibnamefont
  {Haverkort}},\ and\ \bibinfo {author} {\bibfnamefont {G.~A.}\ \bibnamefont
  {Sawatzky}},\ }\bibfield  {title} {\bibinfo {title} {Bond disproportionation
  and dynamical charge fluctuations in the perovskite rare-earth nickelates},\
  }\href {https://doi.org/10.1103/PhysRevB.94.195127} {\bibfield  {journal}
  {\bibinfo  {journal} {Phys. Rev. B}\ }\textbf {\bibinfo {volume} {94}},\
  \bibinfo {pages} {195127} (\bibinfo {year} {2016})}\BibitemShut {NoStop}%
\bibitem [{\citenamefont {Shin}\ and\ \citenamefont
  {Rondinelli}(2022)}]{Shin2022}%
  \BibitemOpen
  \bibfield  {author} {\bibinfo {author} {\bibfnamefont {Y.}~\bibnamefont
  {Shin}}\ and\ \bibinfo {author} {\bibfnamefont {J.~M.}\ \bibnamefont
  {Rondinelli}},\ }\bibfield  {title} {\bibinfo {title} {Magnetic structure of
  oxygen-deficient perovskite nickelates with ordered vacancies},\ }\href
  {https://doi.org/10.1103/PhysRevResearch.4.L022069} {\bibfield  {journal}
  {\bibinfo  {journal} {Phys. Rev. Res.}\ }\textbf {\bibinfo {volume} {4}},\
  \bibinfo {pages} {L022069} (\bibinfo {year} {2022})}\BibitemShut {NoStop}%
\bibitem [{\citenamefont {Haverkort}\ \emph {et~al.}(2004)\citenamefont
  {Haverkort}, \citenamefont {Csiszar}, \citenamefont {Hu}, \citenamefont
  {Altieri}, \citenamefont {Tanaka}, \citenamefont {Hsieh}, \citenamefont
  {Lin}, \citenamefont {Chen}, \citenamefont {Hibma},\ and\ \citenamefont
  {Tjeng}}]{Haverkort2004}%
  \BibitemOpen
  \bibfield  {author} {\bibinfo {author} {\bibfnamefont {M.~W.}\ \bibnamefont
  {Haverkort}}, \bibinfo {author} {\bibfnamefont {S.~I.}\ \bibnamefont
  {Csiszar}}, \bibinfo {author} {\bibfnamefont {Z.}~\bibnamefont {Hu}},
  \bibinfo {author} {\bibfnamefont {S.}~\bibnamefont {Altieri}}, \bibinfo
  {author} {\bibfnamefont {A.}~\bibnamefont {Tanaka}}, \bibinfo {author}
  {\bibfnamefont {H.~H.}\ \bibnamefont {Hsieh}}, \bibinfo {author}
  {\bibfnamefont {H.-J.}\ \bibnamefont {Lin}}, \bibinfo {author} {\bibfnamefont
  {C.~T.}\ \bibnamefont {Chen}}, \bibinfo {author} {\bibfnamefont
  {T.}~\bibnamefont {Hibma}},\ and\ \bibinfo {author} {\bibfnamefont {L.~H.}\
  \bibnamefont {Tjeng}},\ }\bibfield  {title} {\bibinfo {title} {Magnetic
  versus crystal-field linear dichroism in NiO thin films},\ }\href
  {https://doi.org/10.1103/PhysRevB.69.020408} {\bibfield  {journal} {\bibinfo
  {journal} {Phys. Rev. B}\ }\textbf {\bibinfo {volume} {69}},\ \bibinfo
  {pages} {020408} (\bibinfo {year} {2004})}\BibitemShut {NoStop}%
\bibitem [{\citenamefont {Li}\ \emph {et~al.}(2021)\citenamefont {Li},
  \citenamefont {Green}, \citenamefont {Zhang}, \citenamefont {Sutarto},
  \citenamefont {Sadowski}, \citenamefont {Zhu}, \citenamefont {Zhang},
  \citenamefont {Zhou}, \citenamefont {Sun}, \citenamefont {He}, \citenamefont
  {Ramanathan},\ and\ \citenamefont {Comin}}]{Li2021}%
  \BibitemOpen
  \bibfield  {author} {\bibinfo {author} {\bibfnamefont {J.}~\bibnamefont
  {Li}}, \bibinfo {author} {\bibfnamefont {R.~J.}\ \bibnamefont {Green}},
  \bibinfo {author} {\bibfnamefont {Z.}~\bibnamefont {Zhang}}, \bibinfo
  {author} {\bibfnamefont {R.}~\bibnamefont {Sutarto}}, \bibinfo {author}
  {\bibfnamefont {J.~T.}\ \bibnamefont {Sadowski}}, \bibinfo {author}
  {\bibfnamefont {Z.}~\bibnamefont {Zhu}}, \bibinfo {author} {\bibfnamefont
  {G.}~\bibnamefont {Zhang}}, \bibinfo {author} {\bibfnamefont
  {D.}~\bibnamefont {Zhou}}, \bibinfo {author} {\bibfnamefont {Y.}~\bibnamefont
  {Sun}}, \bibinfo {author} {\bibfnamefont {F.}~\bibnamefont {He}}, \bibinfo
  {author} {\bibfnamefont {S.}~\bibnamefont {Ramanathan}},\ and\ \bibinfo
  {author} {\bibfnamefont {R.}~\bibnamefont {Comin}},\ }\bibfield  {title}
  {\bibinfo {title} {Sudden collapse of magnetic order in oxygen-deficient
  nickelate films},\ }\href {https://doi.org/10.1103/PhysRevLett.126.187602}
  {\bibfield  {journal} {\bibinfo  {journal} {Phys. Rev. Lett.}\ }\textbf
  {\bibinfo {volume} {126}},\ \bibinfo {pages} {187602} (\bibinfo {year}
  {2021})}\BibitemShut {NoStop}%
\bibitem [{\citenamefont {Parzyck}\ \emph
  {et~al.}(2024{\natexlab{a}})\citenamefont {Parzyck}, \citenamefont {Anil},
  \citenamefont {Wu}, \citenamefont {Goodge}, \citenamefont {Roddy},
  \citenamefont {Kourkoutis}, \citenamefont {Schlom},\ and\ \citenamefont
  {Shen}}]{Parzyck2024b}%
  \BibitemOpen
  \bibfield  {author} {\bibinfo {author} {\bibfnamefont {C.~T.}\ \bibnamefont
  {Parzyck}}, \bibinfo {author} {\bibfnamefont {V.}~\bibnamefont {Anil}},
  \bibinfo {author} {\bibfnamefont {Y.}~\bibnamefont {Wu}}, \bibinfo {author}
  {\bibfnamefont {B.~H.}\ \bibnamefont {Goodge}}, \bibinfo {author}
  {\bibfnamefont {M.}~\bibnamefont {Roddy}}, \bibinfo {author} {\bibfnamefont
  {L.~F.}\ \bibnamefont {Kourkoutis}}, \bibinfo {author} {\bibfnamefont
  {D.~G.}\ \bibnamefont {Schlom}},\ and\ \bibinfo {author} {\bibfnamefont
  {K.~M.}\ \bibnamefont {Shen}},\ }\bibfield  {title} {\bibinfo {title}
  {Synthesis of thin film infinite-layer nickelates by atomic hydrogen
  reduction: Clarifying the role of the capping layer},\ }\href
  {https://doi.org/10.1063/5.0197304} {\bibfield  {journal} {\bibinfo
  {journal} {APL Materials}\ }\textbf {\bibinfo {volume} {12}},\ \bibinfo
  {pages} {031132} (\bibinfo {year} {2024}{\natexlab{a}})}\BibitemShut
  {NoStop}%
\bibitem [{\citenamefont {Zhang}\ \emph {et~al.}(2017)\citenamefont {Zhang},
  \citenamefont {Botana}, \citenamefont {Freeland}, \citenamefont {Phelan},
  \citenamefont {Zheng}, \citenamefont {Pardo}, \citenamefont {Norman},\ and\
  \citenamefont {Mitchell}}]{Zhang2017}%
  \BibitemOpen
  \bibfield  {author} {\bibinfo {author} {\bibfnamefont {J.}~\bibnamefont
  {Zhang}}, \bibinfo {author} {\bibfnamefont {A.~S.}\ \bibnamefont {Botana}},
  \bibinfo {author} {\bibfnamefont {J.~W.}\ \bibnamefont {Freeland}}, \bibinfo
  {author} {\bibfnamefont {D.}~\bibnamefont {Phelan}}, \bibinfo {author}
  {\bibfnamefont {H.}~\bibnamefont {Zheng}}, \bibinfo {author} {\bibfnamefont
  {V.}~\bibnamefont {Pardo}}, \bibinfo {author} {\bibfnamefont {M.~R.}\
  \bibnamefont {Norman}},\ and\ \bibinfo {author} {\bibfnamefont {J.~F.}\
  \bibnamefont {Mitchell}},\ }\bibfield  {title} {\bibinfo {title} {Large
  orbital polarization in a metallic square-planar nickelate},\ }\href
  {https://doi.org/10.1038/nphys4149} {\bibfield  {journal} {\bibinfo
  {journal} {Nature Physics}\ }\textbf {\bibinfo {volume} {13}},\ \bibinfo
  {pages} {864} (\bibinfo {year} {2017})}\BibitemShut {NoStop}%
\bibitem [{\citenamefont {Ortiz}\ \emph {et~al.}(2025)\citenamefont {Ortiz},
  \citenamefont {Enderlein}, \citenamefont {F\"ursich}, \citenamefont {Pons},
  \citenamefont {Radhakrishnan}, \citenamefont {Schierle}, \citenamefont
  {Wochner}, \citenamefont {Logvenov}, \citenamefont {Cristiani}, \citenamefont
  {Hansmann}, \citenamefont {Keimer},\ and\ \citenamefont
  {Benckiser}}]{Ortiz2025}%
  \BibitemOpen
  \bibfield  {author} {\bibinfo {author} {\bibfnamefont {R.~A.}\ \bibnamefont
  {Ortiz}}, \bibinfo {author} {\bibfnamefont {N.}~\bibnamefont {Enderlein}},
  \bibinfo {author} {\bibfnamefont {K.}~\bibnamefont {F\"ursich}}, \bibinfo
  {author} {\bibfnamefont {R.}~\bibnamefont {Pons}}, \bibinfo {author}
  {\bibfnamefont {P.}~\bibnamefont {Radhakrishnan}}, \bibinfo {author}
  {\bibfnamefont {E.}~\bibnamefont {Schierle}}, \bibinfo {author}
  {\bibfnamefont {P.}~\bibnamefont {Wochner}}, \bibinfo {author} {\bibfnamefont
  {G.}~\bibnamefont {Logvenov}}, \bibinfo {author} {\bibfnamefont
  {G.}~\bibnamefont {Cristiani}}, \bibinfo {author} {\bibfnamefont
  {P.}~\bibnamefont {Hansmann}}, \bibinfo {author} {\bibfnamefont
  {B.}~\bibnamefont {Keimer}},\ and\ \bibinfo {author} {\bibfnamefont
  {E.}~\bibnamefont {Benckiser}},\ }\bibfield  {title} {\bibinfo {title}
  {Oxygen sublattice disorder and valence state modulation in infinite-layer
  nickelate superlattices},\ }\href
  {https://doi.org/10.1103/PhysRevMaterials.9.054801} {\bibfield  {journal}
  {\bibinfo  {journal} {Phys. Rev. Mater.}\ }\textbf {\bibinfo {volume} {9}},\
  \bibinfo {pages} {054801} (\bibinfo {year} {2025})}\BibitemShut {NoStop}%
\bibitem [{\citenamefont {Parzyck}\ \emph
  {et~al.}(2024{\natexlab{b}})\citenamefont {Parzyck}, \citenamefont {Gupta},
  \citenamefont {Wu}, \citenamefont {Anil}, \citenamefont {Bhatt},
  \citenamefont {Bouliane}, \citenamefont {Gong}, \citenamefont {Gregory},
  \citenamefont {Luo}, \citenamefont {Sutarto}, \citenamefont {He},
  \citenamefont {Chuang}, \citenamefont {Zhou}, \citenamefont {Herranz},
  \citenamefont {Kourkoutis}, \citenamefont {Singer}, \citenamefont {Schlom},
  \citenamefont {Hawthorn},\ and\ \citenamefont {Shen}}]{Parzyck2024a}%
  \BibitemOpen
  \bibfield  {author} {\bibinfo {author} {\bibfnamefont {C.~T.}\ \bibnamefont
  {Parzyck}}, \bibinfo {author} {\bibfnamefont {N.~K.}\ \bibnamefont {Gupta}},
  \bibinfo {author} {\bibfnamefont {Y.}~\bibnamefont {Wu}}, \bibinfo {author}
  {\bibfnamefont {V.}~\bibnamefont {Anil}}, \bibinfo {author} {\bibfnamefont
  {L.}~\bibnamefont {Bhatt}}, \bibinfo {author} {\bibfnamefont
  {M.}~\bibnamefont {Bouliane}}, \bibinfo {author} {\bibfnamefont
  {R.}~\bibnamefont {Gong}}, \bibinfo {author} {\bibfnamefont {B.~Z.}\
  \bibnamefont {Gregory}}, \bibinfo {author} {\bibfnamefont {A.}~\bibnamefont
  {Luo}}, \bibinfo {author} {\bibfnamefont {R.}~\bibnamefont {Sutarto}},
  \bibinfo {author} {\bibfnamefont {F.}~\bibnamefont {He}}, \bibinfo {author}
  {\bibfnamefont {Y.~D.}\ \bibnamefont {Chuang}}, \bibinfo {author}
  {\bibfnamefont {T.}~\bibnamefont {Zhou}}, \bibinfo {author} {\bibfnamefont
  {G.}~\bibnamefont {Herranz}}, \bibinfo {author} {\bibfnamefont {L.~F.}\
  \bibnamefont {Kourkoutis}}, \bibinfo {author} {\bibfnamefont
  {A.}~\bibnamefont {Singer}}, \bibinfo {author} {\bibfnamefont {D.~G.}\
  \bibnamefont {Schlom}}, \bibinfo {author} {\bibfnamefont {D.~G.}\
  \bibnamefont {Hawthorn}},\ and\ \bibinfo {author} {\bibfnamefont {K.~M.}\
  \bibnamefont {Shen}},\ }\bibfield  {title} {\bibinfo {title} {Absence of
  3a$_0$ charge density wave order in the infinite-layer nickelate NdNiO$_2$},\
  }\href {https://doi.org/10.1038/s41563-024-01797-0} {\bibfield  {journal}
  {\bibinfo  {journal} {Nature Materials}\ }\textbf {\bibinfo {volume} {23}},\
  \bibinfo {pages} {486} (\bibinfo {year} {2024}{\natexlab{b}})}\BibitemShut
  {NoStop}%
\bibitem [{\citenamefont {Stöhr}\ and\ \citenamefont
  {Siegmann}(2006)}]{Stoehr2006}%
  \BibitemOpen
  \bibfield  {author} {\bibinfo {author} {\bibfnamefont {J.}~\bibnamefont
  {Stöhr}}\ and\ \bibinfo {author} {\bibfnamefont {H.}~\bibnamefont
  {Siegmann}},\ }\href
  {https://doi.org/https://doi.org/10.1007/978-3-540-30283-4} {\emph {\bibinfo
  {title} {Magnetism. - From Fundamentals to Nanoscale Dynamics}}}\ (\bibinfo
  {publisher} {Springer Berlin},\ \bibinfo {year} {2006})\BibitemShut {NoStop}%
\bibitem [{\citenamefont {Haverkort}\ \emph {et~al.}(2012)\citenamefont
  {Haverkort}, \citenamefont {Zwierzycki},\ and\ \citenamefont
  {Andersen}}]{Haverkort2012}%
  \BibitemOpen
  \bibfield  {author} {\bibinfo {author} {\bibfnamefont {M.~W.}\ \bibnamefont
  {Haverkort}}, \bibinfo {author} {\bibfnamefont {M.}~\bibnamefont
  {Zwierzycki}},\ and\ \bibinfo {author} {\bibfnamefont {O.~K.}\ \bibnamefont
  {Andersen}},\ }\bibfield  {title} {\bibinfo {title} {Multiplet ligand-field
  theory using Wannier orbitals},\ }\href
  {https://doi.org/10.1103/PhysRevB.85.165113} {\bibfield  {journal} {\bibinfo
  {journal} {Phys. Rev. B}\ }\textbf {\bibinfo {volume} {85}},\ \bibinfo
  {pages} {165113} (\bibinfo {year} {2012})}\BibitemShut {NoStop}%
\bibitem [{\citenamefont {Krieger}\ \emph {et~al.}(2024)\citenamefont
  {Krieger}, \citenamefont {Sahib}, \citenamefont {Rosa}, \citenamefont {Rath},
  \citenamefont {Chen}, \citenamefont {Raji}, \citenamefont {Pinho},
  \citenamefont {Lefevre}, \citenamefont {Ghiringhelli}, \citenamefont
  {Gloter}, \citenamefont {Viart}, \citenamefont {Salluzzo},\ and\
  \citenamefont {Preziosi}}]{Krieger2024}%
  \BibitemOpen
  \bibfield  {author} {\bibinfo {author} {\bibfnamefont {G.}~\bibnamefont
  {Krieger}}, \bibinfo {author} {\bibfnamefont {H.}~\bibnamefont {Sahib}},
  \bibinfo {author} {\bibfnamefont {F.}~\bibnamefont {Rosa}}, \bibinfo {author}
  {\bibfnamefont {M.}~\bibnamefont {Rath}}, \bibinfo {author} {\bibfnamefont
  {Y.}~\bibnamefont {Chen}}, \bibinfo {author} {\bibfnamefont {A.}~\bibnamefont
  {Raji}}, \bibinfo {author} {\bibfnamefont {P.~V.~B.}\ \bibnamefont {Pinho}},
  \bibinfo {author} {\bibfnamefont {C.}~\bibnamefont {Lefevre}}, \bibinfo
  {author} {\bibfnamefont {G.}~\bibnamefont {Ghiringhelli}}, \bibinfo {author}
  {\bibfnamefont {A.}~\bibnamefont {Gloter}}, \bibinfo {author} {\bibfnamefont
  {N.}~\bibnamefont {Viart}}, \bibinfo {author} {\bibfnamefont
  {M.}~\bibnamefont {Salluzzo}},\ and\ \bibinfo {author} {\bibfnamefont
  {D.}~\bibnamefont {Preziosi}},\ }\bibfield  {title} {\bibinfo {title}
  {Signatures of canted antiferromagnetism in infinite-layer nickelates studied
  by x-ray magnetic dichroism},\ }\href
  {https://doi.org/10.1103/PhysRevB.110.195110} {\bibfield  {journal} {\bibinfo
   {journal} {Phys. Rev. B}\ }\textbf {\bibinfo {volume} {110}},\ \bibinfo
  {pages} {195110} (\bibinfo {year} {2024})}\BibitemShut {NoStop}%
\bibitem [{\citenamefont {Held}\ \emph {et~al.}(2022)\citenamefont {Held},
  \citenamefont {Si}, \citenamefont {Worm}, \citenamefont {Janson},
  \citenamefont {Arita}, \citenamefont {Zhong}, \citenamefont {Tomczak},\ and\
  \citenamefont {Kitatani}}]{Held2022}%
  \BibitemOpen
  \bibfield  {author} {\bibinfo {author} {\bibfnamefont {K.}~\bibnamefont
  {Held}}, \bibinfo {author} {\bibfnamefont {L.}~\bibnamefont {Si}}, \bibinfo
  {author} {\bibfnamefont {P.}~\bibnamefont {Worm}}, \bibinfo {author}
  {\bibfnamefont {O.}~\bibnamefont {Janson}}, \bibinfo {author} {\bibfnamefont
  {R.}~\bibnamefont {Arita}}, \bibinfo {author} {\bibfnamefont
  {Z.}~\bibnamefont {Zhong}}, \bibinfo {author} {\bibfnamefont {J.~M.}\
  \bibnamefont {Tomczak}},\ and\ \bibinfo {author} {\bibfnamefont
  {M.}~\bibnamefont {Kitatani}},\ }\bibfield  {title} {\bibinfo {title} {Phase
  diagram of nickelate superconductors calculated by dynamical vertex
  approximation},\ }\bibfield  {journal} {\bibinfo  {journal} {Frontiers in
  Physics}\ }\textbf {\bibinfo {volume} {9}},\ \href
  {https://doi.org/10.3389/fphy.2021.810394} {10.3389/fphy.2021.810394}
  (\bibinfo {year} {2022})\BibitemShut {NoStop}%
\bibitem [{\citenamefont {Abbate}\ \emph {et~al.}(2002)\citenamefont {Abbate},
  \citenamefont {Zampieri}, \citenamefont {Prado}, \citenamefont {Caneiro},
  \citenamefont {Gonzalez-Calbet},\ and\ \citenamefont
  {Vallet-Regi}}]{Abbate2002}%
  \BibitemOpen
  \bibfield  {author} {\bibinfo {author} {\bibfnamefont {M.}~\bibnamefont
  {Abbate}}, \bibinfo {author} {\bibfnamefont {G.}~\bibnamefont {Zampieri}},
  \bibinfo {author} {\bibfnamefont {F.}~\bibnamefont {Prado}}, \bibinfo
  {author} {\bibfnamefont {A.}~\bibnamefont {Caneiro}}, \bibinfo {author}
  {\bibfnamefont {J.~M.}\ \bibnamefont {Gonzalez-Calbet}},\ and\ \bibinfo
  {author} {\bibfnamefont {M.}~\bibnamefont {Vallet-Regi}},\ }\bibfield
  {title} {\bibinfo {title} {Electronic structure and metal-insulator
  transition in ${\mathrm{LaNiO}}_{3\ensuremath{-}\ensuremath{\delta}}$},\
  }\href {https://doi.org/10.1103/PhysRevB.65.155101} {\bibfield  {journal}
  {\bibinfo  {journal} {Phys. Rev. B}\ }\textbf {\bibinfo {volume} {65}},\
  \bibinfo {pages} {155101} (\bibinfo {year} {2002})}\BibitemShut {NoStop}%
\bibitem [{\citenamefont {Krieger}\ \emph {et~al.}(2023)\citenamefont
  {Krieger}, \citenamefont {Raji}, \citenamefont {Schlur}, \citenamefont
  {Versini}, \citenamefont {Bouillet}, \citenamefont {Lenertz}, \citenamefont
  {Robert}, \citenamefont {Gloter}, \citenamefont {Viart},\ and\ \citenamefont
  {Preziosi}}]{Krieger2023}%
  \BibitemOpen
  \bibfield  {author} {\bibinfo {author} {\bibfnamefont {G.}~\bibnamefont
  {Krieger}}, \bibinfo {author} {\bibfnamefont {A.}~\bibnamefont {Raji}},
  \bibinfo {author} {\bibfnamefont {L.}~\bibnamefont {Schlur}}, \bibinfo
  {author} {\bibfnamefont {G.}~\bibnamefont {Versini}}, \bibinfo {author}
  {\bibfnamefont {C.}~\bibnamefont {Bouillet}}, \bibinfo {author}
  {\bibfnamefont {M.}~\bibnamefont {Lenertz}}, \bibinfo {author} {\bibfnamefont
  {J.}~\bibnamefont {Robert}}, \bibinfo {author} {\bibfnamefont
  {A.}~\bibnamefont {Gloter}}, \bibinfo {author} {\bibfnamefont
  {N.}~\bibnamefont {Viart}},\ and\ \bibinfo {author} {\bibfnamefont
  {D.}~\bibnamefont {Preziosi}},\ }\bibfield  {title} {\bibinfo {title}
  {Synthesis of infinite-layer nickelates and influence of the capping-layer on
  magnetotransport},\ }\href {https://doi.org/10.1088/1361-6463/aca54a}
  {\bibfield  {journal} {\bibinfo  {journal} {Journal of Physics D: Applied
  Physics}\ }\textbf {\bibinfo {volume} {56}},\ \bibinfo {pages} {024003}
  (\bibinfo {year} {2023})}\BibitemShut {NoStop}%
\bibitem [{\citenamefont {Virtanen}\ \emph {et~al.}(2020)\citenamefont
  {Virtanen}, \citenamefont {Gommers}, \citenamefont {Oliphant}, \citenamefont
  {Haberland}, \citenamefont {Reddy}, \citenamefont {Cournapeau}, \citenamefont
  {Burovski}, \citenamefont {Peterson}, \citenamefont {Weckesser},
  \citenamefont {Bright}, \citenamefont {{van der Walt}}, \citenamefont
  {Brett}, \citenamefont {Wilson}, \citenamefont {Millman}, \citenamefont
  {Mayorov}, \citenamefont {Nelson}, \citenamefont {Jones}, \citenamefont
  {Kern}, \citenamefont {Larson}, \citenamefont {Carey}, \citenamefont {Polat},
  \citenamefont {Feng}, \citenamefont {Moore}, \citenamefont {{VanderPlas}},
  \citenamefont {Laxalde}, \citenamefont {Perktold}, \citenamefont {Cimrman},
  \citenamefont {Henriksen}, \citenamefont {Quintero}, \citenamefont {Harris},
  \citenamefont {Archibald}, \citenamefont {Ribeiro}, \citenamefont
  {Pedregosa}, \citenamefont {{van Mulbregt}},\ and\ \citenamefont {{SciPy 1.0
  Contributors}}}]{2020SciPy-NMeth}%
  \BibitemOpen
  \bibfield  {author} {\bibinfo {author} {\bibfnamefont {P.}~\bibnamefont
  {Virtanen}}, \bibinfo {author} {\bibfnamefont {R.}~\bibnamefont {Gommers}},
  \bibinfo {author} {\bibfnamefont {T.~E.}\ \bibnamefont {Oliphant}}, \bibinfo
  {author} {\bibfnamefont {M.}~\bibnamefont {Haberland}}, \bibinfo {author}
  {\bibfnamefont {T.}~\bibnamefont {Reddy}}, \bibinfo {author} {\bibfnamefont
  {D.}~\bibnamefont {Cournapeau}}, \bibinfo {author} {\bibfnamefont
  {E.}~\bibnamefont {Burovski}}, \bibinfo {author} {\bibfnamefont
  {P.}~\bibnamefont {Peterson}}, \bibinfo {author} {\bibfnamefont
  {W.}~\bibnamefont {Weckesser}}, \bibinfo {author} {\bibfnamefont
  {J.}~\bibnamefont {Bright}}, \bibinfo {author} {\bibfnamefont {S.~J.}\
  \bibnamefont {{van der Walt}}}, \bibinfo {author} {\bibfnamefont
  {M.}~\bibnamefont {Brett}}, \bibinfo {author} {\bibfnamefont
  {J.}~\bibnamefont {Wilson}}, \bibinfo {author} {\bibfnamefont {K.~J.}\
  \bibnamefont {Millman}}, \bibinfo {author} {\bibfnamefont {N.}~\bibnamefont
  {Mayorov}}, \bibinfo {author} {\bibfnamefont {A.~R.~J.}\ \bibnamefont
  {Nelson}}, \bibinfo {author} {\bibfnamefont {E.}~\bibnamefont {Jones}},
  \bibinfo {author} {\bibfnamefont {R.}~\bibnamefont {Kern}}, \bibinfo {author}
  {\bibfnamefont {E.}~\bibnamefont {Larson}}, \bibinfo {author} {\bibfnamefont
  {C.~J.}\ \bibnamefont {Carey}}, \bibinfo {author} {\bibfnamefont
  {{\.I}.}~\bibnamefont {Polat}}, \bibinfo {author} {\bibfnamefont
  {Y.}~\bibnamefont {Feng}}, \bibinfo {author} {\bibfnamefont {E.~W.}\
  \bibnamefont {Moore}}, \bibinfo {author} {\bibfnamefont {J.}~\bibnamefont
  {{VanderPlas}}}, \bibinfo {author} {\bibfnamefont {D.}~\bibnamefont
  {Laxalde}}, \bibinfo {author} {\bibfnamefont {J.}~\bibnamefont {Perktold}},
  \bibinfo {author} {\bibfnamefont {R.}~\bibnamefont {Cimrman}}, \bibinfo
  {author} {\bibfnamefont {I.}~\bibnamefont {Henriksen}}, \bibinfo {author}
  {\bibfnamefont {E.~A.}\ \bibnamefont {Quintero}}, \bibinfo {author}
  {\bibfnamefont {C.~R.}\ \bibnamefont {Harris}}, \bibinfo {author}
  {\bibfnamefont {A.~M.}\ \bibnamefont {Archibald}}, \bibinfo {author}
  {\bibfnamefont {A.~H.}\ \bibnamefont {Ribeiro}}, \bibinfo {author}
  {\bibfnamefont {F.}~\bibnamefont {Pedregosa}}, \bibinfo {author}
  {\bibfnamefont {P.}~\bibnamefont {{van Mulbregt}}},\ and\ \bibinfo {author}
  {\bibnamefont {{SciPy 1.0 Contributors}}},\ }\bibfield  {title} {\bibinfo
  {title} {{{SciPy} 1.0: Fundamental Algorithms for Scientific Computing in
  Python}},\ }\href {https://doi.org/10.1038/s41592-019-0686-2} {\bibfield
  {journal} {\bibinfo  {journal} {Nature Methods}\ }\textbf {\bibinfo {volume}
  {17}},\ \bibinfo {pages} {261} (\bibinfo {year} {2020})}\BibitemShut
  {NoStop}%
\bibitem [{\citenamefont {Lu}\ \emph {et~al.}(2014)\citenamefont {Lu},
  \citenamefont {H\"oppner}, \citenamefont {Gunnarsson},\ and\ \citenamefont
  {Haverkort}}]{Lu2014}%
  \BibitemOpen
  \bibfield  {author} {\bibinfo {author} {\bibfnamefont {Y.}~\bibnamefont
  {Lu}}, \bibinfo {author} {\bibfnamefont {M.}~\bibnamefont {H\"oppner}},
  \bibinfo {author} {\bibfnamefont {O.}~\bibnamefont {Gunnarsson}},\ and\
  \bibinfo {author} {\bibfnamefont {M.~W.}\ \bibnamefont {Haverkort}},\
  }\bibfield  {title} {\bibinfo {title} {Efficient real-frequency solver for
  dynamical mean-field theory},\ }\href
  {https://doi.org/10.1103/PhysRevB.90.085102} {\bibfield  {journal} {\bibinfo
  {journal} {Phys. Rev. B}\ }\textbf {\bibinfo {volume} {90}},\ \bibinfo
  {pages} {085102} (\bibinfo {year} {2014})}\BibitemShut {NoStop}%
\bibitem [{\citenamefont {Haverkort}\ \emph {et~al.}(2014)\citenamefont
  {Haverkort}, \citenamefont {Sangiovanni}, \citenamefont {Hansmann},
  \citenamefont {Toschi}, \citenamefont {Lu},\ and\ \citenamefont
  {Macke}}]{Haverkort2014}%
  \BibitemOpen
  \bibfield  {author} {\bibinfo {author} {\bibfnamefont {M.~W.}\ \bibnamefont
  {Haverkort}}, \bibinfo {author} {\bibfnamefont {G.}~\bibnamefont
  {Sangiovanni}}, \bibinfo {author} {\bibfnamefont {P.}~\bibnamefont
  {Hansmann}}, \bibinfo {author} {\bibfnamefont {A.}~\bibnamefont {Toschi}},
  \bibinfo {author} {\bibfnamefont {Y.}~\bibnamefont {Lu}},\ and\ \bibinfo
  {author} {\bibfnamefont {S.}~\bibnamefont {Macke}},\ }\bibfield  {title}
  {\bibinfo {title} {Bands, resonances, edge singularities and excitons in core
  level spectroscopy investigated within the dynamical mean-field theory},\
  }\href {https://doi.org/10.1209/0295-5075/108/57004} {\bibfield  {journal}
  {\bibinfo  {journal} {EPL (Europhysics Letters)}\ }\textbf {\bibinfo {volume}
  {108}},\ \bibinfo {pages} {57004} (\bibinfo {year} {2014})}\BibitemShut
  {NoStop}%
\end{thebibliography}

\clearpage


\begin{thebibliography}{8}%
\makeatletter
\providecommand \@ifxundefined [1]{%
 \@ifx{#1\undefined}
}%
\providecommand \@ifnum [1]{%
 \ifnum #1\expandafter \@firstoftwo
 \else \expandafter \@secondoftwo
 \fi
}%
\providecommand \@ifx [1]{%
 \ifx #1\expandafter \@firstoftwo
 \else \expandafter \@secondoftwo
 \fi
}%
\providecommand \natexlab [1]{#1}%
\providecommand \enquote  [1]{``#1''}%
\providecommand \bibnamefont  [1]{#1}%
\providecommand \bibfnamefont [1]{#1}%
\providecommand \citenamefont [1]{#1}%
\providecommand \href@noop [0]{\@secondoftwo}%
\providecommand \href [0]{\begingroup \@sanitize@url \@href}%
\providecommand \@href[1]{\@@startlink{#1}\@@href}%
\providecommand \@@href[1]{\endgroup#1\@@endlink}%
\providecommand \@sanitize@url [0]{\catcode `\\12\catcode `\$12\catcode
  `\&12\catcode `\#12\catcode `\^12\catcode `\_12\catcode `\%12\relax}%
\providecommand \@@startlink[1]{}%
\providecommand \@@endlink[0]{}%
\providecommand \url  [0]{\begingroup\@sanitize@url \@url }%
\providecommand \@url [1]{\endgroup\@href {#1}{\urlprefix }}%
\providecommand \urlprefix  [0]{URL }%
\providecommand \Eprint [0]{\href }%
\providecommand \doibase [0]{https://doi.org/}%
\providecommand \selectlanguage [0]{\@gobble}%
\providecommand \bibinfo  [0]{\@secondoftwo}%
\providecommand \bibfield  [0]{\@secondoftwo}%
\providecommand \translation [1]{[#1]}%
\providecommand \BibitemOpen [0]{}%
\providecommand \bibitemStop [0]{}%
\providecommand \bibitemNoStop [0]{.\EOS\space}%
\providecommand \EOS [0]{\spacefactor3000\relax}%
\providecommand \BibitemShut  [1]{\csname bibitem#1\endcsname}%
\let\auto@bib@innerbib\@empty
\bibitem [{\citenamefont {Sahib}\ \emph {et~al.}(2025)\citenamefont {Sahib},
  \citenamefont {Raji}, \citenamefont {Rosa}, \citenamefont {Merzoni},
  \citenamefont {Ghiringhelli}, \citenamefont {Salluzzo}, \citenamefont
  {Gloter}, \citenamefont {Viart},\ and\ \citenamefont {Preziosi}}]{Sahib2025S}%
  \BibitemOpen
  \bibfield  {author} {\bibinfo {author} {\bibfnamefont {H.}~\bibnamefont
  {Sahib}}, \bibinfo {author} {\bibfnamefont {A.}~\bibnamefont {Raji}},
  \bibinfo {author} {\bibfnamefont {F.}~\bibnamefont {Rosa}}, \bibinfo {author}
  {\bibfnamefont {G.}~\bibnamefont {Merzoni}}, \bibinfo {author} {\bibfnamefont
  {G.}~\bibnamefont {Ghiringhelli}}, \bibinfo {author} {\bibfnamefont
  {M.}~\bibnamefont {Salluzzo}}, \bibinfo {author} {\bibfnamefont
  {A.}~\bibnamefont {Gloter}}, \bibinfo {author} {\bibfnamefont
  {N.}~\bibnamefont {Viart}},\ and\ \bibinfo {author} {\bibfnamefont
  {D.}~\bibnamefont {Preziosi}},\ }\bibfield  {title} {\bibinfo {title}
  {Superconductivity in PrNiO$_2$ infinite-layer nickelates},\ }\href
  {https://doi.org/https://doi.org/10.1002/adma.202416187} {\bibfield
  {journal} {\bibinfo  {journal} {Advanced Materials}\ }\textbf {\bibinfo
  {volume} {37}},\ \bibinfo {pages} {2416187} (\bibinfo {year}
  {2025})}\BibitemShut {NoStop}%
\bibitem [{\citenamefont {Parzyck}\ \emph {et~al.}(2024)\citenamefont
  {Parzyck}, \citenamefont {Gupta}, \citenamefont {Wu}, \citenamefont {Anil},
  \citenamefont {Bhatt}, \citenamefont {Bouliane}, \citenamefont {Gong},
  \citenamefont {Gregory}, \citenamefont {Luo}, \citenamefont {Sutarto},
  \citenamefont {He}, \citenamefont {Chuang}, \citenamefont {Zhou},
  \citenamefont {Herranz}, \citenamefont {Kourkoutis}, \citenamefont {Singer},
  \citenamefont {Schlom}, \citenamefont {Hawthorn},\ and\ \citenamefont
  {Shen}}]{Parzyck2024aS}%
  \BibitemOpen
  \bibfield  {author} {\bibinfo {author} {\bibfnamefont {C.~T.}\ \bibnamefont
  {Parzyck}}, \bibinfo {author} {\bibfnamefont {N.~K.}\ \bibnamefont {Gupta}},
  \bibinfo {author} {\bibfnamefont {Y.}~\bibnamefont {Wu}}, \bibinfo {author}
  {\bibfnamefont {V.}~\bibnamefont {Anil}}, \bibinfo {author} {\bibfnamefont
  {L.}~\bibnamefont {Bhatt}}, \bibinfo {author} {\bibfnamefont
  {M.}~\bibnamefont {Bouliane}}, \bibinfo {author} {\bibfnamefont
  {R.}~\bibnamefont {Gong}}, \bibinfo {author} {\bibfnamefont {B.~Z.}\
  \bibnamefont {Gregory}}, \bibinfo {author} {\bibfnamefont {A.}~\bibnamefont
  {Luo}}, \bibinfo {author} {\bibfnamefont {R.}~\bibnamefont {Sutarto}},
  \bibinfo {author} {\bibfnamefont {F.}~\bibnamefont {He}}, \bibinfo {author}
  {\bibfnamefont {Y.~D.}\ \bibnamefont {Chuang}}, \bibinfo {author}
  {\bibfnamefont {T.}~\bibnamefont {Zhou}}, \bibinfo {author} {\bibfnamefont
  {G.}~\bibnamefont {Herranz}}, \bibinfo {author} {\bibfnamefont {L.~F.}\
  \bibnamefont {Kourkoutis}}, \bibinfo {author} {\bibfnamefont
  {A.}~\bibnamefont {Singer}}, \bibinfo {author} {\bibfnamefont {D.~G.}\
  \bibnamefont {Schlom}}, \bibinfo {author} {\bibfnamefont {D.~G.}\
  \bibnamefont {Hawthorn}},\ and\ \bibinfo {author} {\bibfnamefont {K.~M.}\
  \bibnamefont {Shen}},\ }\bibfield  {title} {\bibinfo {title} {Absence of
  3a$_0$ charge density wave order in the infinite-layer nickelate NdNiO$_2$},\
  }\href {https://doi.org/10.1038/s41563-024-01797-0} {\bibfield  {journal}
  {\bibinfo  {journal} {Nature Materials}\ }\textbf {\bibinfo {volume} {23}}
  (\bibinfo {year} {2024})}\BibitemShut {NoStop}%
\bibitem [{\citenamefont {Haverkort}\ \emph {et~al.}(2012)\citenamefont
  {Haverkort}, \citenamefont {Zwierzycki},\ and\ \citenamefont
  {Andersen}}]{Haverkort2012S}%
  \BibitemOpen
  \bibfield  {author} {\bibinfo {author} {\bibfnamefont {M.~W.}\ \bibnamefont
  {Haverkort}}, \bibinfo {author} {\bibfnamefont {M.}~\bibnamefont
  {Zwierzycki}},\ and\ \bibinfo {author} {\bibfnamefont {O.~K.}\ \bibnamefont
  {Andersen}},\ }\bibfield  {title} {\bibinfo {title} {Multiplet ligand-field theory using Wannier orbitals},\ }\href
  {https://doi.org/10.1103/PhysRevB.85.165113} {\bibfield  {journal} {\bibinfo
  {journal} {Phys. Rev. B}\ }\textbf {\bibinfo {volume} {85}},\ \bibinfo
  {pages} {165113} (\bibinfo {year} {2012})}\BibitemShut {NoStop}%
\bibitem [{\citenamefont {Begum-Hudde}\ \emph {et~al.}(2023)\citenamefont
  {Begum-Hudde}, \citenamefont {Lojewski}, \citenamefont {Rothenbach},
  \citenamefont {Eggert}, \citenamefont {Eschenlohr}, \citenamefont {Ollefs},
  \citenamefont {Gruner},\ and\ \citenamefont {Pentcheva}}]{BegumHudde2023S}%
  \BibitemOpen
  \bibfield  {author} {\bibinfo {author} {\bibfnamefont {V.}~\bibnamefont
  {Begum-Hudde}}, \bibinfo {author} {\bibfnamefont {T.}~\bibnamefont
  {Lojewski}}, \bibinfo {author} {\bibfnamefont {N.}~\bibnamefont
  {Rothenbach}}, \bibinfo {author} {\bibfnamefont {B.}~\bibnamefont {Eggert}},
  \bibinfo {author} {\bibfnamefont {A.}~\bibnamefont {Eschenlohr}}, \bibinfo
  {author} {\bibfnamefont {K.}~\bibnamefont {Ollefs}}, \bibinfo {author}
  {\bibfnamefont {M.~E.}\ \bibnamefont {Gruner}},\ and\ \bibinfo {author}
  {\bibfnamefont {R.}~\bibnamefont {Pentcheva}},\ }\bibfield  {title} {\bibinfo
  {title} {Nature of excitons in the $\mathrm{Ti}~L$ and $\mathrm{O}~K$ edges of x-ray absorption spectra in bulk ${\mathrm{SrTiO}}_{3}$ from a combined first principles and many-body theory approach},\ }\href
  {https://doi.org/10.1103/PhysRevResearch.5.013199} {\bibfield  {journal}
  {\bibinfo  {journal} {Phys. Rev. Res.}\ }\textbf {\bibinfo {volume} {5}},\
  \bibinfo {pages} {013199} (\bibinfo {year} {2023})}\BibitemShut {NoStop}%
\bibitem [{\citenamefont {Harrison}\ and\ \citenamefont
  {Straub}(1987)}]{Harrison1987S}%
  \BibitemOpen
  \bibfield  {author} {\bibinfo {author} {\bibfnamefont {W.~A.}\ \bibnamefont
  {Harrison}}\ and\ \bibinfo {author} {\bibfnamefont {G.~K.}\ \bibnamefont
  {Straub}},\ }\bibfield  {title} {\bibinfo {title} {Electronic structure and properties of d- and f-shell-metal compounds},\ }\href
  {https://doi.org/10.1103/PhysRevB.36.2695} {\bibfield  {journal} {\bibinfo
  {journal} {Phys. Rev. B}\ }\textbf {\bibinfo {volume} {36}},\ \bibinfo
  {pages} {2695} (\bibinfo {year} {1987})}\BibitemShut {NoStop}%
\bibitem [{\citenamefont {Haverkort}(2005)}]{Haverkort2005S}%
  \BibitemOpen
  \bibfield  {author} {\bibinfo {author} {\bibfnamefont {M.}~\bibnamefont
  {Haverkort}},\ }\emph {\bibinfo {title} {Spin and orbital degrees of freedom in transition metal oxides and oxide thin films studied by soft x-ray absorption spectroscopy}},\ \href {https://kups.ub.uni-koeln.de/1455/} {Ph.D.
  thesis},\ \bibinfo  {school} {Universit{\"a}t zu K{\"o}ln} (\bibinfo {year}
  {2005})\BibitemShut {NoStop}%
\bibitem [{\citenamefont {de~Groot}\ \emph {et~al.}(1990)\citenamefont
  {de~Groot}, \citenamefont {Fuggle}, \citenamefont {Thole},\ and\
  \citenamefont {Sawatzky}}]{deGroot1990S}%
  \BibitemOpen
  \bibfield  {author} {\bibinfo {author} {\bibfnamefont {F.~M.~F.}\
  \bibnamefont {de~Groot}}, \bibinfo {author} {\bibfnamefont {J.~C.}\
  \bibnamefont {Fuggle}}, \bibinfo {author} {\bibfnamefont {B.~T.}\
  \bibnamefont {Thole}},\ and\ \bibinfo {author} {\bibfnamefont {G.~A.}\
  \bibnamefont {Sawatzky}},\ }\bibfield  {title} {\bibinfo {title} {$2p$ x-ray absorption of $3d$ transition-metal compounds: An atomic multiplet
  description including the crystal field},\ }\href
  {https://doi.org/10.1103/PhysRevB.42.5459} {\bibfield  {journal} {\bibinfo
  {journal} {Phys. Rev. B}\ }\textbf {\bibinfo {volume} {42}},\ \bibinfo
  {pages} {5459} (\bibinfo {year} {1990})}\BibitemShut {NoStop}%
\bibitem [{\citenamefont {Mizokawa}\ and\ \citenamefont
  {Fujimori}(1996)}]{Mizokawa1996S}%
  \BibitemOpen
  \bibfield  {author} {\bibinfo {author} {\bibfnamefont {T.}~\bibnamefont
  {Mizokawa}}\ and\ \bibinfo {author} {\bibfnamefont {A.}~\bibnamefont
  {Fujimori}},\ }\bibfield  {title} {\bibinfo {title} {Electronic structure and orbital ordering in perovskite-type $3d$ transition-metal oxides studied by Hartree-Fock band-structure calculations},\ }\href
  {https://doi.org/10.1103/PhysRevB.54.5368} {\bibfield  {journal} {\bibinfo
  {journal} {Phys. Rev. B}\ }\textbf {\bibinfo {volume} {54}},\ \bibinfo
  {pages} {5368} (\bibinfo {year} {1996})}\BibitemShut {NoStop}%
\end{thebibliography}
\end{document}